\documentclass[prd,aps,10pt,a4paper,showpacs,superscriptaddress,nofootinbib,nobibnotes,floatfix,showkeys,notitlepage]{revtex4-1}

\usepackage{graphicx}
\usepackage{bm}

\graphicspath{{IMG/}{FIGS/}{W/}}
\newcommand{\OBS}{{\rm obs}}
\newcommand{\FO}{{\rm FO}}
\newcommand{\FF}{{\rm ff}}
\newcommand{\SHELL}{{\rm S}}

\newcommand{\NULL}{{\rm null}}
\newcommand{\MAX}{{\rm max}}
\newcommand{\RAD}{{\rm rad}}
\newcommand{\MIN}{{\rm min}}

\newcommand{\STAT}{{\rm stat}}
\newcommand{\EXTR}{{\rm extr}}
\newcommand{\EE}{{\rm e}}
\newcommand{\SSS}{{\rm s}}
\newcommand{\CRIT}{{\rm crit}}
\newcommand{\TURN}{{\rm turn}}
\newcommand{\HOR}{{\rm hor}}

\newcommand{\PER}{{\rm per}}

\newcommand{\IN}{{\rm in}}
\newcommand{\OUT}{{\rm out}}

\newcommand{\ddd}{{\rm d}}
\newcommand{\eqdef}{:=}

\begin{document}

\title{Seeing relativity -- II. Revisiting and visualizing the
  Reissner-Nordstr\"om metric}

\author{Alain Riazuelo} 

\affiliation{Sorbonne Universit\'e, CNRS, UMR~7095, Institut
  d’Astrophysique de Paris, 98 bis boulevard Arago, 75014~Paris,
  France}

\email{riazuelo@iap.fr}

\begin{abstract}
  In this paper we study some features of the Reissner-Nordstr\"om
  metric both from an analytic and a visual point of view. We perform
  an accurate ray tracing and study of null geodesics in various
  situations. Among the issues we focus on are (i) the comparison with
  the Schwarzschild case, (ii) the naked singularity case, where, if
  the electric charge is not too large, some dark shell appears on
  images despite there is no horizon in the metric, and (iii) the
  wormhole crossing case, i.e., a visual exploration of the maximal
  analytic extension of the metric.
\end{abstract}

\date{1st December 2018}



\maketitle

\section{Introduction}

The Reissner-Nordstr\"om (hereafter RN) metric is the second exact
spherically symmetric solution to the Einstein equation that has been
discovered independently by Hans Reissner~\cite{reissner16} and Gunnar
Nordstr\"om~\cite{nordstrom18}, soon after the discovery of General
Relativity. It describes the gravitational field of a pointlike mass
$M$ possessing an electric charge $Q$. It is therefore the
generalization of the Schwarzschild metric to the case of an
electrically charged body. An exhaustive study of this metric has
already been performed by various authors (see,
e.g.,~\cite{chandrasekhar83}), but most of these studies lacked to
make an emphasis on the visual aspect of black hole, in contrast with
the Schwarzschild (together with Kerr) metrics which have deserved a lot
of attention because of their obvious astrophysical interest, see,
e.g. Refs.~\cite{fukue88,viergutz93,marck95,fanton97,falcke00,hamilton04,beckwith05}.
These early papers simulated the view of the black hole and its
accretion disk seen by a distant observer and were then transformed
into what could be observed by some astronomical device, i.e.,
converted into a single pixel, with spectral and photometric
informations. From an observational point of view, only two black
holes exhibit a sufficiently large angular size to allow to go beyond the
single pixel threshold: Sgr~A*, the supermassive black hole of our
Galaxy, and that of M87. For the latter, the much larger distance is
almost compensated by a much larger mass and both are expected to have
an angular diameter of order of 50~$\mu$as. However, even in this
case, only two dedicated instruments, the event Horizon
Telescope~\cite{eht} which observes at millimetric wavelengths, and
GRAVITY~\cite{gravity17}, in the infrared K~band, are able to directly
probe structures of size comparable of the black hole size, either by
direct imaging in the case of the Event Horizon Telescope, or by
astrometry in the case of GRAVITY, both having given some promising
results recently~\cite{eht12,eht18,gravity18a,gravity18b}.

In contrast with the Schwarzschild and Kerr metrics, the
Reissner-Nordstr\"om solution did not receive much attention are is
rarely studied on an equal footing with respect to the former
(Ref.~\cite{chandrasekhar83} being an exception in this respect). This
is especially the case regarding its visual aspect, despite the fact
that this metric exhibits a much larger variety of phenomena than the
Schwarzschild metric and is technically simpler (because of spherical
symmetry) than the Kerr metric. One may argue that since RN~black
holes do not exist in nature, or, more precisely, black hole
electrical charges are expected to be so small that any astrophysical
charged black hole is unlikely to exhibit features that are related to
its charge. This is likely to be true, however, we argue that
understanding any exact solution of general relativity deserves to be
studied and their unlikeliness should not preclude us from doing
so. Moreover, visualizing space-time is one of the best tools to
understand some of their properties~\cite{muller08,muller10,muller15},
and the more complicated the metric is, the more useful will the
visualization tools be.  This paper is therefore aimed at filling this
gap, as well as exploring the variety of phenomena that arise when one
considers the full set of parameters of the metric as well as its
maximal analytic extension.

This paper is organized as follows. In \S\ref{sec_metr}, we recall the
form of the metric and perform a classification of null geodesics as a
function of the charge-to-mass ratio of the RN~metric. In
\S\ref{sec_rayt}, we describe the tools that we need to perform our
simulations. We then perform (\S\ref{sec_feat}) a simple analytical
study of the features that are expected to be seen in this metric by
comparing with the Schwarzschild case.  We then describes three
interesting aspects related to the Reissner-Nordstr\"om metric. First
(\S\ref{sec_naked}), we explain the visual aspect of a naked
singularity in such metric. Secondly (\S\ref{sec_darkshell}), we
describe the visual consequences of bounded null trajectories in the
absence of any horizon. Thirdly, we simulate in \S\ref{sec_crosswh} the
crossing of a Reissner-Nordstr\"om wormhole which is significantly
more complicated than the more standard (but no less unrealistic)
horizonless Morris-Thorne wormhole.

\section{Description of the metric and classification of null
  geodesics}
\label{sec_metr} 

\subsection{A few notations} 

The Reissner-Nordstr\"om line element can be written in a diagonal form
\begin{equation}
\label{def_RN}
\ddd s^2
 =   A(r) \ddd t^2 - \frac{1}{A(r)} \ddd r^2
   - r^2(\ddd \theta^2 + \sin^2 \theta \ddd \varphi^2) ,
\end{equation}
where the function $A(r) = g_{tt}$ is defined as
\begin{equation}
\label{def_ARN}
A(r) = 1 - \frac{2 M}{r} + \frac{Q^2}{r^2} ,
\end{equation}
and where we consider a unit system in which $c = G = 1$, as well as
$4 \pi \varepsilon_0 = 1$. The quantity $M$ represents as usual the
mass of the black hole and $Q$ its electrical charge, both in general
relativistic units. The fact that $Q$ is the electric charge of the
black hole comes from the fact that the electromagnetic vector
potential $A_\mu$ has a non zero time component given by
$A_t = Q / r$.  If we use correct units, the actual charge $\tilde q$,
expressed in Coulombs, relates to the length-normalized charge $Q$
through the formula
\begin{equation}
  \frac{Q}{1\;{\rm m}} = 8.6 \times 10^{-18} \frac {\tilde q}{1\;{\rm C}} .
\end{equation}
Standard arguments (see~\cite{zajacek18}) say that an astrophysical,
non rotating black hole in unlikely to have a charge larger than
$200 (M / M_\odot)\;{\rm C}$, since this would imply that the
electrostatic repulsion between a positively charged black hole and a
proton would be larger than their gravitational attraction. Such bound
translates into $|Q| / M < 6 \times 10^{-21}$. This bound can be
raised by six orders of magnitude by considering a spinning
(Kerr-Newmann) black hole because in the presence of a magnetic field,
the frame dragging effect twists the magnetic field and allows for a
larger (necessarily positive) charge, but this is by insufficient to
have a non negligible $|Q| / M$ ratio. This is the reason why the
Reissner-Nordstr\"om metric has not gotten much attention from an
astrophysical point of view, and some of its properties are still
underexplored (for example, papers like
Refs.~\cite{pugliese11,pugliese17} could in principle have been
written decades ago).

The coordinates $(t, r, \theta, \varphi)$ reduce to the spherical
coordinates of the Minkowski space when $r$ is sufficiently large,
i.e., the metric is asymptotically flat. Moreover, being of diagonal
form with the function $A$ depending only on the $r$ coordinate, this
metric is static everywhere $A (r)$ is positive. It possesses horizons
whenever $A(r) = 0$, which happens only when $M \leq |Q|$ at
$r = r_\pm$, with
\begin{equation}
r_\pm = M \pm \sqrt{M^2 - Q^2} .
\end{equation}
For any non zero value of $M$ or $Q$, the region $r = 0$ is a
curvature singularity. Consequently, one has a black hole when
horizon(s) surround the singularity, i.e., when $M \geq |Q|$. In the
opposite case ($M < |Q|$), one has a naked singularity, whose visual
aspect shall be studied in the next sections together with the black
hole case.

\subsection{Classifying null geodesics}
\label{sec_class}

Being spherically symmetric, any test particle of four-velocity or
four-momentum experiences a planar geodesic motion in the metric, so
that we can reduce our analysis to the case where the particle is
confined within the plane $\theta = \pi / 2$. Also, the metric being
static, one can extract two constants of motion for a massive test
particle of four-velocity $u^\mu$ or massless particle of
four-wavevector $k^\mu$. Those are
\begin{itemize}

\item The particle total energy per unit of mass or its total energy
  $E \eqdef g_{\mu t} u^\mu$, or $E \eqdef g_{\mu t} k^\mu$

\item The particle projected angular momentum per unit of mass, or its
  projected total angular momentum,
  $L \eqdef - g_{\mu \varphi} u^\mu$, or
  $L \eqdef - g_{\mu \varphi} k^\mu$.

\end{itemize}

From the fact that the particle four-velocity is of constant norm
$\kappa$ ($\kappa = 1$ for massive particles, and $0$ for massless
particles), one has
\begin{equation}
\kappa = \frac{E^2 - \dot r}{A(r)} - \frac{L^2}{r^2} ,
\end{equation}
where a dot is denoting a derivative of the particle coordinates with
respect to an affine parameter $p$ (i.e.,
$u^\mu, k^\mu = \ddd x^\mu / \ddd p$). Consequently, if one considers
only null geodesics, which we shall do from now on, one has
\begin{equation}
\label{eq_rdot}
E^2 - A(r) \frac{L^2}{r^2} = \dot r^2 ,
\end{equation}
and the radial motion of the particle corresponds formally to that of
an abstract massive particle experiencing an effective potential
$V_\NULL(r)$ (normalized to its mass) of the form
\begin{equation}
\label{Vnull}
V_\NULL(r) = \frac{1}{2} \frac{L^2}{r^2} A(r)
 =   \frac{1}{2} \frac{L^2}{r^2} 
   - \frac{M L^2}{r^3} + \frac{1}{2} \frac{Q^2 L^2}{r^4} ,
\end{equation}
and endowed with a total energy per unit of mass
${\cal E} \eqdef E^2 / 2$.

Except in the case where $L = 0$, i.e., a purely radial motion of the
real particle, the function $V_\NULL(r)$ is positive and diverges when
$r$ tends to $0$, which means that $r = 0$ is never reached by any
particle, unless $L = 0$ (radial null geodesic). Also, $V_\NULL(r)$
decreases both when $r$ is sufficiently small or large (all powers of
$r$ involved are negative and the coefficients of the largest and
smallest powers of $r$ are positive), so that any particle starting
from infinity will bounce at some point on the potential and go back
to infinity. But because of the negative term $- M L^2 / r^3$, the
potential $V_\NULL$ is not necessarily always decreasing and may
possess a local minimum, which implies that some particles may be
trapped locally within the potential. This can happen only when there is
actually a local minimum in $V_\NULL(r)$. This local minimum exists if
there are two roots to the equation $\ddd V_\NULL / \ddd r = 0$, which
corresponds to
\begin{equation}
\label{eq_rextr}
(r_\EXTR^\pm)^2 - 3 M r_\EXTR^\pm + 2 Q^2 = 0 .
\end{equation}
Such roots exist where the discriminant $9 M^2 - 8 Q^2$ is positive,
which happens for any value of $M \geq |Q|$ (black hole case) as well
as a small interval $M < |Q| < 3 M / 2 \sqrt{2}$ in the naked
singularity case. The first of these two situations was to be expected
since $V(r)$ is proportional to $A(r)$ and thus possesses zeroes at
$r = r_\pm$, which means that it has at least one local extremum
(actually a minimum) in between those two values.

\begin{figure}[htbp]
\includegraphics*[angle=270,width=3.2in]{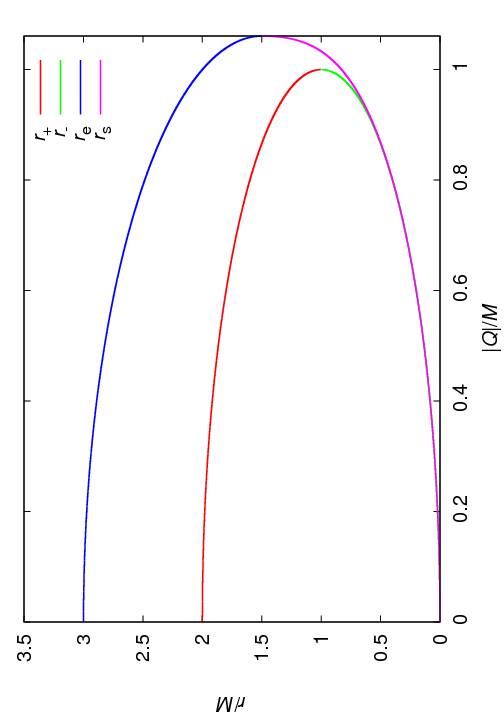}
\includegraphics*[angle=270,width=3.2in]{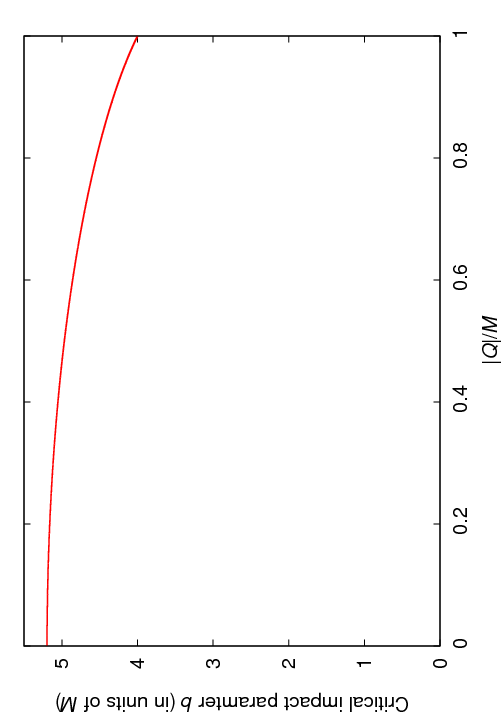}
\caption{[Left panel] The inner and outer horizons, $r_-$ (green) and
  $r_+$ (red), as well as the inner and outer limit of the dark shell
  phenomenon described in the next Section, $r_\SSS$ (pink) and
  $r_\EE$ (blue), as a function of charge $|Q|$. As explained in the
  Appendix, $r_\SSS$ is extremely close to $r_-$ for low or even
  moderate values of $|Q|$ because the potential is extremely steep
  around $r_-$, whereas $V_\NULL(r_\SSS) \leq L^2 / 27$ is close to
  $V_\NULL(r_-) = 0$. [Right panel] The critical impact parameter of
  null geodesics delineating the black hole angular size (see
  \S\ref{sec_feat} later). In a Newtonian approximation where light
  would not be deflected, a black hole of mass $M$ would be delineated
  by light ray with impact parameter equal to the black hole radius,
  that is, $r_+$.}
\label{fig_rrrr}
\label{fig_b_crit}
\end{figure}

When they exist, the roots of the potential derivative $V'_\NULL$ are
situated at
\begin{equation}
r_\EXTR^\pm = \frac{3 M \pm \sqrt{9 M^2 - 8 Q^2}}{2} .
\end{equation}
When $r_\pm$ exist, since $V$ is decreasing for small $r$ and reaches
$0$ for $r = r_\pm$, it is clear that its minimum has to lie between
$r_-$ and $r_+$ and that its maximum is larger than $r_+$, so that one
has $r_- < r_\EXTR^- < r_+ < r^+_\EXTR$ as illustrated in
Fig.~\ref{fig_rrrr}. Because, in the formal analogy, the abstract
particle has a positive energy, so must be $V(r)$ when $\dot r = 0$.
But because the turning points of the trajectory must surround the
local minimum of $V_\NULL$, i.e., $r_\EXTR^-$, one of these turning
point must be smaller than $r_-$, whereas the other is necessary
larger than $r_+$ (since $V_\NULL$ must be positive at any turning
point). The largest extent of those bound geodesics are obtained when
the far turning point of the trajectory is at
$r_\SHELL^+ \eqdef r_\EXTR^+$, in which case the other turning point
of the trajectory is the only other $r_\SHELL^-$ such that
$V_\NULL(r_\SHELL^-) = V_\NULL(r_\EXTR^+)$. After some algebra, we
find that
\begin{equation}
\label{rshellmoins}
r_\SHELL^-
 = r_\EXTR^+ \left( -1 + \frac{1}{\sqrt{1 - \frac{Q^2}{M r_\EXTR^+}}}\right) ,
\end{equation}
which we also show on Fig.~\ref{fig_rrrr}.  (A proof of
Eq.~(\ref{rshellmoins}) is given in the Appendix.) Because the
potential is very steep around $r_\SHELL^-$ and because
$V(r_\SHELL^-) - V(r_-)$ is never very large, $r_\SHELL^-$ and $r_-$
are usually very close to each others.  As an example, for
$Q / M = 0.5$, we have $r_\EXTR^+ / M = \frac{3 + \sqrt{7}}{2}$, from
which we obtain
$r_\SHELL^- / M = \frac{3 + \sqrt{7}}{2} \left(-1 + \sqrt{\frac{2
      (\sqrt{7} - 1)}{3}}\right) \simeq 0.133967$,
whereas $r_- / M = 1 - \sqrt{3}/2 \simeq 0.133974$, that is a 0.0056\%
difference. We shall come to the observational consequences of this
fact in \S\ref{sec_crosswh}.  In the discussion that follows the value
$r_\EXTR^+$ and $r_\SHELL^-$ will be often needed, therefore, we shall
use simplified notations $r_\EE \eqdef r_\EXTR^+$ and
$r_\SSS \eqdef r_\SHELL^-$.

The above discussion allows us to classify null geodesics into three
categories:
\begin{enumerate}

\item {\bf Unbound null geodesics of the first kind}. These geodesics
  start from infinity, then bounce on the potential at some $r$
  greater than $r_\EE$ and go back to infinity. Those are the analog
  of unbound null geodesics in the Schwarzschild metric which exist as
  soon as their impact parameter $b \eqdef |L| / E$ is larger than
  $3 \sqrt{3} M$.

\item {\bf Unbound null geodesics of the second kind}. These geodesics
  start from infinity, cross the two horizons, then bounce on the
  potential at some $r$ smaller than $r_-$, then cross again the two
  horizons and go back to infinity. No such analog of those exist in
  the Schwarzschild metric since in this case, geodesics crossing the
  horizon originate from or end up at the singularity.

\item {\bf Bound null geodesics}. These geodesics are trapped. When
  the metric possesses horizons, these geodesics cross them twice per
  cycle. In the other case, their are bound in the same (and only)
  asymptotic region that forms the metric.

\end{enumerate}

The existence of each of these geodesics depend both on the existence
of horizons (for bound geodesics and unbound geodesics of the second
kind) and the fact that $V(r)$ is monotonous or not (for bound
geodesics only). Given the previous discussion regarding these
properties, we can summarize which values of the parameters (here, the
ratio $|Q| / M$) allow or not the existence of each type. This is done
in Table~\ref{table_class_geod}.
\begin{table}
\begin{tabular}{|c|c|c|c|}
\hline
Geodesic type & $|Q| / M \leq 1$
              & $1 < |Q| / M < 3 / 2 \sqrt{2} $ 
              & $3 / 2 \sqrt{2} \leq |Q| / M$ \\ \hline
Unbound of the first kind & Yes & Yes & Yes \\ \hline
Unbound of the second kind & Yes & No & No \\ \hline
Bound & Yes & Yes & No \\ \hline
\end{tabular}
\caption{Type of geodesics that exist as a function of the $|Q| / M$ ratio.}
\label{table_class_geod}
\end{table}

\section{Ray tracing within a Reissner-Nordstr\"om metric}
\label{sec_rayt}

If we want to perform some ray tracing in any metric, we need to
compute the geodesic equation of null geodesics and compute the
structure of the null geodesics that cross the observer's worldline.

Since the Reissner-Nordstr\"om metric is spherically symmetric, null
geodesics depend only on one parameter, which, for example, may
correspond to the ratio $b \eqdef |L| / E$, i.e., the impact factor of
the geodesic at null infinity (in case it reaches it; we shall however
keep the name even when it is not the case). We shall therefore closely
follow the technique explained in Ref.~\cite{riazuelo15}.

First, we need to write the geodesic equation. As long as one can stay
within the $(t, r, \theta, \varphi)$ coordinates, i.e., as long as
there is no horizon crossing, the geodesic equation reads
\begin{eqnarray}
\label{rrn_deb}
\frac{\ddd k^t}{\ddd p} & = & - \frac{A'}{A} k^r k^t ,\\
\frac{\ddd k^r}{\ddd p} & = &
 - \frac{1}{2} A A' (k^t)^2 + \frac{1}{2} \frac{A'}{A} (k^r)^2 
 + A r \left( (k^\theta)^2 + \sin^2 \theta (k^\varphi)^2 \right),\\
\frac{\ddd k^\theta}{\ddd p} & = &
 - \frac{2}{r} k^r k^\theta + \sin \theta \cos \theta (k^\varphi)^2,\\
\label{rrn_fin}
\frac{\ddd k^\varphi}{\ddd p} & = &
 - \frac{2}{r} k^r k^\varphi 
 - 2 \frac{\cos \theta}{\sin \theta} k^\theta k^\varphi ,
\end{eqnarray}
where $k^\mu$ is the particle four-velocity or four-wavevector, $p$ is
an affine parameter, i.e., $k^\mu = \ddd x^\mu / \ddd p$ and where a
prime denotes a derivative with respect to $r$. This set of eight
equations has exactly the same structure as that of the Schwarzschild
metric, except that the function $A (r) = g_{tt}$ is not the same. As
for the Schwarzschild metric, it can be reduced to a set of six
equations in case one assumes that $\theta = \pi / 2$, and can even be
simplified further by introducing the constants of motion $E$ and
$L$. In any case, a standard $4^{\rm th}$ order Runge-Kutta method
such as the one in Ref.~\cite{press92} is sufficient to solve it.

\subsection{Defining fundamental observers}

Although what an observer sees depends both on its position and
velocity, we have shown in Ref.~\cite{riazuelo15} that for this
purpose, it suffices to solve the geodesic equation in a plane for a
fiducial observer that shares the same position but not the same
velocity as the true observer, since any direction of observation of
the true observer can be matched to another direction of the fiducial
observer by performing a Lorentz transform. We therefore have to
choose a class of observers that is defined everywhere in the metric.

As long as one is outside the horizon, a natural choice is a static
observer, whose four-velocity is given by
\begin{equation}
\label{def_u_stat}
u_\STAT^\mu = \left(\begin{array}{c} \frac{1}{\sqrt{A(r)}} \\ 
                                    0 \\ 0 \\ 0 
                   \end{array} \right) .
\end{equation}
However this is insufficient to explore the whose metric since they
are not defined between the two horizons where $A(r) < 0$. Therefore,
we need a second class of observers which are freely falling onto the
black hole with a zero velocity and zero angular momentum at
infinity. We can define both ingoing and outgoing observers of
respective velocity $u_{\FF,-}^\mu$ and $u_{\FF,+}^\mu$. These
observer's radial coordinate velocity $\ddd r / \ddd \tau$ is same as
that of fiducial, Newtonian observers who would have zero total energy
and who would experience a radial effective potential defined as
\begin{equation}
\label{V_red_obs}
V_\RAD^\OBS (r) \eqdef - \frac{M}{r} + \frac{1}{2} \frac{Q^2}{r^2} .
\end{equation}
Such observers shall only reach a radial coordinate distance
$r^\FF_\MIN = Q^2 / 2$. Obviously, when $|Q| < M$, we have
$r^\FF_\MIN < r_-$, so that $r^\FF_\MIN$ is situated within the inner
horizon (when it exists), and where $t$ is timelike. The freely
falling observer velocity's components are then.
\begin{equation}
\label{def_u_ff}
u_{\FF, \pm} = \left(\begin{array}{c} \frac{1}{A(r)} \\ 
                                    \pm \sqrt{1 - A(r)} \\ 0 \\ 0 
                   \end{array}
             \right) .
\end{equation}
None of these classes of observers are defined everywhere on the
manifold, so that we shall define everywhere a class of fundamental
observer by mixing the two already defined. Except in one specific
case that arise when studying the maximal analytic extension of the
metric, we choose fundamental observers' velocity $T_\FO^\mu$ as
\begin{eqnarray}
T_\FO^\mu & = & u_{\FF, -}^\mu
                 \qquad {\rm for}\;r > r_\MIN^\FF , \\
T_\FO^\mu & = & u_\STAT^\mu
                 \qquad {\rm for}\;r \leq r_\MIN^\FF .
\end{eqnarray}
We shall then define an orthonormal tetrad
$(T_\FO^\mu, R_\FO^\mu, \Theta^\mu, \Phi^\mu)$ so that $\Theta^\mu$
and $\Phi^\mu$ are only spanned by $\partial / \partial \theta$ and
$\partial / \partial \varphi$, respectively. Those two last vectors do
not depend on whether one considers freely falling or static
observers, but the other vector, $R^\mu_\FO$, does , and is
consistently chosen as
\begin{equation}
R_{\FF,\pm}^\mu = \left(\begin{array}{c} \mp \frac{\sqrt{1 - A(r)}}{A(r)} \\ 
                                           1 \\ 0 \\ 0 
                          \end{array}
                    \right) \qquad,\qquad
\label{def_r_stat}
R_\STAT^\mu = \left(\begin{array}{c} 0 \\ \sqrt{A(r)} \\ 0 \\ 0 
                     \end{array}
               \right),
\end{equation}
and we shall define $R_\FO^\mu$ either by $R_{\FF,\pm}^\mu$ or
$R_\STAT^\mu$ under the same conditions as for $T_\FO^\mu$. In what
follows, we shall drop the $\FO$ subscript unless it brings some
confusion.

If we consider a null geodesic passing at some coordinate $r$, we can,
without loss of generality, consider that case where it propagates in
the horizontal plane $\theta = \pi / 2$, therefore the null geodesic
four-wavevector $k^\mu$ can be written
\begin{equation}
k^\mu = \omega_\FO (T^\mu + \cos \delta R^\mu + \sin \delta \Phi^\mu) ,
\end{equation}
where $\omega_\FO$ is the angular frequency of the corresponding wave
measured by the observer. Then, there exists an unambiguous relation
between $\delta$ and the geodesic impact parameter $b \eqdef L / E$
because we have in addition
\begin{eqnarray}
\label{defL}
L & = & \omega_\FO r \sin \delta , \\
\label{defE}
E & = & k^\mu g_{\mu t} = \omega_\FO (T_t + R_t \cos \delta) .
\end{eqnarray}
As a consequence, any null geodesic originating from to infinity can be
traced back from the observer once one computes by how much a geodesic
passing at the observer's position and making an angle $\delta$ with
respect to the radial direction has been deflected prior to reaching
the observer, see Ref.~\cite{riazuelo15} for details. In other words,
all the information regarding null geodesics seen from a given
coordinate distance $r$ is encoded within a deviation function
$\varphi_\infty (\delta)$, where $\delta$ is, as explained above, the
angle the geodesic makes with respect to the radial direction, and
where $\varphi_\infty$ is the angle with respect to the same radial
direction it originates from. In case we consider a static observer
and if there is no deviation at all, then $\phi_\infty$ differs from
$\delta$ by $\pi$ as a straight trajectory expressed in polar
coordinates originated from the opposite direction it is heading
toward. Just as in the Schwarzschild case, once this function is
computed, the deformation along the whole celestial sphere and/or for
all pixel of some image can be computed by the rather simple steps:
\begin{enumerate}
\item Associating to the observed pixel a direction, $N^\mu$, i.e., a
  unit spacelike vector orthogonal to the observer's velocity,
  $u^\mu_\OBS$. The corresponding null geodesic possesses therefore
  four-wavevector given by $k^\mu = \omega_\OBS (u_\OBS^\mu - N^\mu)$;

\item Projecting the four-wavevector on the fundamental observer's
  tetrad, i.e. rewriting $k^\mu$ under the form
  $k^\mu = \omega_\FO (T^\mu + N'^\mu)$;

\item Defining the plane containing the radial direction $R^\mu$ and
  $N'^\mu$  in the vicinity of the observer. This plane contains
  the whole geodesic and is spanned by $R^\mu$ and $N'^\mu$;

\item Computing the angle $\delta$ between $R^\mu$ and $N'^\mu$ and
  interpolating the deviation function so that to find the angle
  $\varphi_\infty$ with respect to $R^\mu$ from which the geodesic
  originates from;

\item Deducing which direction of the celestial sphere the geodesic
  originates from.

\end{enumerate}

\section{A few rather simple features of the Reissner-Nordstr\"om
  metric}
\label{sec_feat}

The procedure outlined above does not differ from the one detailed
in~\cite{riazuelo15}, only the function $\varphi_\infty (\delta)$ 
does, but this leads to many differences in the behaviour of the
metric.

Firstly, the function $-A' / 2 = - M / r^2 + Q^2 / r^3$ is not
negative everywhere, which means that the coordinate center does not
always act as an attractive gravitational source. Rather, in the
vicinity of it, when $r < Q^2 / M$ it exhibits gravitationally
repulsive properties. An interpretation of this phenomenon is that a
Reissner-Nordstr\"om black hole mass is made by a ``bare'' infinite,
pointlike, negative mass plus a positive contribution of electrostatic
origin, whose contribution is also infinite but which is not localized
at the center of the coordinate system. Instead it is spread
everywhere (in classical physics, one would say that there is some
energy density proportional to the square of the electric field).  By
virtue of Gauss' theorem, the mass felt at some distance $r$ is given
of the bare mass plus the contribution of the electrostatic field
within the sphere of radius $r$. As one approaches the origin of the
coordinate system, this contribution decreases, and the apparent mass
may eventually become negative, thus behaving as a gravitationally
repulsive entity at small distance. Let us add that this repulsive
gravitational effect is probably never felt in practice. Indeed, the
region where it occurs lies at $r < Q^2 / M$, which puts it within the
inner horizon situated at $r = r_-$. Furthermore, the inner horizon
being a Cauchy horizon is unstable as pointed out in the sixties by
R.~Penrose~\cite{penrose68}~(see also~\cite{poisson90}), so that is
probably transforms into a singularity and no such region within the
inner horizon ever exists. Nevertheless, assuming that such metric can
exist, it exhibits many interesting features that we shall study
below.

As already mentioned, because of this repulsive gravitational
behaviour, almost no geodesic can reach the singularity at $r = 0$.
All timelike geodesics and any non radial ingoing null geodesic
bounces on a steep potential slope before reaching it and then escape
from the black hole (which is this case is not a black hole, but a
wormhole instead).

\begin{figure}[htbp]
\includegraphics*[width=3.2in]{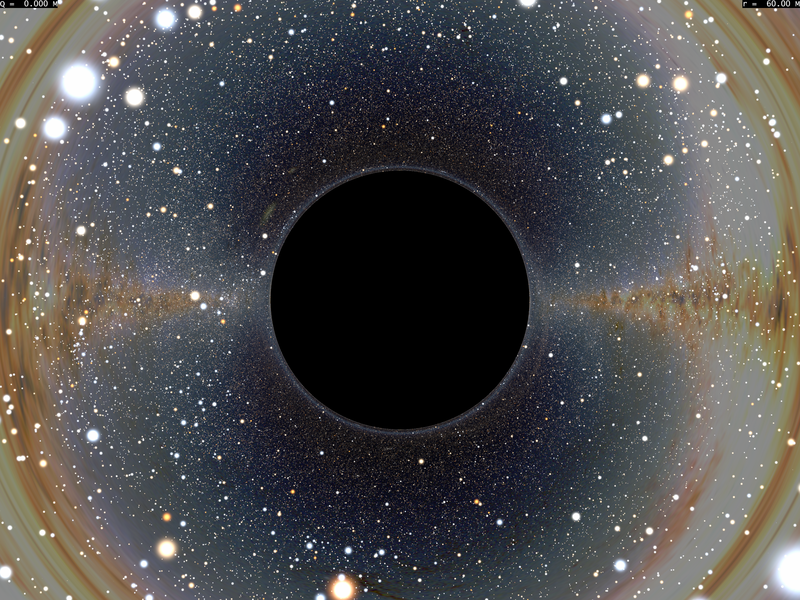}
\includegraphics*[width=3.2in]{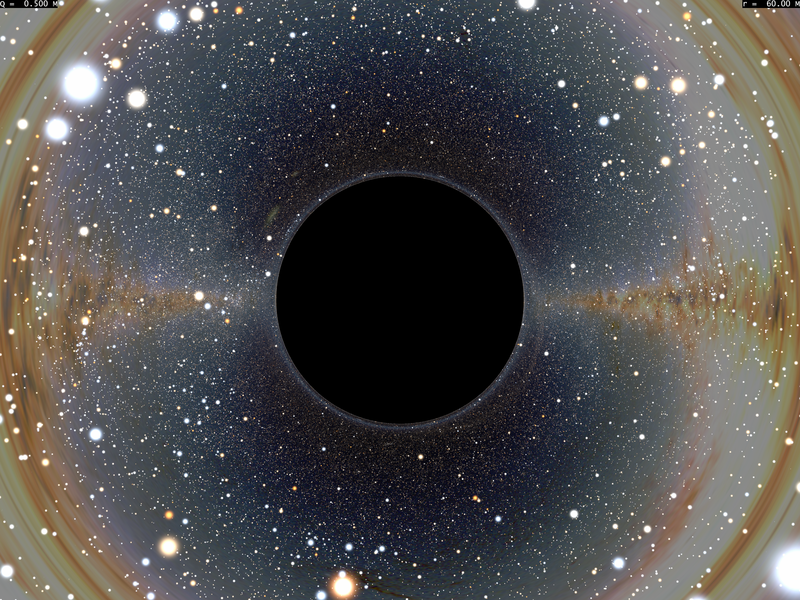}
\vskip 0.12cm
\includegraphics*[width=3.2in]{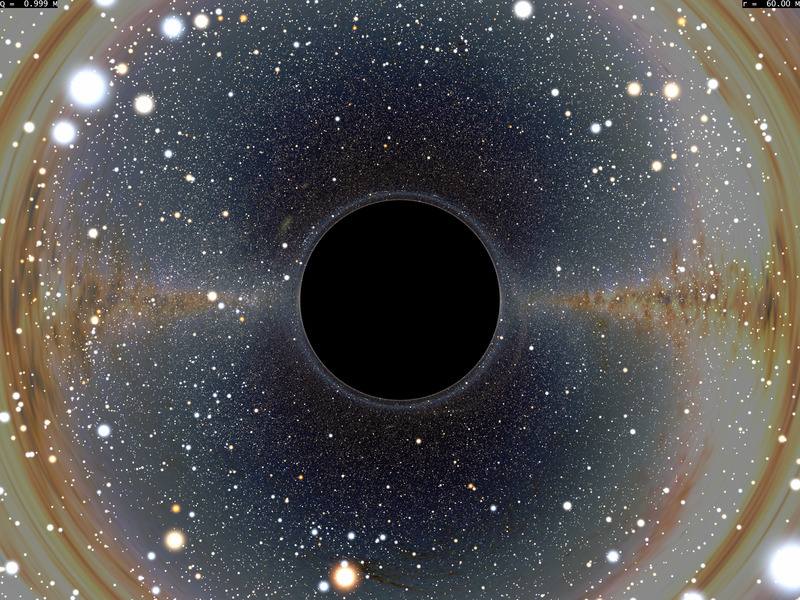}
\includegraphics*[width=3.2in]{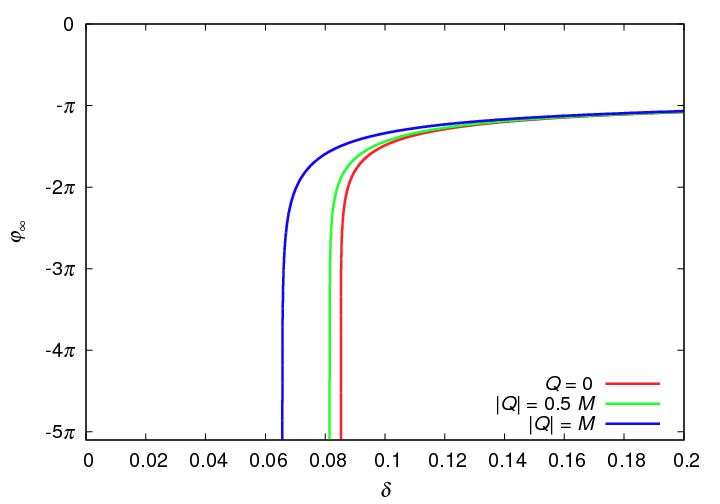}
\caption{Comparison between a Schwarzschild black hole (top left), an
  intermediate, $|Q|/M = 0.5$ (top right) and a nearly extremal
  $|Q|/M = 0.999$ Reissner-Nordstr\"om black hole (bottom left). Apart
  from the change in the angular size of the black hole silhouette,
  whose value is proportional to those given by Fig.~\ref{fig_b_crit},
  the three images all look qualitatively the same. All images show
  the distortion of a background star field corresponding to the Milky
  Way seen from Earth, toward Galactic coordinates
  $l = 355\,{\rm deg}, b = 0\,{\rm deg}$. Above the Einstein ring, one
  distinguishes a flattened version of the head and claws of Scorpius,
  and the bright two stars on the right of the pictures are $\alpha$
  and $\beta$ Centauri. Sagittarius is also visible (mostly under the
  Einstein ring) but far too deformed to be recognizable.  The bottom
  right graph shows the deviation functions for the three images. The
  asymptotes of each curves, which represents null geodesics which
  spent a long time orbiting close to $r_\EE$, correspond to the
  angular size of the black hole, which fit into the ratio predicted
  in Eq.~(\ref{eq_bcrit}).}
\label{fig_compq}
\end{figure}

Still, as long as one considers the exterior of a Reissner-Nordstr\"om
black hole, the differences with the Schwarzschild case are
quantitatively rather weak, as shown in Fig.~\ref{fig_compq}.
The reason for this qualitative similarity comes from two facts we
shall explain:
\begin{enumerate}
\item There exists a critical value of the impact parameter under
  which a null geodesic crosses the horizon;

\item The deviation function diverges as the impact factor reaches the
  critical value.
\end{enumerate}

The first point above is a direct consequence that, just in the
Schwarzschild case, the effective potential experienced by null
geodesics, $V_\NULL$, exhibits a local maximum for some $r$ larger
than the outer horizon. This maximum, which is at $r = 3 M$, is
shifted to $r = r_\EE$ in this Reissner-Nordstr\"om case. Moreover,
the corresponding extremal value of the effective potential, which was
$L^2 / 54 M^2$ in the Schwarzschild case, is shifted to 
\begin{equation}
\label{V_null_re}
V_\NULL^\EE \eqdef V_\NULL (r_\EE)
 = \frac{L^2 A(r_\EE)}{2 r_\EE^2}
 = \frac{L^2}{2} 
   \frac{r_\EE^3 - 2 M r_\EE + Q^2}
       {r_\EE^4} ,
\end{equation}
which can be rewritten in several way using the fact that $r_\EE$
satisfies Eq.~(\ref{eq_rextr}).

Among the null geodesics starting from infinity those that delineate
the silhouette of the black hole are those whose impact parameter is
given by
\begin{equation}
\label{eq_bcrit}
b_\CRIT^2 = \frac{L^2}{2 V_\NULL^\EE}
 = \frac{r_\EE^4}{r_\EE^2 - 2 M r_\EE + Q^2 } 
 = \frac{2 r_\EE^3}{r_\EE - M} ,
\end{equation}
where for the last equality we have used the fact that $r_\EE$ is
solution of Eq.~(\ref{eq_rextr}). Since the function
$f(x) = x^3 / (x - 1)$ is monotonous increasing for $x \geq 1.5$,
$b_\CRIT$ is an increasing function of $r_\EE$ (since its smallest
value is $3 M / 2$), and is therefore a decreasing function of
$|Q|/M$. Its maximum value is $3 \sqrt{3} M$ for the Schwarzschild
case. It decreases to $4 M$ for the extremal case and further to
$3 \sqrt{3} M / 2$ for the largest value of $|Q| / M$ for which it is
defined. This is shown in second panel of Fig.~\ref{fig_b_crit}. The
fact that $b_\CRIT$ decreases as $|Q| / M$ increases means that a
charged black hole angular diameter appears smaller in the charged
case than in the uncharged one, although this decrease is not
proportional to the black hole coordinate radius $r_+$.

As for the second point, the divergence of the deviation experienced
by null geodesics of impact parameter $b_\CRIT$ arises from the fact
that the variation of the azimuthal angle of the geodesic with respect
to the radial coordinate is given by the well-known results (here
adapted of the Reissner-Nordstr\"om case),
\begin{equation}
\frac{\ddd \varphi}{\ddd r}
 = \frac{L}{r^2 \ddd r / \ddd p} = \frac{L}{\sqrt{r^4 E^2 - r^2 L^2 A (r)}}
 = \frac{L}{\sqrt{2 r^4 \left(V_\NULL^\EE - V_\NULL(r) \right) }} .
\end{equation}
In the neighbourhood of $r_\EE$, the term in the square root can be
expanded as $- (r_\EE)^4 (r - r_\EE)^2 V''_\NULL (r_\EE) / 2$, so that
the integral $\int _{r_\EE} \ddd \varphi$ diverges logarithmically. On
the other hand, null geodesics with impact parameter close to
$b_\CRIT$, one can expand the above equation in the vicinity of the
turning point that we shall note $r_\PER$
\begin{equation}
\frac{\ddd \varphi}{\ddd r} \sim
 \frac{L}{\sqrt{- r_\PER^4 (r - r_\PER) V'_\NULL (r_\PER) }} .
\end{equation}
The integral of the expression no longer diverges, and is proportional
to $r_\PER^{-2} \sqrt{ -V'(r_\PER)}$. The first term of this
expression is bounded by $r_\EE$ and hence finite, but the second one
can be made as small as possible, thus allowing for an arbitrarily
large deviation, which completes the proof. 

These two properties are not very interesting since they already existed
in the Schwarzschild metric. Things becomes more interesting either
when one crosses the horizon (because one becomes sensitive to the
inner part of the metric which very different from the Schwarzschild
case), or when there is no horizon at all. It is on this last case that
we shall focus now.

\section{Looking at a naked singularity}
\label{sec_naked}

When $|Q| > M$, there are no horizons and the singularity is
naked. However, this singularity is pointlike, and among null
geodesics only radial ones can originate from it. Consequently, the
naked singularity occupies only one pixel of the celestial sphere,
whose spectral properties are unknown since they arise in principle
from some quantum theory of gravity. We shall therefore ignore them
and assume that no photons emerge from this singularity.  

If $|Q| > (3 / 2\sqrt{2}) M$ or if $r \not \in ]r_\SSS, r_\EE[$ then
all null non radial geodesics crossing the observer's worldline
originate from null infinity and end at null infinity. Consequently,
the whole field of view, up to the pixel corresponding to the
singularity, shows the background celestial sphere. In the other case,
there are bound geodesics (see Table~\ref{table_class_geod}) which,
assuming that no photon fill any of those geodesics, will appear as a
dark, spherical shell (since the metric is spherically symmetric)
surrounding the singularity (see next Section).

\subsection{Distortion of the celestial sphere}

Regarding the deformation of the celestial sphere, the deviation
function is plotted at the bottom of Fig.~\ref{fig_QN_dev}. Contrarily
to the Schwarzschild case, it shows a maximum for some value of
$\delta$ since any null geodesic hits the effective potential when its
first derivative is non zero, which implies that deviation is always
finite. Moreover, the deviation decays to a finite value ($0$) for
$\delta = 0$.

\begin{figure}[htbp]
\includegraphics*[width=3.2in]{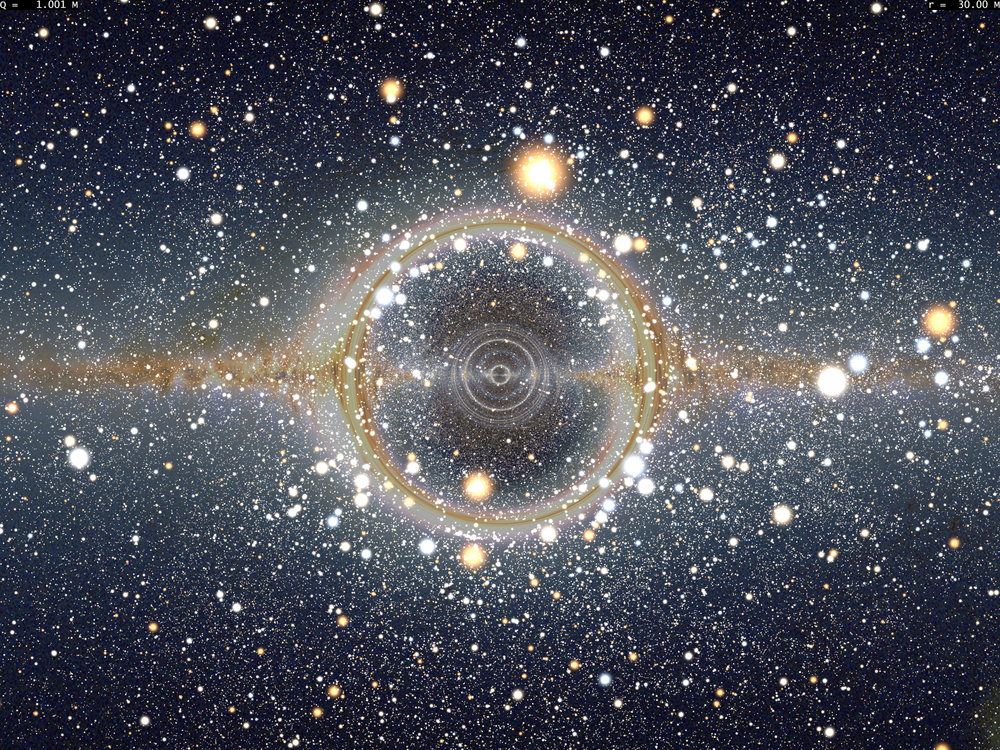}
\includegraphics*[width=3.2in]{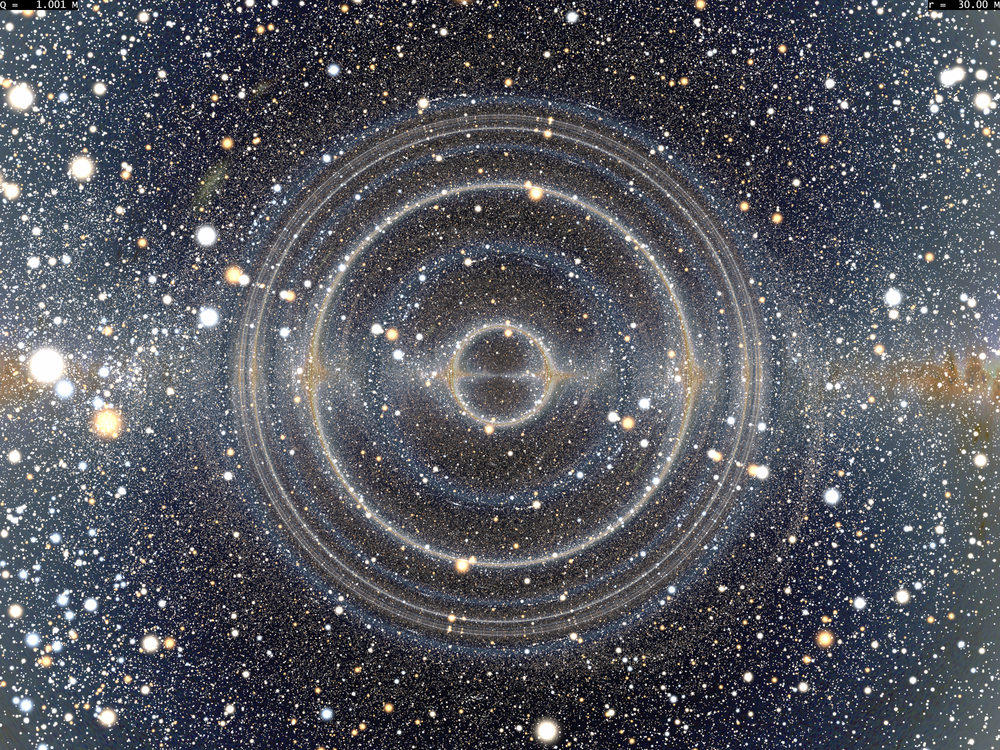}
\vskip 0.12cm
\includegraphics*[width=3.2in]{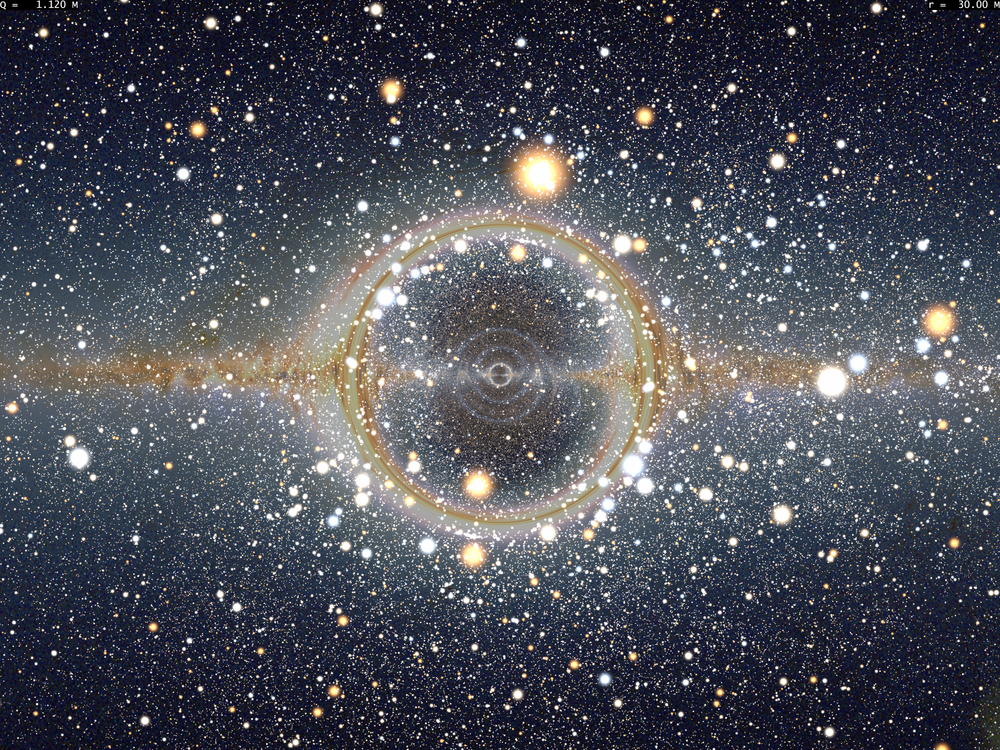}
\includegraphics*[width=3.2in]{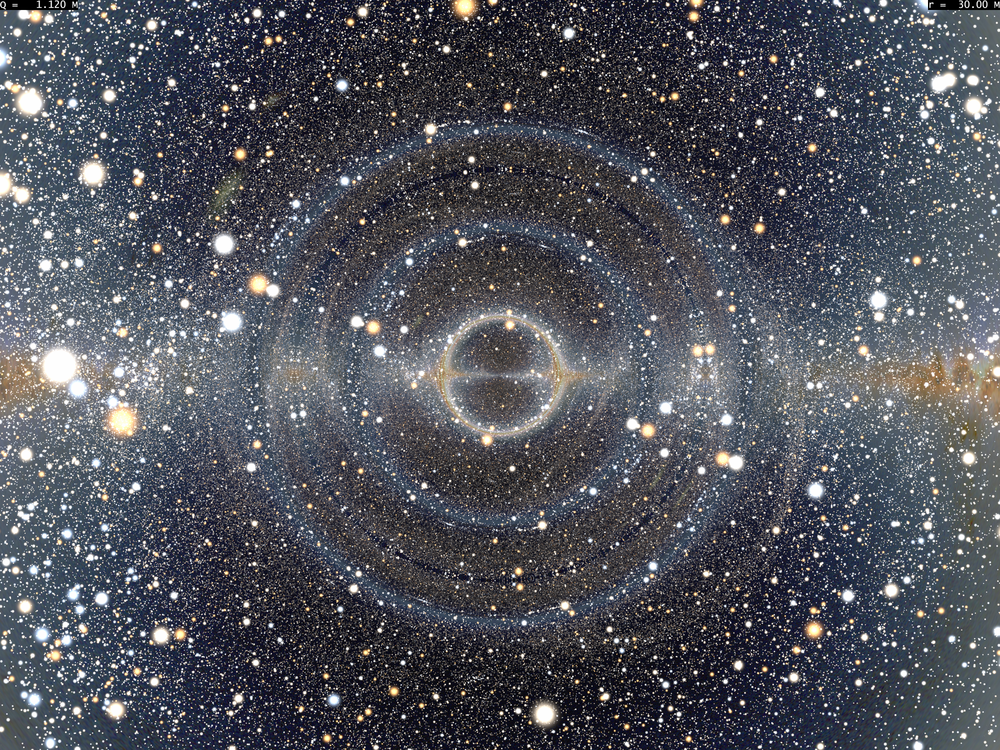}
\vskip 0.12cm
\includegraphics*[width=3.2in]{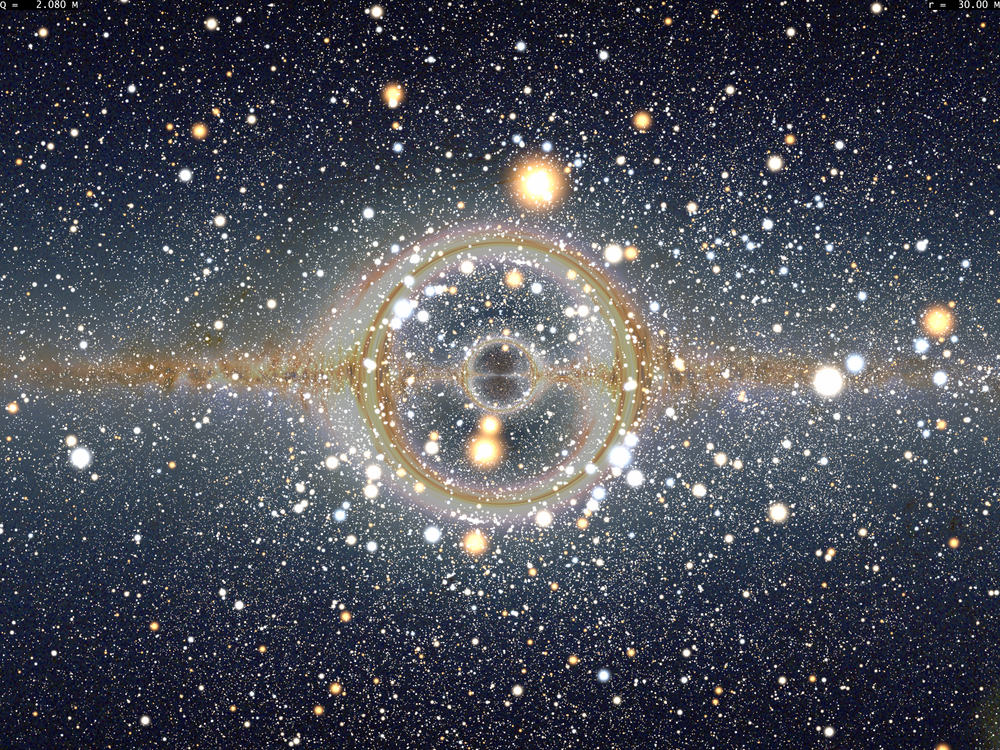}
\includegraphics*[width=3.2in]{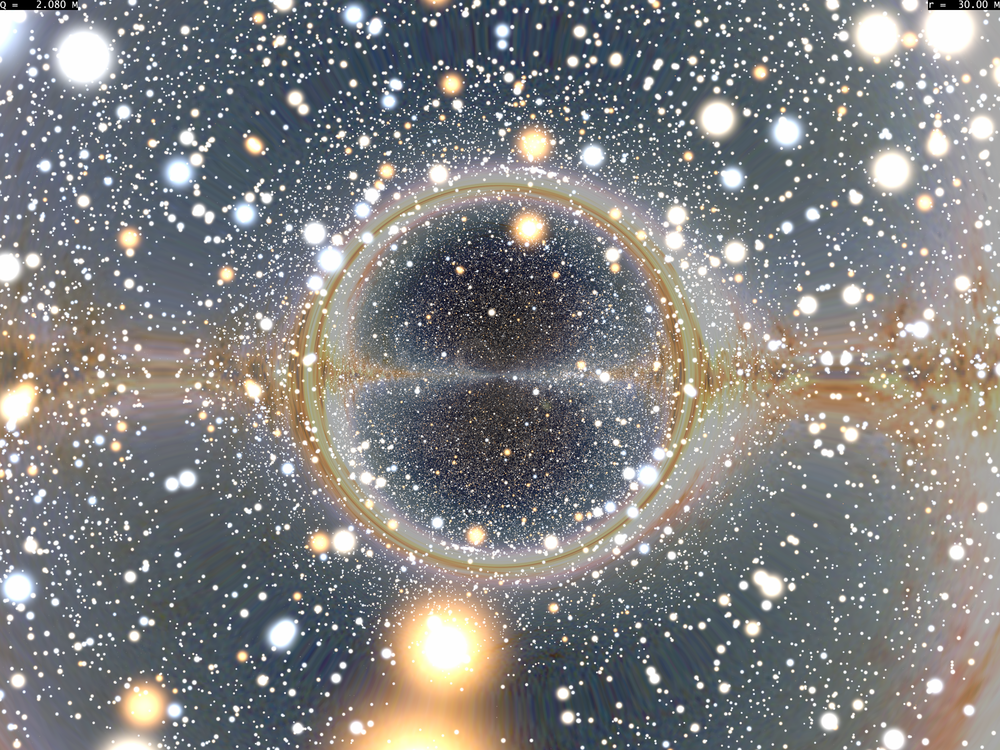}
\caption{Three views of a naked Reissner-Nordstr\"om singularity for a
  ratio $|Q|/M = 1.001$ (top), $1.12$ (middle) and $2.08$ (bottom)
  from some distance ($r = 30 M$). When the $|Q| / M$ ratio increases,
  the maximum amount of light deflection decreases, and the number of
  multiple images decreases as well as seen clearly on the series of
  three pictures. In order to outline the number of multiple images of
  any object, we have set the observer within the Galactic plane, so
  that copies of the Galactic disk are readily visible as bright
  circles.  The picture shows a region close to the Galactic center
  ($l = 355\,{\rm deg}, b = 0\;{\rm deg}$). Left view as seen at a
  large opening angle, whereas right views are shown with a $5 \times$
  zoom, showing better the decrease of multiple images due to the
  lesser extent of the deflection function. The corresponding
  deviation function are shown in Figure~\ref{fig_QN_dev}.}
\label{fig_QN_img}
\end{figure}

\begin{figure}[htbp]
\includegraphics*[angle=270,width=3.2in]{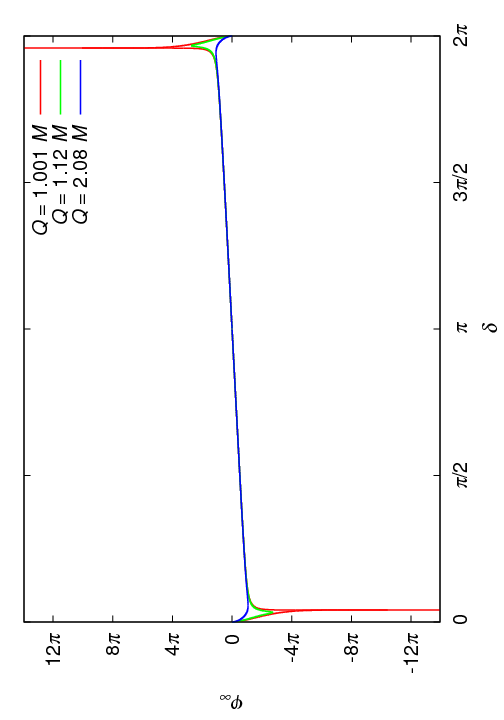}
\includegraphics*[angle=270,width=3.2in]{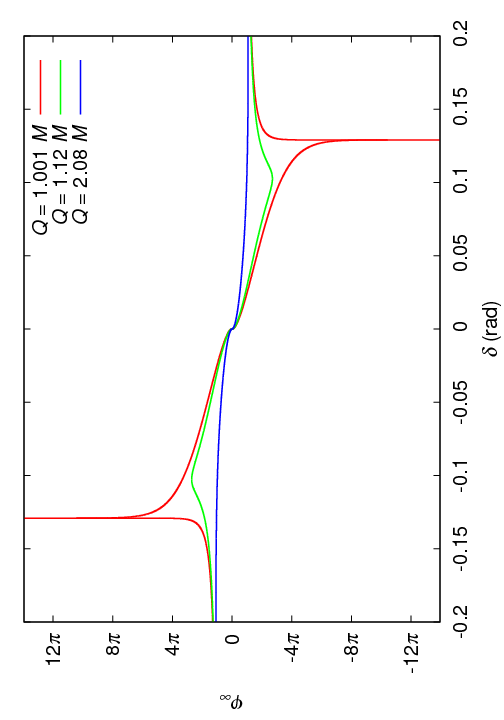}
\caption{Deviation functions of the three views of
  Fig.~\ref{fig_QN_img}. A zoom on the these is shown on the right so
  as to better show that the largest deviation decreases as the ratio
  $|Q|/M$ increases.}
\label{fig_QN_dev}
\end{figure}

These two features induce several differences with respect to the
Schwarzschild metric. Firstly, because the deviation function does not
diverge, there is only a finite number of multiple images of a given
background object. Secondly, because the deviation function derivative
cancels at the maximum of deviation, the background image distortion
will produce radial shear rather than tangential shear one is used to
observe at an Einstein ring. Because of the spherical symmetry of the
metric, this radial shear region will lie along a circle surrounding
the singularity. Thirdly, along this radial shear phenomenon, a set of
two multiple images of a given object will exist inside and outside
the radial shear ring. When the amount of maximum deviation varies,
for example when the observer's distance changes with respect to the
black hole, these images appear of disappear by pairs.  In
Figure~\ref{fig_QN_img}, we show three realizations of a naked
singularity for increasing values of $|Q|$, together with the
corresponding deviation functions.

\begin{figure}[htbp]
\includegraphics*[width=4.0in]{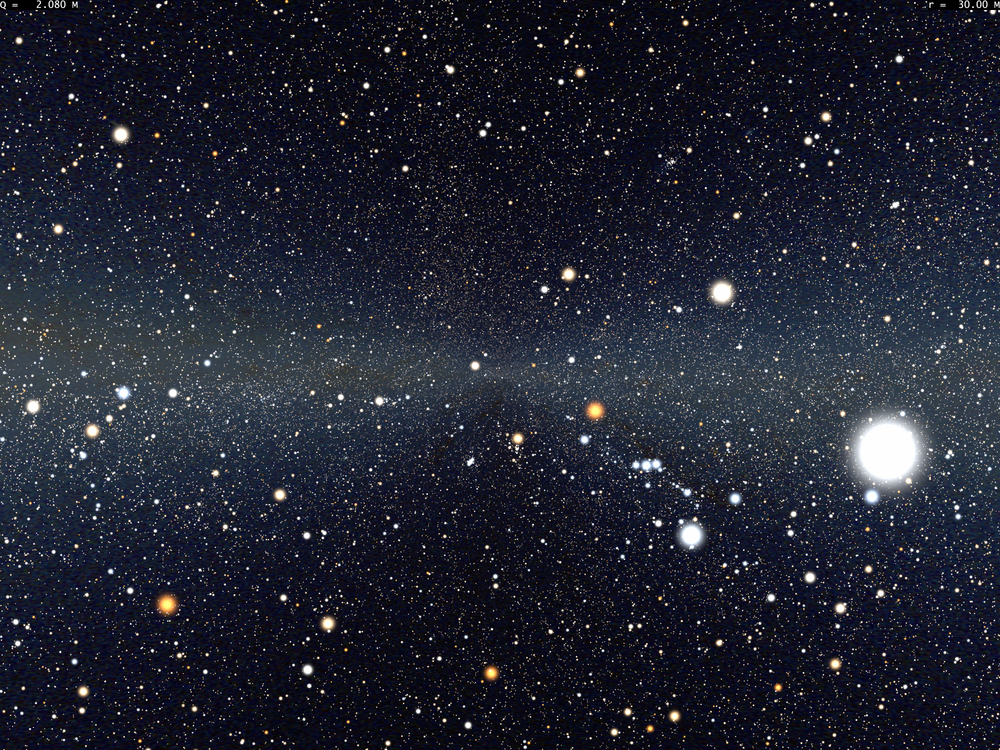}
\caption{Zooming inside the bottom part of Fig.~\ref{fig_QN_img}
  (opening angle slightly above one degree). As explained in the text,
  geodesics we see close to the direction of the singularity are
  approached it almost straight on and bounced on it, thus going away
  in almost the opposite direction from which they
  originated. Therefore, whereas one looks at the Galactic center, one
  sees a(n elongated and flipped) view of Orion constellation in the
  lower right quadrant of the picture, together with bright Sirius.}
\label{fig_QN_zoom}
\end{figure}

An interesting feature of the deflection function is that whatever the
value of $|Q| / M$ it tends to 0 as $\delta$ tends to 0. This means
that what is seen when looking toward the singularity is actually what
is exactly behind the observer. In other words, if one looks toward
the Galactic center with a naked singularity in between, then one
shall see the Galactic anticenter close to the singularity, as shown
in Fig.~\ref{fig_QN_zoom}.
This effect is somewhat counter intuitive, for a radial, null geodesic
going toward the singularity will reach it. But apart from those
exactly radial geodesics, any other which exhibits a slight deviation
from the purely radial case will bounce and go back (almost) in the
opposite direction it was travelling prior to the bounce. 

This can be understood as follows. When the impact parameter is very
small, Eq.~(\ref{eq_rdot}) can be approximately rewritten by saying
that $\dot r \simeq \pm E$ everywhere it is defined, that is, for any
$r > r_\MIN$ where $r_\MIN$ is the turning point of the
trajectory. Consequently, the total deviation experienced by such
geodesic can be rewritten
\begin{equation}
\Delta \varphi \simeq 2 \int_{r_\MIN}^\infty \frac{L}{r^2 | \dot r |} \ddd r .
\end{equation}
(We assume that the observer sits at infinity. Dropping this
assumption does not change the conclusions.) The value of this
expression is obviously,
\begin{equation}
\Delta \varphi \simeq 2 \frac{b}{r_\MIN} .
\end{equation}
Now, in the limit where the impact parameter is small, the turning
point of the trajectory occurs when
$1 / b^2 = 2 V_\NULL (r_\MIN) / L^2$ which can occur only at very low
values of $r$ since $V_\NULL$ must be very large. In this case,
$V_\NULL$ is well approximated by the leading term in $1 / r$, so that
one has $r_\MIN \simeq \sqrt{b Q}$ and the deviation of the geodesic
is therefore of order
\begin{equation}
\Delta \varphi \simeq 2 \sqrt{b / Q} ,
\end{equation}
which means that it is very small. Very roughly speaking, this amounts
to say that the vicinity of the singularity acts as a spherical mirror
on which geodesics are reflected back, except for those which are
exactly radial. This analogy, although crude, allows to understand
that the view of the celestial sphere shown by these geodesics will be
flipped, as also shown on Fig.~\ref{fig_QN_zoom}. Also,
Fig.~\ref{fig_QN_dev} show that even for large value of $|Q|/M$, the
deviation function changes rather abruptly around $\delta = 0$, which
means that geodesics with similar, almost 0 impact parameter will be
deflected by a significantly different amount. This translates into
the fact that a small beam of geodesics reaching the observer after
having traveled in the immediate vicinity of the singularity come from
a larger patch of the celestial sphere, which shall therefore appear
elongated along the radial direction from the point of view of the
observer, a feature also obvious in Fig.~\ref{fig_QN_zoom}.

\subsection{Frequency shift close to the singularity}

\begin{figure}[htbp]
\includegraphics*[width=4.0in]{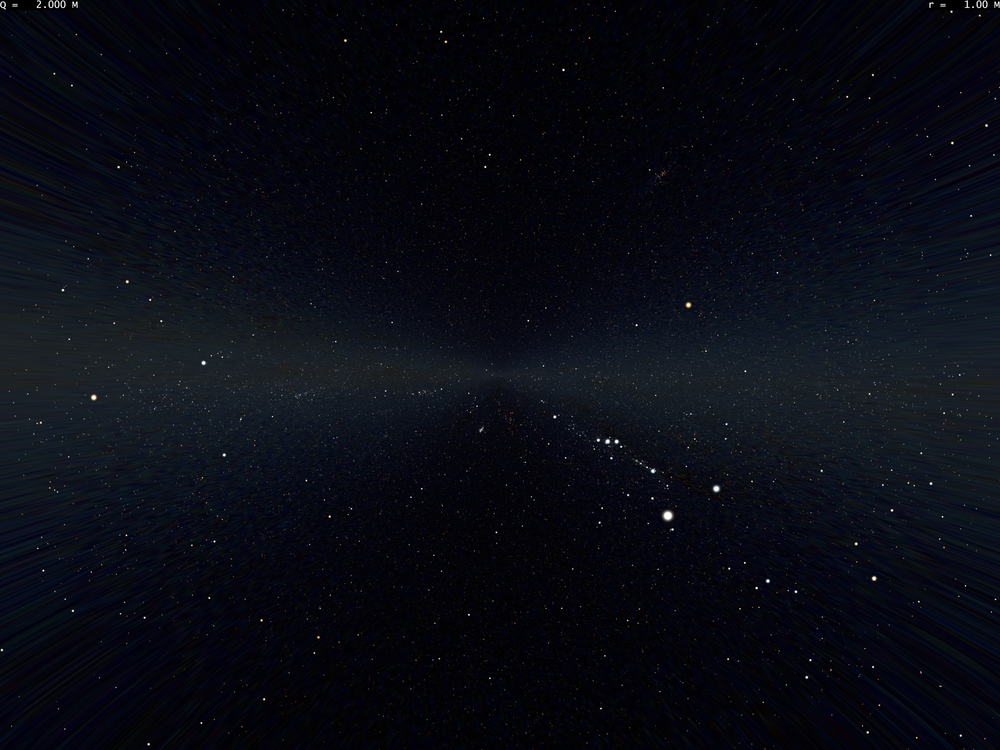}
\caption{A view of a naked Reissner-Nordstr\"om singularity
  $|Q| = 2 M$ for a static observer very near to it ($r = M$). In this
  case, the whole celestial sphere is highly redshifted because the
  singularity is gravitationally repulsive and lies on top of an
  infinitely high potential hill. Note that the fact that star field
  is similar to that of Fig.~\ref{fig_QN_zoom} is partly coincidental
  here. Fig.\ref{fig_QN_zoom} was a zoomed in ($\sim 1\,{\rm deg}$
  opening angle) view of the singularity seen from large distance
  ($r = 30 M$), whereas here one is very close to the singularity with
  a much wider (110~deg) opening angle. Also, the fact that the
  picture is not only distorted but also redshifted with respect to
  what one is used to see is more spectacular for cool star such as
  Betelgeuse ($\alpha$~Ori) which is almost invisible here.}
\label{fig_ns_red}
\end{figure}

Since the metric of a naked singularity is everywhere static, it is
natural to consider static observers as well. At large distances
(where the $Q^2 / r^2$ is negligible in the metric~(\ref{def_RN},
\ref{def_ARN}), such observers are in a weak potential well and
therefore see any radiation coming from past null infinity with a weak
blueshift, similarly to the Schwarzschild case, but the situation
changes when one is near the singularity. The reason goes as
follows. A photon whose frequency is $\omega_0$ as seen by a static
observer far from the singularity has a four-wavevector whose time
component is therefore $k^t = \omega_0 / A$ (so that $g_{tt} k^t$ is a
constant that reduces to $\omega_0$). Consequently a static observer
catching this photon will measure a frequency given by
\begin{equation}
\omega_0^\STAT = k^\mu u_\STAT^\nu g_{\mu\nu}
              = \omega_0 u_\STAT^t 
              = \omega_0 / \sqrt{A} .
\end{equation}
Sufficiently far from the singularity, $A$ is close to 1, although
slightly smaller, consequently a weak gravitational blueshift is
observed as we just said. Conversely, very near to the singularity,
$A$ is larger than 1 and the whole celestial sphere seen by a static
observer would appear redshifted redshifted, as shown in
Fig.~\ref{fig_ns_red}, the redshift
$z \eqdef \omega_0 / \omega_0^\STAT - 1$ being
\begin{equation}
z = \sqrt{A} - 1 ,
\end{equation}
which correspond to $\sim 0.732$ in Fig~\ref{fig_ns_red}.
In the limit where the observer is arbitrarily close to the
singularity, the celestial sphere redshift tends to infinity, a mere
consequence of the fact that in this case the observer lies on a very
high potential hill. As for what the observer would see if one could
neglect redshift, it would correspond to a tiny portion of the
celestial sphere. The reason is that only null geodesics with a small
impact factor can reach very small values of $r$ before bouncing on
the singularity. But those geodesics experience only a weak deviation,
so that only the part of the celestial sphere that is situated in a
narrow beam centered on the half-line joining the singularity and the
observer can be seen.

\section{The ``dark shell'' phenomenon}
\label{sec_darkshell}

Since there exists bounded null trajectories, an observer lying within
the region where such trajectories extend into will see nothing when
looking toward the directions they propagate. Of course, when a
horizon exists, those trajectory can be mistaken by the horizon
itself. But such trajectories also exists when there is no horizon,
for a moderate range of $|Q|$, i.e., when
$1 < |Q| / M < 3 / 2 \sqrt{2}$. For a static observer, those bound
null geodesics will be seen around the perpendicular direction with
respect to the radial one, giving an overall aspect of a thin, dark
shell surrounding the singularity, which appears in front of a fairly
bright (i.e., blueshifted) background of stars. The reason why the
stars are blueshifted goes as follows. The frequency shift between
radiation seen by a static observer at infinity and a static observer
near to the singularity is given by
$1 / (1 + z_\STAT) = 1 / \sqrt{A(r)}$.  It then suffices that $A$ is
smaller than $1$ for the radiation to be blueshifted, which can be
shown rather easily. Solving $A(r) = 1$ gives
$r_\FF^\MIN = Q^2 / 2 M$, which we label this way since it corresponds
to the minimal coordinate distance a freely falling observer would
reach. For any $r > r_\MIN^\FF$, one has $A(r) < 1$. Moreover, we have
$V_\NULL(r_\MIN^\FF) = 2 L^2 M^2 / Q^4$, whose minimum value when
$1 < |Q| / M < 3 / 2 \sqrt{2}$ is $32 L^2 / 81$, which is always
larger than $V_\NULL(r_\EE) = V_\NULL (r_\SSS)$. Consequently,
$r_\MIN^\FF$ is situated at some smaller value than $r_\SSS$ and the
celestial sphere is blueshifted.

More quantitatively, the minimal value of $A$ is obtained for
$r = Q^2 / M$. Given the interval where $Q$ is allowed here, this
value of $r$ does not exceed $9 M / 8$, which is smaller than the
smallest value of $r_\EE$, therefore $Q^2 / M < r_\EE$. Since $A(r)$
is a decreasing function for $r > Q^2 / M$ and a decreasing one
otherwise, the largest value of $A$ within the interval
$[r_\SSS, r_\EE]$ is either $A(r_\SSS)$ or A$(r_\EE)$, but the fact
that by definition $V_\NULL (r_\SSS) = V_\NULL (r_\EE)$ means that
$A (r_\SSS) = (r_\SSS / r_\EE)^2 A (r_\EE) < A(r_\EE)$, therefore the
maximal value of $A$ in the above mentioned interval is $A(r_\EE)$,
whose value can be written, using Eq.~(\ref{eq_rextr}),
\begin{equation}
A (r_\EE) = \frac{1}{2} \left(1 - \frac{M}{r_\EE} \right)
         = \frac{1}{3} \left(1 - \frac{Q^2}{r_\EE^2} \right) .
\end{equation}
The first formula ensures that $A < 1 / 4$ as soon as $|Q| \geq M$
since then, the smallest value of $r_\EE$ is $2 M$, which means that
the frequency shift seen by a static observer within the dark shell
region is at least $2$. As for the maximal frequency shift, it is
obtained when $A$ is close to $0$, which occurs in the limiting case
of an extremal black hole with an observer close to $r = M$.

\begin{figure}[htbp]
\includegraphics*[angle=270,width=3.2in]{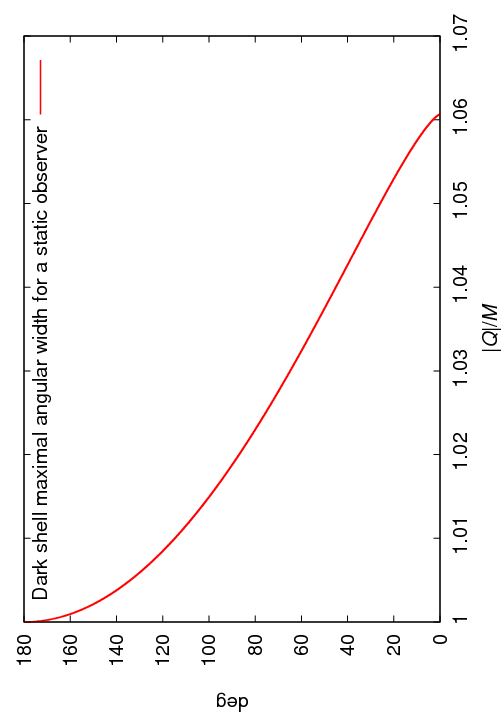}
\includegraphics*[angle=270,width=3.2in]{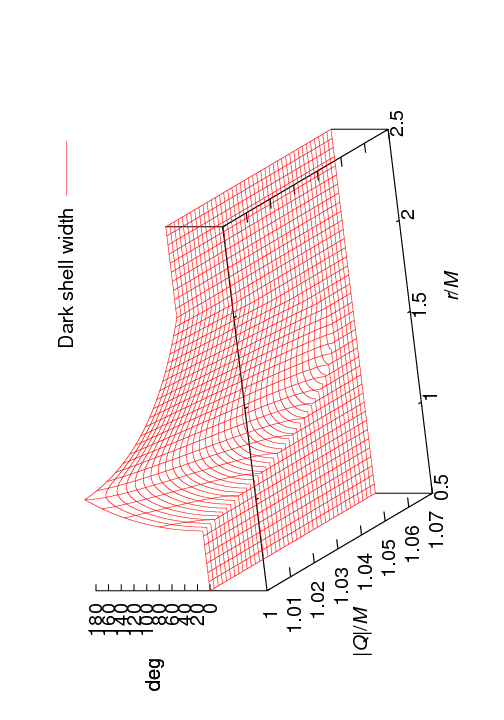}
\caption{The dark shell maximal width as a function of $|Q|$ (top
  figure) and its width as function of both $r$ and $|Q|$
  (bottom). The shell exists only for $r_\SSS < r < r_\EE$,
  and such interval exists only if $|Q|/M < 3/2\sqrt{2} \simeq 1.061$,
  hence the delineation of the hill in this plot.}
\label{fig_DSmax}
\end{figure}

We now come to the thickness of this shell seen from a static
observer. From a static observer, one can attach a reference tetrad
with one vector corresponding to the observer four-velocity,
$u_\STAT^\mu$, i.e., the vector defined in Eq.~(\ref{def_u_stat}), one
to an orthonormal vector spanned by $\partial/\partial r$, i.e., the
vector defined in Eq.~(\ref{def_r_stat}), $R_\STAT^\mu$, and two
normal vectors spanned by $\partial / \partial \varphi$ and
$\partial / \partial \theta$, $\Phi^\mu$ and $\Theta^\mu$. By suitably
choosing an orientation of the coordinate system, a null geodesic is
of the form
\begin{equation}
k^\mu = \omega_\STAT (u_\STAT^\mu + \cos S R_\STAT^\mu + \sin S \Phi^\mu) .
\end{equation}
From this geodesic, one can compute the constant of motion $E$ and $L$
similarly to Eqns.~(\ref{defL}, \ref{defE}). A null geodesic
participates to the dark shell phenomenon if it is bound, i.e., if
$r \in [r_\SSS, r_\EE]$ and if its effective energy $E^2/2$ is
smaller than the local extremum of $V_\NULL$. This amounts to says that 
\begin{equation}
\frac{A(r)}{r^2} < \frac{E^2}{L^2} < \frac{A(r_\EE)}{r_\EE^2} ,
\end{equation}
which translates into 
\begin{equation}
\sin^2 S > \frac{A (r)}{r^2} \frac{r_\EE^2}{A(r_\EE)} .
\end{equation}
The dark shell is at its widest when $A (r) / r^2$ is at its smallest
value within the range of $r$ one considers, i.e., at $r = r_\EXTR^-$.
The dark shell maximal angular width $\Delta S_\MAX$ is therefore
\begin{equation}
 \Delta S_\MAX = 2 \arccos \left(\sqrt{\frac{A(r_\EXTR^-)}{(r_\EXTR^-)^2} 
                                      \frac{r_\EE^2}{A(r_\EE)}} 
                          \right) .
\end{equation}
In the extremal case $|Q| = M$ one has $r_\EXTR^- = M$ so that
$A(r_\EXTR^-) = 0$. Therefore, $\Delta S_\MAX = \pi$ and the dark
shell occupies all the celestial sphere, however no static observer
can lie at $r = M$ besides an extremal black hole. Should one take
$r = M(1 + \epsilon)$ instead of $r_\EXTR^-$ (or, alternatively,
$|Q| = M (1 + \epsilon)$), with $\epsilon \ll 1$, then the dark shell
would have a width close to but smaller than $\pi$ and would fill
almost but not all the celestial sphere. The maximal dark shell width
seen by a static observer is shown in Fig.~\ref{fig_DSmax}. Its width
as a function of both $r$ and $|Q|$ is plotted in the same Figure.
A similar phenomenon arises when $|Q| \leq M$ but it is then less
spectacular since there is no distinction between the dark shell
itself (i.e. bound null geodesics in the sense described above) and
other, unbound, geodesics that also cross the black hole past
horizon. The only way to differentiate the two is to consider the
maximal analytic extension of the metric, something we shall consider
in next Section.

Let us add that the width of the shell of course depends on the
observer's velocity. Should one consider a freely falling observer
along a radial trajectory, then aberration will push and shrink the
shell toward the singularity. Also, the freely falling observer
induces a large redshift and blueshift with respect to the static case
which can be computed as follows. The Lorentz boost $\gamma$ to go
from a static observer with four-velocity $u_\STAT^\mu$ to a freely
falling observer with four-velocity $u_\FF^\mu$ is given by
\begin{equation}
\gamma = g_{\mu\nu} u_\STAT^\mu u_{\FF, -}^\nu = \frac{1}{\sqrt{A (r)}} , 
\end{equation}
the relative velocity $v$ between the two frame being therefore
$v = \sqrt{1 - A (r)}$. If we suppose light rays coming from infinity
along an almost radial ingoing or outgoing trajectory, their
corresponding four-wavevector components are
\begin{equation}
k^\mu_\pm = \left(\begin{array}{c} 
                 \frac{\omega}{A} \\ \pm \omega \\ 0 \\ 0
                 \end{array} 
           \right) ,
\end{equation}
where $\omega$ is the light ray frequency measured by an observer at
infinity. The frequency $\omega_{\FF, -}$ measured by a freely falling
observer is then
\begin{equation}
\omega_{\FF, -}
 = k^\mu_\pm u_{\FF, -}^\nu g_{\mu\nu}
 = \frac{\omega}{A} \left(1 \pm \sqrt{1 - A} \right) .
\end{equation}
Using the notations $\gamma$ and $v$ defined above, we have
\begin{equation}
\omega_{\FF, -} = \gamma^2 \omega (1 \mp v) .
\end{equation}
This formula differs from the special relativistic formula of redshift
by a factor $\gamma$, which corresponds to the extra contribution of
gravitational blueshift experienced when within a potential well. It
is then easy to show that this quantity reaches its extrema whichever
the $\mp$ sign is when $A$ does, i.e. at
$r = Q^2 / M = 2 r_{\FF, \MIN}$.  Then, one has
$A(r) = 1 - M^2 / Q^2$, which implies that
$\omega_{\FF, -} = 1 / (1 \pm M / |Q|)$. The maximum blueshift can
therefore be fairly large and even tends to infinity when $|Q| \to M$.

\begin{figure}[htbp]
\includegraphics*[width=3.2in]{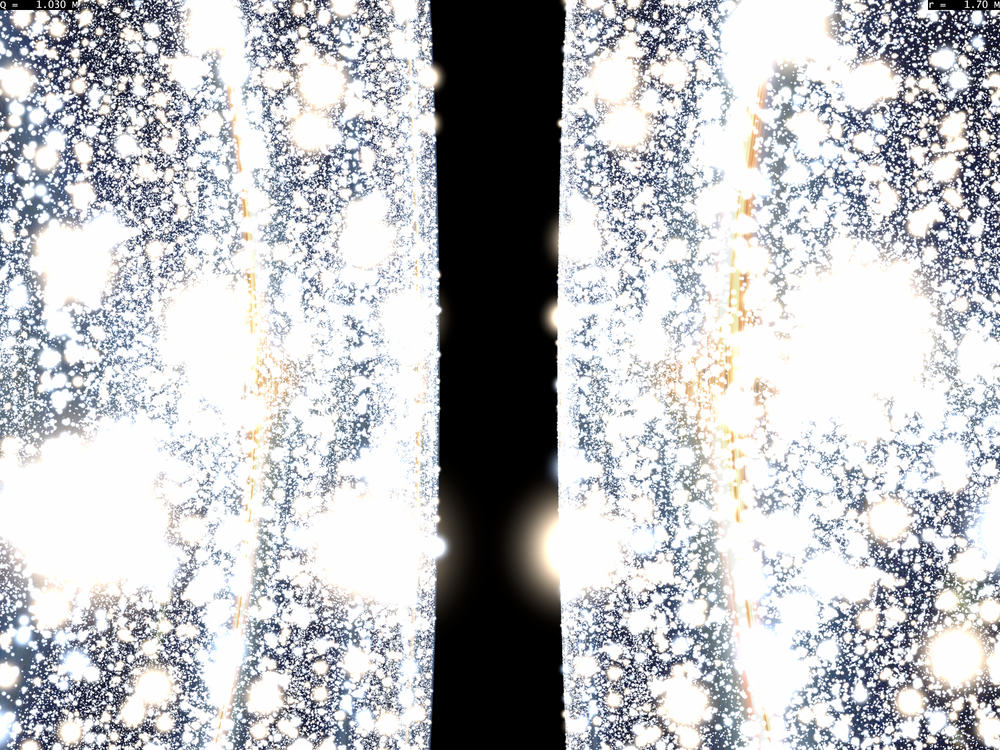}
\includegraphics*[width=3.2in]{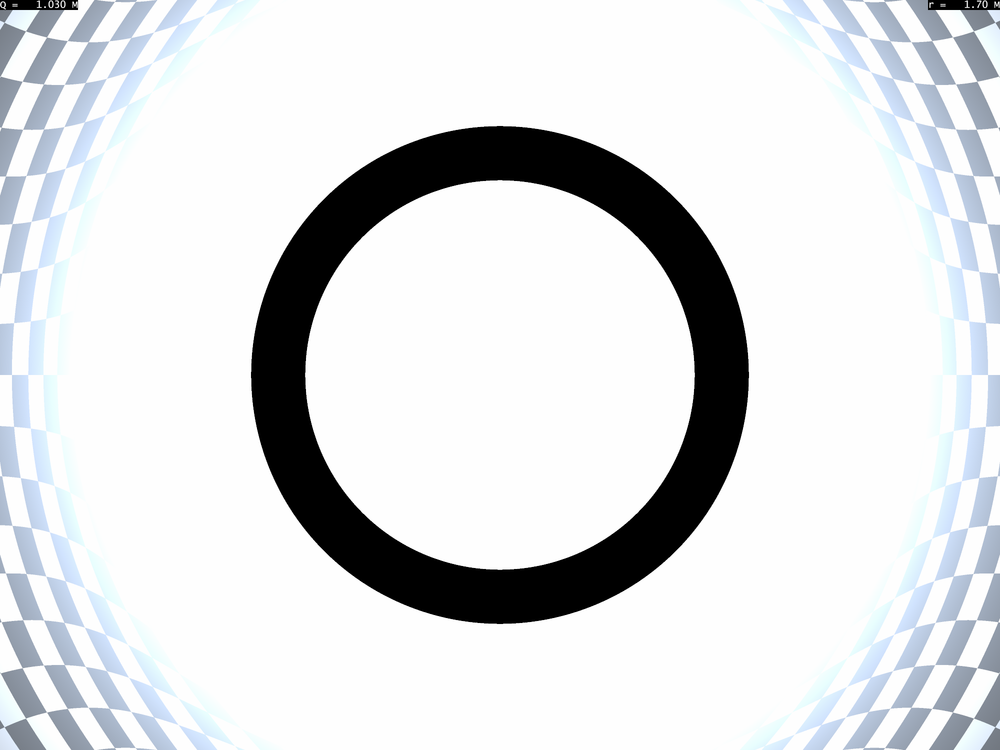}
\caption{A view of a Reissner-Nordstr\"om naked singularity
  $|Q| = 1.03 M$ from close distance ($r = 1.7 M$) and a static
  observer, where the ``dark shell'' phenomenon arises (left
  view). This view is shown perpendicularly to the singularity
  direction (90~degrees to the left of the centre of the
  image). Overall blueshift is defined as $1 + B = 1 / (1 + z)$ is
  $\simeq 1.290$ with the value chosen here.  In case one considers a
  freely falling observer (right view), the shell is seen at a smaller
  angular size surrounding the singularity. The background sky in that
  direction is fairly brighter than above because it is blueshifted
  both by gravitational effect (already seen in the top view) and the
  observer motion, which has a $\gamma = 1 / \sqrt{A}$ boost with
  respect to a static observer. For this reason, realistic celestial
  sphere and stars were removed and replaced by a coordinate grid
  showing an attenuated version of blueshift.}
\label{DS_stat_ff}
\end{figure}

Regarding the shell angular width, it has then has a fairly cumbersome
form and is always smaller than in the static case because of
aberration. In Fig.~\ref{DS_stat_ff}, we show how the dark shell
aspect transforms when one goes from a static observer to a freely
falling observer with zero velocity at infinity.

\section{Looking at and crossing a Reissner-Nordstr\"om wormhole}
\label{sec_crosswh}

The maximal analytic extension of the Reissner-Nordstr\"om metric
possesses an infinite tower of pairs of asymptotic regions similar to
those of the maximal extension of the Schwarzschild metric. Its
Carter-Penrose diagram can be described as follows. Regions 1 and 3
(labels are chosen so as to match those the Schwarzschild metric in
Ref.~\cite{riazuelo15}) are asymptotic regions, with $r > r_+$ being a
spacelike coordinate and $t$ being a timelike coordinate, which is
future-oriented in 1 and past oriented in 3. The inter-horizon part of
the metric, i.e., $r_- < r < r_+$, in region~1 (or 3) future is
labelled 2. There, $r$ is a past-oriented timelike
coordinate. Similarly to the Schwarzschild case, there exists another
such inter-horizon region which is in regions~1 and 3 causal
past. Such a region will be labelled 4 so as to follow the
Schwarzschild case.

\begin{figure}[htbp]
\includegraphics*[width=4.0in]{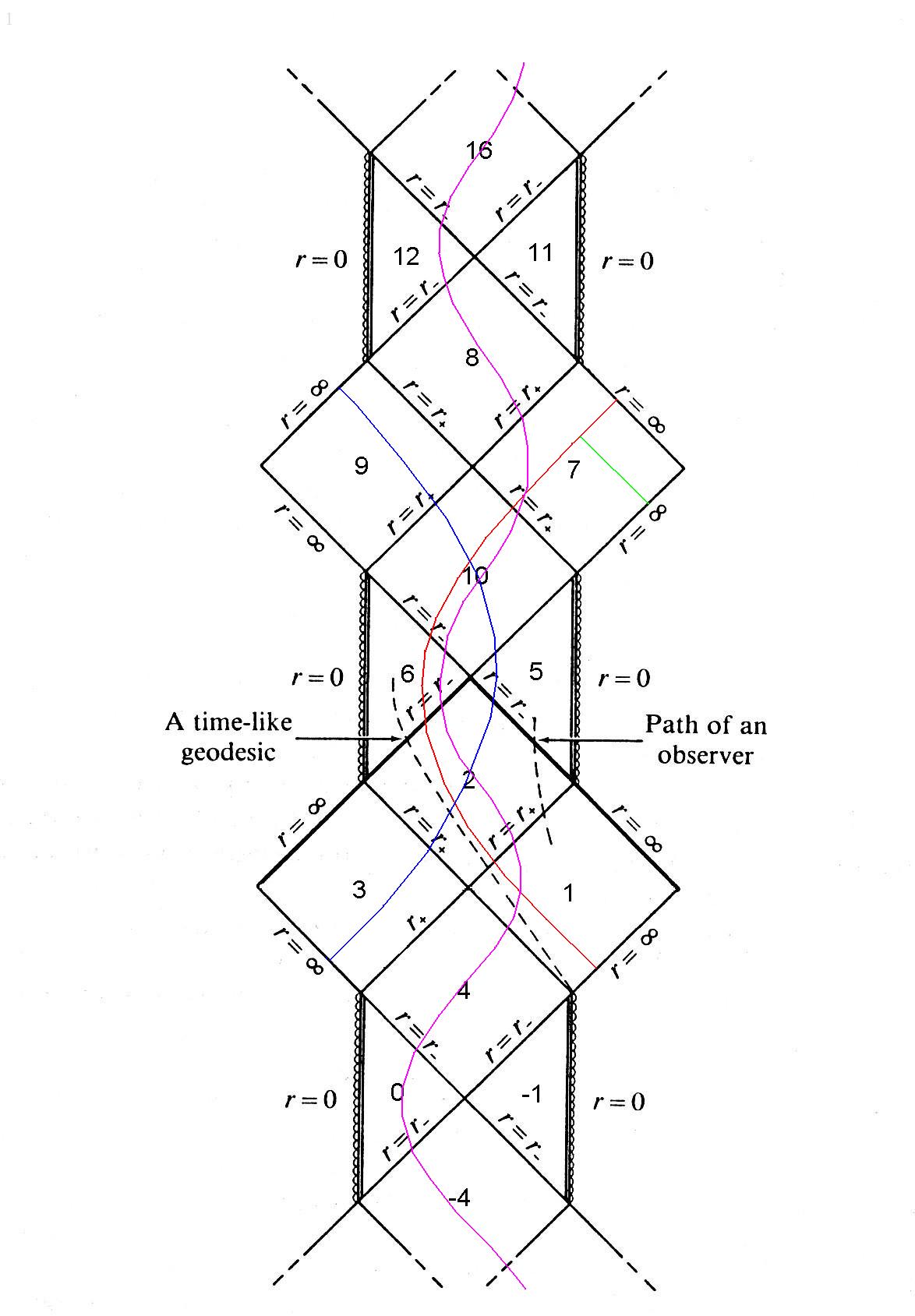}
\caption{(A part of) The maximal analytic extension of the
  Reissner-Nordstr\"om metric, which consists of an infinite tower of
  six diamond- or half-diamond-shaped regions. The Figure is adapted
  from~\cite{chandrasekhar83}.  }
\label{cp_diag}
\end{figure}

The inner region ($r < r_-$) part of the metric is split into two
regions, 5 and 6, where, as in 1 and 3, $t$ is a timelike coordinate
and $r$ is a spacelike coordinate. We choose to label 5 the region
directly above 1 (i.e., to the right of the diagram), and 6 is the one
above 3.  Obviously, a radial null geodesics that goes through the
black hole travels along 1, 2 and 6 since it is by construction a
straight line going toward an upper left quadrant. Since $\pi_t$ is
constant along this (or any) geodesics and since $A(r)$ is positive in
both 1 and 6, $\dot t$ has to be positive in 6 as well. Consequently,
$t$ is a future-oriented timelike coordinate in 6 and similar
reasoning with a radial null geodesic starting from 3 shows that $t$
is past-oriented in 5. Since $\pi_t$ is conserved for any geodesic,
timelike geodesics will also travel along the 1-2-6 or the 3-2-5
sequence. There, they will bounce at some distance of the singularity,
and will follow a symmetric outgoing trajectory. They will first
travel along another inter-horizon region where this time $r$ is
future-oriented. This region is similar to region 4 defined above and
will be labelled 10. Then geodesics will exit a new asymptotic region
that we shall label 7 (above 5 and 1) and 9 (above 6 and 3). There is
therefore an infinite tower whose basic blocks are a series of six
regions in the Carter-Penrose diagram as summarized in
Fig.~\ref{cp_diag}. Timelike or null geodesics (except in the case of
radial null geodesics) originating from past null infinity of
regions~1 or 3 will therefore travel along the 1-2-6-10-7 or
3-2-5-10-9 sequence. Also, because the effective potential $V_\NULL$
admits a local minimum, there also exist bound timelike of null
geodesics which endlessly cross and exists the numerous horizons of
the metric. Those passing in region 1 then have followed the sequence
...4-1-2-6-10-7-8-12-16..., when the $n$-th region is deduced from the
$n-4$-th one by adding 6.
Even though the analysis of the Carter-Penrose diagram allows to know
which regions are seen along the wormhole hole crossing (which we
shall perform from a radially infalling observer from region~1, thus
following the 1-2-6-10-7 sequence, see above), it is not intuitive to
guess what such observer will actually see. Also it is not much easier
to guess how region~-5 might look as seen from region~1, even when far
from the wormhole.

We shall therefore address this first issue, and then address the
whole problem of what is seen not only outside a wormhole, but when
traveling through it.  

\subsection{Orbiting around a wormhole}

Outside a Reissner-Nordstr\"om wormhole the metric is the same as in
the black hole case and therefore allows for circular timelike
geodesics, at least as long as $r$ is not too close to the horizon. It
is therefore legitimate to consider the view seen from a timelike
circular orbit.

\begin{figure}[htbp]
\includegraphics*[width=4.0in]{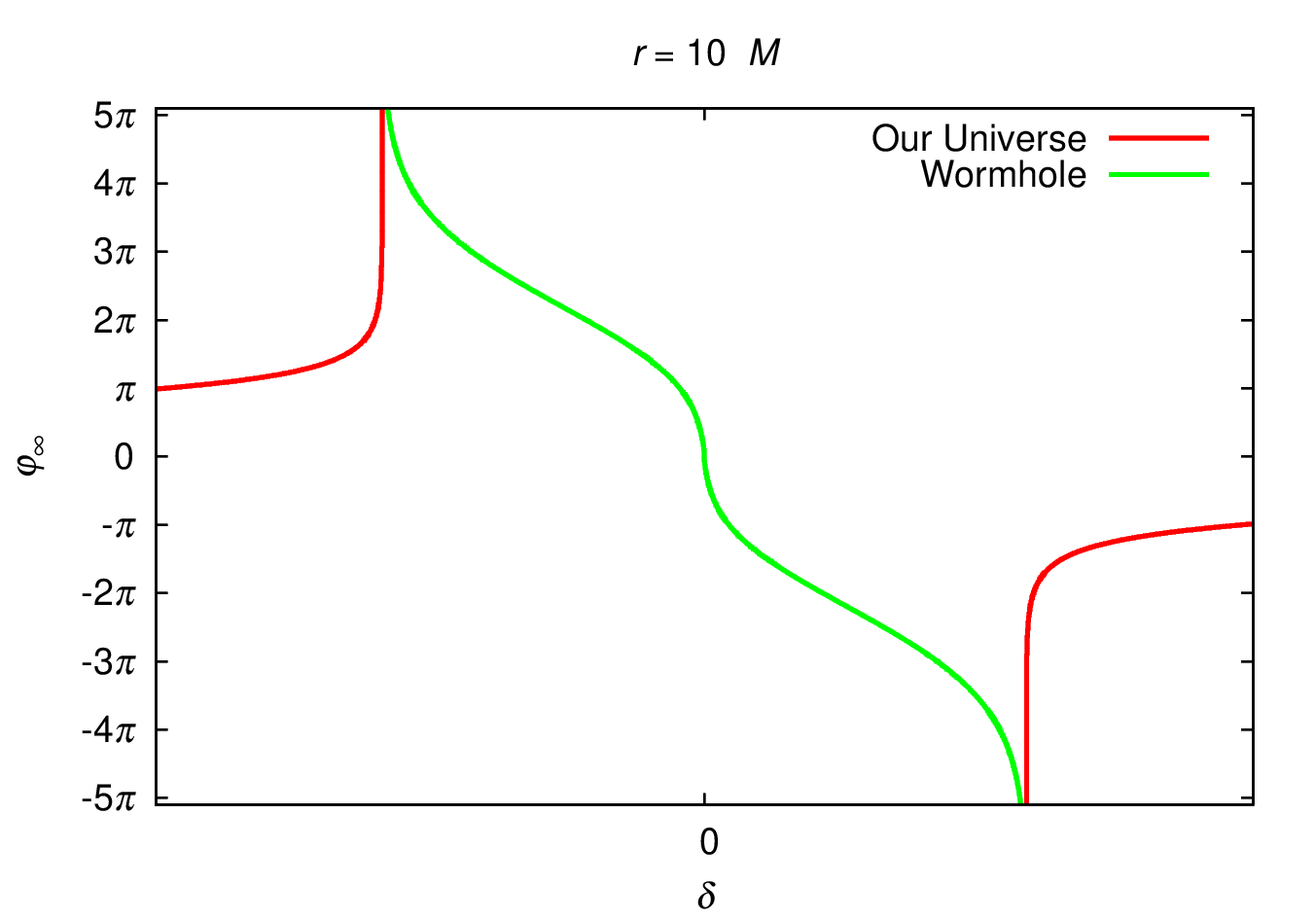}
\caption{Deviation function seen by a static observer at $r = 10 M$ of
  a Reissner-Nordstrom wormhole. The edge of the wormhole, which would
  correspond to the edge of a black hole silhouette of identical mass
  and charge, is delineated by geodesics of impact parameter close to
  the critical one hence an infinite deviation. Close to the inner
  edge of the wormhole, geodesics also experienced a large deviation,
  actually twice, both before and after wormhole crossing.}
\label{fig_WH_dev}
\end{figure}

We start from the view seen by a static observer. The deviation
function is shown in Fig.~\ref{fig_WH_dev}.
The external part of the deviation function is unsurprising since it
correspond to that one sees in a standard black hole case and it
diverges for an angular separation which corresponds to the critical
impact parameter. The other region is seen through outgoing null
geodesics which have an impact parameter smaller than the critical
one. Close to the edge of the wormhole, the impact parameter is
vanishingly close than the critical one and geodesics experience a
diverging deviation. Close to the geometric center of the wormhole,
geodesics bounced on the central region of the wormhole before exiting
it, just as in the white hole case. Consequently, one expects that the
deviation function is decreasing from $+\infty$ to $-\infty$ when
looking from left to right at the wormhole throat. Such a feature
implies that any part of the region seen though the wormhole throat
will be seen an infinite number of time and that a double series of
Einstein ring shall outline the region on the other side of the wormhole
directly ahead of the observer and that on the opposite side. 

Looking at Fig.~\ref{fig_WH_dev}, it is clear that the angular
intervals where the deviation function is not very step are somewhat
extended. This will translate into the fact that the angular distance
between the Einstein rings will not be small as it is the case for the
(outer part of the) Schwarzschild solution, which means that several
Einstein rings will be rather easy to spot as concentric, circular
regions with sparse but bright stars because of gravitational
lensing. A star that appears lensed in one Einstein ring will also
appear as such in the other associated rings.  Since Einstein rings
alternate between what can be dubbed as the rear direction (i.e.,
trajectory very close to the center of the wormhole) and the opposite
one, a given lensed star will be seen every two Einstein rings.

Moreover, since the deviation function is zero of radial trajectories,
the center of the image seen through the wormhole does not correspond
to deflected photons passing through the wormhole, but to photon that
bounced at some distance of the singularity. In other words, whereas a
Morris-Thorne wormhole does not show significant amount of distortion
in the center of its silhouette, a Reissner-Nordstrom wormhole does,
and it is difficult to recognize anything whose image has traveled
through the wormhole as this image will always appear distorted. An
example of this, with the Milky Way seen through the wormhole given be
shown in Fig.~\ref{fig_cross_8}.

\begin{figure}[htbp]
\includegraphics*[width=4.8in]{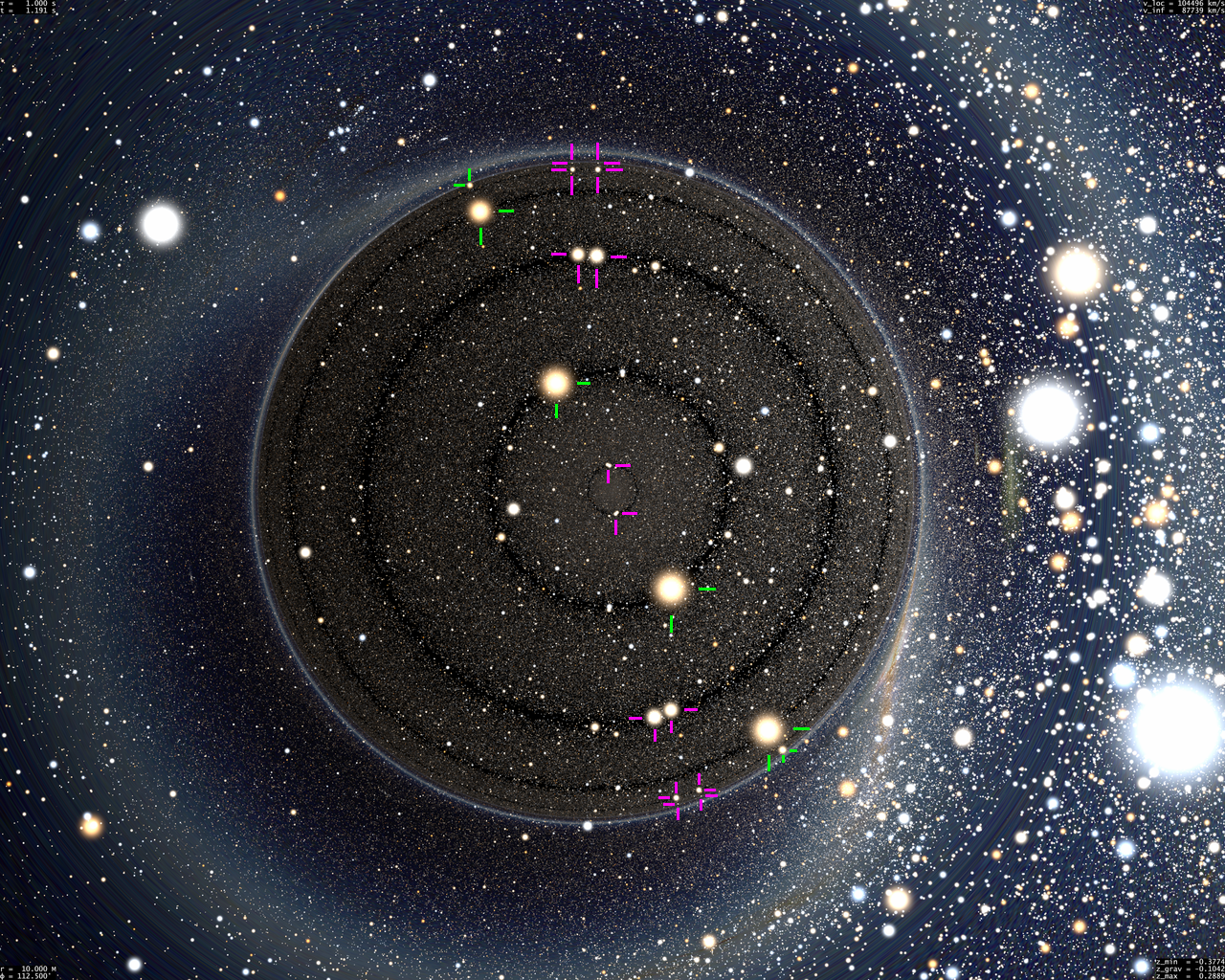}
\caption{A view of a Reissner-Nordstr\"om wormhole from a circularly
  orbiting observer at $r = 10 M$. The observer is orbiting
  counterclockwise and is looking 90 degrees on its left, toward the
  center of the wormhole that we define by the null radial outgoing
  geodesic. Hence, the front direction lies to the right of the
  picture which therefore is brighter due to Doppler shift. Because of
  aberration, the black hole silhouette is slightly off-center with
  respect to the screen because of aberration which shifts the
  silhouette in a different manner than its center. \\
  Numerous Einstein rings are visible within the wormhole, which
  alternatively zoom in on the region that lies somehow behind the
  singularity from observer's point of view and the opposite
  region. They are easily spotted as circular thin shells devoid of
  stars (regardless of aberration which transforms circles into
  circles). One bright star and its multiple images are outlined with
  green segments. There are two images per Einstein rings, on each
  even ring when counted from the center. Seen by a static observer,
  all these images would lie on a straight line, which is transformed
  into a circle by aberration. Four bright images of the star are
  seen, plus two fainter ones very close to the edge of the
  wormhole. We have also shown with purple segments the multiple
  images of a pair of stars on odd-numbered circles. Eight images of
  these stars can be seen, although the outermost ones are barely
  visible. Many other multiple images of stars within the wormhole can
  be seen.}
\label{fig_WH_img}
\end{figure}

Including the observer's circular motion introduces two extra features
(that are also present in the standard Schwarzschild black hole case,
but at a lesser extent because of the visual structure of Einstein
rings). Firstly, lens images of stars will no longer be aligned as
aberration will transform great circles into circles. Secondly,
Doppler shift will affect both color and magnitude of stars. Such
features do not prevent from identifying lensed star image alignments,
however, as shown in Fig.~\ref{fig_WH_img}.

\subsection{Crossing the wormhole -- Infalling part}

When one lies outside the wormhole, its angular size varies in a
similar fashion as in the Schwarzschild case, but the combination of
special and general relativistic effects make it complicated to
interpret. In Refs.~\cite{riazuelo15}, some counterpointed aspects
were detailed, such as the fact that a static observer close to by
outside a Schwarzschild black hole will have the visual impression to
actually be inside the black hole, whereas on the contrary, an
infalling observer soon after horizon crossing will be tempted to
think that he/she is still outside. Considering that the maximal
analytic extension of the metric adds some extra layers that can be
seen by an observer, e.g., dark shell and other asymptotic regions,
it is very difficult to guess how a wormhole crossing would be
visually felt by an observer. Furthermore, the simplest wormhole one
may think of, the Morris-Thorne wormhole~\cite{morris88}, possesses
only two regions, the entrance and the exit of the wormhole, so that
wormhole crossing is not spectacular as no intermediate region is
actually visible~\cite{james15}.

This motivates the study of a more complicated wormhole crossing such
as the RN~wormhole.  In practice, the most obvious visual features of
some landscape are the large-scale ones, so that we shall here first
address the problem of the angular size of each regions that are seen
by the observer. We shall split this problem into two steps, the
infalling part (this subsection), where the observer starts from
infinity, crosses the outer then the inner horizons, and bounces at
$r = r_\MIN^\FF = Q^2 / 2 M$ (see discussion after
Eq.~(\ref{V_red_obs})). The outgoing part of the travel will be
described in the following subsection.

\subsubsection{Size of region~1}

The edge of the observer's initial region, 1, is delineated by null
geodesics of impact parameter $b_\CRIT$. Let us consider an observer
of four-velocity $T^\mu = u_{\FF,-}^\mu$ (see
Eq.~(\ref{def_u_ff})). Given the definitions of the constants of
motion $L$ and $E$ (see Eqns.~(\ref{defL}, \ref{defE})), one can write
the impact parameter of a geodesic making and angle $\theta$ with
respect to the radial outgoing direction. This gives
\begin{equation}
b = \frac{r \sin \theta}{1 + (\cos \theta) \sqrt{1 - A (r)} } .
\end{equation}

Writing that a null geodesic with impact parameter $b_\CRIT$ is making
an angle $\delta_1^\IN$ with respect to the radial direction is therefore 
equivalent to
\begin{equation}
\label{meq_hor}
\beta = \frac{2 x \tan (\delta_1^\IN / 2)}
             {  (1 + \sqrt{1 - A(x)}) \tan^2 (\delta_1^\IN / 2)
              + (1 - \sqrt{1 - A(x)})} ,
\end{equation}
where we have introduced the simplifying notations
\begin{equation}
x \eqdef \frac{r }{M} , \quad \beta \eqdef \frac{b_\CRIT}{M} .
\end{equation}
Moreover, we shall consider only the case where $\delta > 0$ since the
negative values of $\delta$ are expected to give the same results, up
to change the sign of $b$.

We now have to solve this equation, taking into the fact that it must
also satisfy the following criteria:
\begin{enumerate}

\item The geodesic must originate from region~1, where $t$ is a
  future-oriented time coordinate, which amounts to impose that
  $E > 0$

\item In order to delineate the edge of the wormhole, it must have
  reached the value $r = r_\EE$ (where it experienced an arbitrarily
  large deviation) {\it prior} to reaching the observer. This amounts
  to say that the geodesic must be outgoing for $r > r_\EE$ and
  ingoing for $r < r_\EE$

\item A last tweak must be taken into account when the observer
  reaches region~6. In this case, both ingoing and outgoing geodesics
  do delineate the edge of region~6.

\end{enumerate}
Forgetting about the last point above, the general solution of the
equation is given by (see Ref.~\cite{riazuelo15} for the derivation):
\begin{equation}
\label{delta_1_in}
\tan (\delta_1^\IN / 2)
 = \frac{x - (x - x_\EE) \sqrt{\displaystyle 
                              1 + \frac{2 x_\EE}{x} 
                              - \left(\frac{\beta q}{x x_\EE}\right)^2}}
        {\beta \left(1 + \sqrt{\displaystyle  \frac{2}{x} - \frac{q^2}{x^2}} 
               \right)} ,
\end{equation}
where we have set
\begin{equation}
q \eqdef \frac{Q}{M} , \quad x_\EE \eqdef \frac{r_\EE}{M} ,
\end{equation}
and we have made use of the two equalities
\begin{equation}
\label{int_beta}
\frac{q^2}{x_\EE^2}  + 3 \frac{x_\EE^2}{\beta^2} = 1 , \quad 
\frac{q^2}{x_\EE}  + \frac{x_\EE^3}{\beta^2} = 1 ,
\end{equation}
which result from Eqns.~(\ref{eq_rextr},\ref{eq_bcrit}).  This angle
corresponds to the silhouette of the black hole, since any value of
$\delta$ larger than $\delta_1^\IN$ obviously correspond to geodesics
that can reach the observer after originating from past null
infinity. At large distances, this expression reduces to
$\delta_1^\IN \simeq \beta / u = b_\CRIT / r$, as expected from an
object of cross section $b_\CRIT$. Also, the angular size of the
wormhole increases (albeit slowly) as the distance to the horizon
decreases. 

\begin{figure}[htbp]
\includegraphics*[angle=270,width=3.2in]{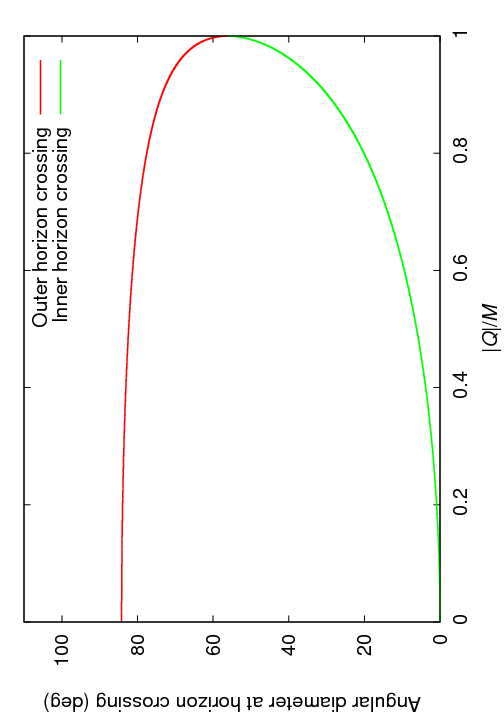}
\caption{Angular diameter of a Reissner-Nordstr\"om black hole at
  outer horizon crossing, as well as inner horizon crossing. Because
  both $x_+$ and $\beta$ are weakly varying functions of $|Q| / M$ for
  small values of this parameter, so is the angular size at outer
  horizon crossing. The fact that both angular sizes becomes
  increasingly similar as $|Q|/M$ gets close to $1$ comes from the
  fact that these two horizons fuse into the same horizon in the
  extremal case. }
\label{fig_sizehor}
\end{figure}

One can also evaluate this expression at outer horizon crossing,
i.e. when the observer actually enters into the black hole (or, in
this context, the wormhole, but it does not matter here). One obtains
\begin{equation}
\label{d1_outh_cross}
\left.\tan (\delta_1^\IN / 2) \right|_{x = x_+} = \frac{x_+}{\beta} ,
\end{equation}
a formula which has the same form as that of the Schwarzschild case
(see Refs.~\cite{muller08,riazuelo15}). The quantity $x_+ / \beta$ is
a decreasing function of $q$, which means that, just as for a static,
distant observer (see Fig.~\ref{fig_compq}), a charged black hole has
a smaller angular diameter with respect to an uncharged one when a
freely falling observer enter into it. It angular diameter varies from
$2 \arccos (23/31) \sim 84.2\,{\rm deg}$ for the Schwarzschild case to
$2 \arccos (15/17) \sim 56.1\,{\rm deg}$ for the extremal case. The
angular size as a function of the charge-to-mass ratio is shown in
Figure~\ref{fig_sizehor}. Incidentally, a similar formula holds for
inner horizon crossing, i.e.,
\begin{equation}
\label{d1_innh_cross}
\left.\tan (\delta_1^\IN / 2) \right|_{x = x_-} = \frac{x_-}{\beta} ,
\end{equation}
which immediately shows that the wormhole or black hole angular size
begins to shrink at some point as the radial coordinate $r$
decreases. We did not however find any analytic expression of the
value of $x$ for which the angular size reaches its maximum.

In fact, solution~(\ref{delta_1_in}) is only defined for
$r \geq r_\SSS$ since the square root of the denominator can be
rewritten as
\begin{equation}
\sqrt{1 + \frac{2 x_\EE}{x} - \left(\frac{\beta q}{x x_\EE}\right)^2}
 = \sqrt{\frac{(x - x_\SSS) (x + 2 x_\EE - x_\SSS)}{x^2}} .
\end{equation}
For values of $r$ smaller than $r_\SSS$, there is no solution to
Eq.~(\ref{meq_hor}) which means that region~1 fills the whole
celestial sphere although no geodesics with impact factor as large as
$b_\CRIT$ reach such low values of $r$. This is because a null
geodesic turning point $r_\TURN$ is determined by equation
$1 / b^2 = A(r_\TURN) / r_\TURN^2$.

In the rather small interval $r_\SSS < r < r_\EE$, one must consider
both ingoing and outgoing solutions of Eq.~(\ref{meq_hor}). This
ingoing solution has been expressed above, and the outgoing one
amounts to change the sign in front of the square root of the
numerator, which gives
\begin{equation}
\label{delta_1p_in}
\tan ({\delta'}_1^\IN / 2)
 = \frac{x + (x - x_\EE) \sqrt{\displaystyle 
                              1 + \frac{2 x_\EE}{x} 
                              - \left(\frac{\beta q}{x x_\EE}\right)^2}}
        {\beta \left(1 + \sqrt{\displaystyle  \frac{2}{x} - \frac{q^2}{x^2}} 
               \right)} .
\end{equation}
This solution goes to $0$ at $r = r_-$ and catches
$\tan (\delta_1^\IN / 2)$ at $r = r_\SSS$. Its interpretation is that
during most of the infalling phase, region~1 does not appear when one
looks exactly in the center of the wormhole. When far from it, one
sees region~-5 in this direction, and when that region disappears at
outer horizon crossing, it is replaced by region~3. It is only when
one crosses the inner horizon that the observer can see (almost)
radial null geodesics from region~1 that have already bounced close to
the singularity and that now travel outward. Such geodesics to the not
belong to the same region as those that made the main image of
region~1 which are all ingoing once outer horizon has been
crossed. This second copy of region~1 first appears as a dot at inner
horizon crossing and further grows so as to touch the other, main,
view of region~1 (seen along directions $\delta > \delta_1^\IN$) when
the observer reaches $r = r_\SSS$. Then, only a single copy of
region~1 is seen filling the whole celestial sphere, although is it
still made of two types of geodesics: those that are ingoing and which
show a distorted but unflipped view of region~1 and those that are
outgoing and which show a flipped view of region~1 for the same
reasons that were explained in the naked singularity case
(\S\ref{sec_naked}).

\subsubsection{Size of region~3}

A similar reasoning allows to compute the angular radius,
$\delta_3^\IN$ of region~3, which becomes visible in region~2, between
the two horizons. The only difference is that in this case, the $t$
coordinate is past-oriented in region~3, which implies that the
constant of motion $E$ is negative. This amounts to change the sign
of the numerator in Eq.~(\ref{meq_hor}). The result is then
\begin{equation}
\label{delta_3_in}
\tan (\delta_3^\IN / 2)
 = \frac{- x - (x - x_\EE) 
               \sqrt{\displaystyle  1 + \frac{2 x_\EE}{x}
                                    - \left(\frac{\beta q}{x x_\EE}\right)^2}}
        {\beta \left(1 + \sqrt{\displaystyle  \frac{2}{x} - \frac{q^2}{x^2}} 
               \right)} 
 = \frac{x}{\beta} 
   \frac{-1 + \sqrt{\displaystyle 1 - \frac{\beta^2 A(x)}{x^2}}}
        {1 + \sqrt{\displaystyle  \frac{2}{x} - \frac{q^2}{x^2}}} .
\end{equation}
(Again, the derivation is adapted from the Schwarzschild case whose
derivation is given in Ref.~\cite{riazuelo15}.) This time, region~3
corresponds to geodesics endowed with an angle
$\delta < \delta_3^\IN$. The second formula shows that $\delta_3^\IN$
is defined (i.e., again, is positive) only when $A$ is negative, which
means that $r_- < r < r_+$. This means that region is visible only
from region~2. Also, $\delta_3^\IN$ and goes to $0$ at those two
values and close to those, one has
\begin{equation}
\label{size_reg3_hor}
\left. \delta_3^\IN \right|_{x = x_\pm \mp \Delta x}
 \sim \frac{\beta (x_+ - x_-) \Delta x}{2 x_\pm^3} .
\end{equation}
Seen as a function of coordinate distance $r$, the angular size of
this region shrinks faster as the observer gets near the inner horizon
than it had grown soon after outer horizon crossing. This is also true
in term of the observer's proper time since for the observers we are
considering, one has $\dot r^2 = 1 - A(r)$, so that
$\ddd r / \ddd \tau$ takes the same value (i.e., $1$) at both horizon
crossings.

\subsubsection{Size of region~-5}

The last bit of solution one might be interested in are geodesics from
region~-5. The derivation is the same as for region~1, except that all
these geodesics have bounced in the inner horizon region, which means
that they are always outgoing, regardless $r$ is larger or smaller
than $r_\EE$. The solution for these geodesics is then, after a few
manipulations,
\begin{equation}
\tan (\delta_{-5}^\IN / 2)
 = \frac{x + |x - x_\EE| \sqrt{\displaystyle 
                              1 + \frac{2 x_\EE}{x} 
                              - \left(\frac{\beta q}{x x_\EE}\right)^2}}
        {\beta \left(1 + \sqrt{\displaystyle  \frac{2}{x} - \frac{q^2}{x^2}} 
               \right)} ,
\end{equation}
It is unsurprisingly the same formula as $\delta_1^\IN$ for $r > r_\EE$
since there is no dark shell for those values of $r$, so that images
of region~-5 and 1 touch each others. The two formulae differ for
smaller values of $r$ as expected from the fact that the dark shell
phenomenon exists. For $r < r_+$, there are no solutions to this
equation, which means that region~-5 ceases to be visible when the
observer crosses the outer horizon, again as expected from the
Carter-Penrose diagram.

\subsection{Crossing the wormhole -- Outgoing part}

When considering an outgoing observer, the above procedure is still
valid, except that in the definition of $E$, we must consider an
outgoing observer, i.e., with a four-velocity $u_{\FF,+}^\mu$, which
amounts to change the sign in front of the square root of
Eq.~(\ref{meq_hor}). Moreover, any almost radial null geodesic coming
from region~1 will bounce before reaching the singularity and will
eventually intersect the observer's worldline. Therefore, directions
close to the ingoing radial direction will still show region~1, and
region~3 shall appear in the opposite direction when reaching
region~10. When leaving region~10, region~3 will disappear forever and
will be replaced by the new region the observer's journey ends to,
region~7. In order to compute angular size of all the observable
regions, we now need to consider an observer with velocity
$u_{\FF,+}$. The same machinery as before can be used, and the
simplest region to study is region~3. Taking care of all the signs,
one obtains that it is delineated by angle $\delta_3^\OUT$ given by
\begin{equation}
\tan (\delta_3^\OUT / 2)
 = \frac{- x + (x - x_\EE) 
               \sqrt{\displaystyle  1 + \frac{2 x_\EE}{x}
                                    - \left(\frac{\beta q}{x x_\EE}\right)^2}}
        {\beta \left(1 - \sqrt{\displaystyle  \frac{2}{x} - \frac{q^2}{x^2}} 
               \right)} .
\end{equation}
It is rather straightforward to check that
$\tan (\delta_3^\IN / 2) \tan (\delta_3^\OUT / 2) = 1$ (for this
purpose, one needs, again to use Eqns.~(\ref{int_beta})), so that we
have the very simple result
\begin{equation}
\delta_3^\OUT = \pi - \delta_3^\IN .
\end{equation}
In other words, when the outgoing observer lies between the two
horizons, region~3 appears not only at the opposite side but also with
the same angular size as it had for the same value of $r$ during the
ingoing phase. This result is valid for an observer whose initial
velocity is $u_{\FF,-}^\mu$ and who further follows a
geodesic. However, this is also true for any observer who follows any
geodesic, whether it is radial or not, or that it reaches infinity or
not. Indeed, considering another freely-falling observer with a
different initial velocity, he/she will have, when entering region~2,
a different view of region~3 because of aberration. However, once this
second observer will bounce back in region~6 and enter region~10, thus
seeing again region~3, the second observer's relative velocity with
respect to the first one will be exactly the same as in the infalling
phase, because in both case, the two observer's velocity differ only
by changing the sign of their $r$ component. Therefore, it is the same
Lorentz transform that allows to go from the first to the second
observer's velocity during the infalling and outgoing phase, for a
given $r$. Therefore, at given $r$, region~3 will have the same size
for any observer following a geodesic during the ingoing and outgoing
phases.

Regarding region~7, we have to solve Eq.~(\ref{meq_hor}), after having
flipped signs in front of the square roots, by keeping in mind that
what delineates region~7 are ingoing null geodesics as long as
$r < r_\EE$ and outgoing null geodesics afterwards. After a few
manipulations, one obtains
\begin{equation}
\tan (\delta_7^\OUT / 2)
 = \frac{x - (x - x_\EE) 
               \sqrt{\displaystyle  1 + \frac{2 x_\EE}{x}
                                    - \left(\frac{\beta q}{x x_\EE}\right)^2}}
        {\beta \left(1 - \sqrt{\displaystyle  \frac{2}{x} - \frac{q^2}{x^2}} 
               \right)} ,
\end{equation}
whose positive values are, as expected, defined only for $r > r_+$. As
for $\delta_1^\IN$, the expression tends toward
$\beta / u = b_\CRIT / r$ at large $r$.

The last bit of solution corresponds of what one sees of region~1
during the outgoing phases. When the observer is in region~6, the most
deflected null geodesics coming from region~1 are those that are
outgoing, and therefore delineate the edge of this region. Moreover,
as soon as the observer leaves region~6, all null geodesics that reach
him/her are outgoing, therefore it is those geodesics, with impact
parameter $b_\CRIT$, that need to be taken into account. Solving for
the last time Eq.~(\ref{meq_hor}) finally gives, for $r \geq r_\SSS$,
\begin{equation}
\label{delta_1_out}
\tan (\delta_1^\OUT / 2)
 = \frac{x - |x - x_\EE| 
               \sqrt{\displaystyle  1 + \frac{2 x_\EE}{x}
                                    - \left(\frac{\beta q}{x x_\EE}\right)^2}}
        {\beta \left(1 - \sqrt{\displaystyle  \frac{2}{x} - \frac{q^2}{x^2}} 
               \right)}   .
\end{equation}
This solution is identical to that of $\delta_7^\OUT$ for
$r \geq r_\EE$ since in that case, the dark shell phenomenon no longer
arise and images of regions~1 and 7 touch each other (this is the same
reasoning we outlined for regions~-5 and 1 during the ingoing
phase). For smaller values of $r$, they differ because of the dark
shell phenomenon, however, in this region we have, just as in the case
of region~3,
\begin{equation}
\tan (\delta_1^\IN / 2) \tan (\delta_1^\OUT / 2)
 = 1 \;,\qquad {\rm for\;} r < r_\EE , 
\end{equation}
so that
\begin{equation}
\delta_1^\OUT = \pi - \delta_1^\IN  \;,\qquad {\rm for\;} r < r_\EE .
\end{equation}
This means that the angular size of region~1 is identical in the
ingoing and outgoing phases, as long as region~1 and region~7 do not
touch each others. The first part of this result can be shown exactly
as we did for region~3. Regarding the last part, the fact that
region~1, and, hence, wormhole angular size does not match between the
infalling and outgoing phases is a mere consequence of the fact that in
both case a static observer would see the same thing, but here ingoing
and outgoing observers see of modified angular size with respect to
the static case because of aberration.

\begin{figure}[htbp]
\includegraphics*[angle=270,width=3.2in]{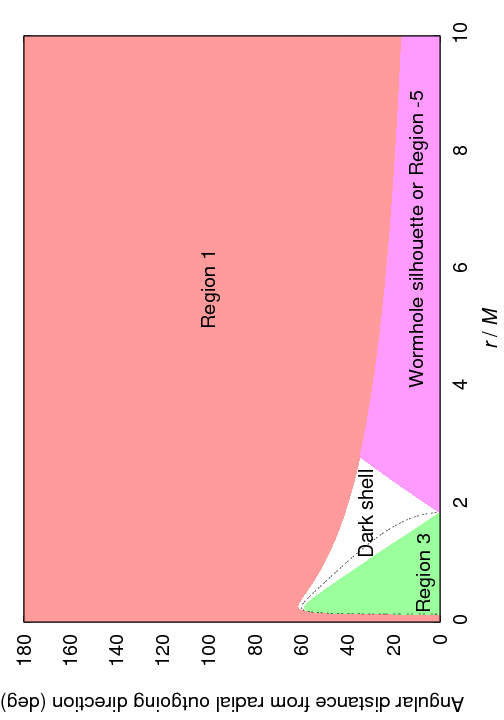}
\includegraphics*[angle=270,width=3.2in]{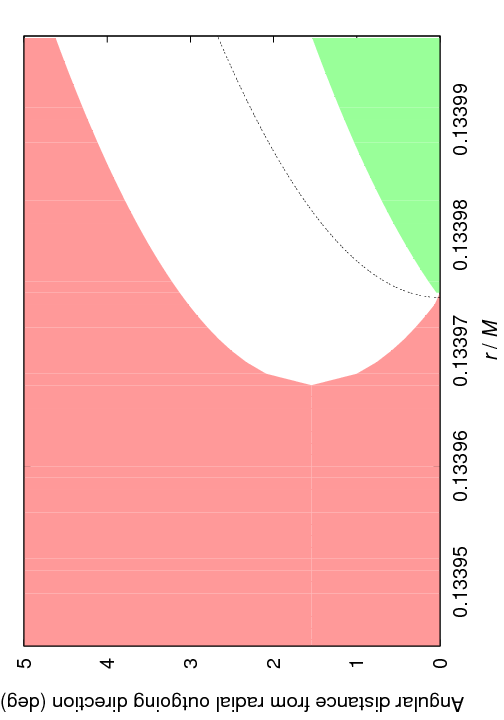}
\caption{Visibility and angular size of the different regions of the
  Carter Penrose diagram during a crossing of a Reissner-Nordstr\"om
  wormhole, in the ingoing phase ($\dot r < 0$). Angles separating the
  different regions are shown as a function of the radial coordinate
  $r$. The dark shell, that exists for $r_\SSS < r < r_\EE$ is always
  black. If we assume that nothing initially emerges from the wormhole
  at the beginning of the journey, then the ``region~-5'' part of the
  diagram is also black, and so is region~3 if one assumes that no
  light comes from there.  When this is not the case, then, as the
  observer enters the outer horizon, region~3 becomes visible, but its
  edges do not correspond to those of region~1 because of the dark
  shell phenomenon. Then, as the observer crosses the inner horizon,
  region~3 disappears, and almost immediately after (when $r$ goes
  below $r_\SSS$), so does the dark shell, leaving region~1 as the
  only thing that is visible, encompassing the whole celestial
  sphere. Right panel shows a zoom-in version in the vicinity of
  $r_\SSS$, $r_-$ whose values in units of $M$ are $0.133967$ and
  $0.133974$, respectively (we have chosen $|Q| / M$ = 0.5 here). In
  both panels, the dashed lines represents the angles toward which
  some radiation coming from past null infinity of any region would be
  seen with some infinite blueshift, a situation that never occurs,
  except along the radial outgoing direction at both horizon
  crossings.}
\label{fig_cross_in}
\end{figure}

\begin{figure}[htbp]
\includegraphics*[angle=270,width=4.0in]{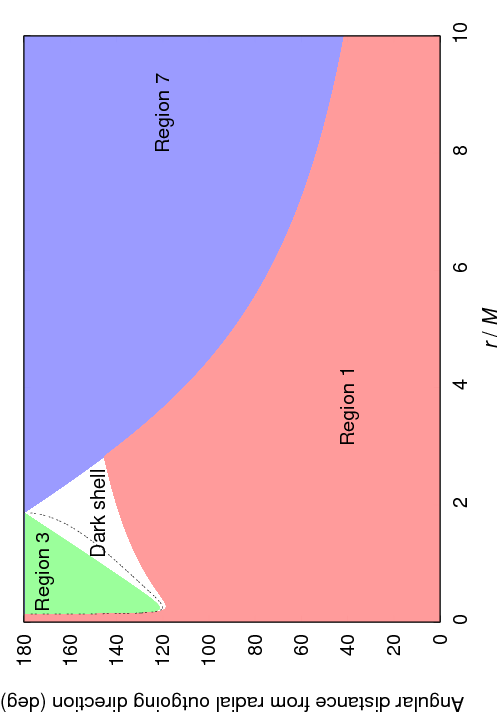}
\caption{Same as in Fig.~\ref{fig_cross_in}, but for the outgoing part
  of the wormhole crossing.  When the observer starts the outgoing
  part of the crossing, region~3 appears at $r = r_-$ (which
  corresponds to region~10 in Fig.~\ref{cp_diag}), together with the
  dark shell a hair earlier. Region~3 disappears when the observer
  exits the outer horizon, hence entering into region~7. At the exact
  moment region~3 disappears, region~7, which is the observer's final
  destination appears in the direction where region~3 was. Region~1 is
  still visible and, when entering region~7, separated from region~7
  by the dark shell, and joins region~7 at $r = r_\EE$. As the observer
  goes away from the wormhole, region~7 occupies a larger and larger
  portion of the sky, whose remaining part is occupied by region~1,
  now in the observer's causal past (which was the case from the
  moment the observer had entered the wormhole, actually).}
\label{fig_cross_out}
\end{figure}

All these results can be summarized in Fig.~\ref{fig_cross_in} for the
ingoing phase and Fig.~\ref{fig_cross_out} for the outgoing one.

\subsection{Computing redshifts}

The other quantity of importance is the redshift or blueshift of the
radiation as a function of the direction. From Eq.~(\ref{defE}), the
constant of motion $E$ of a photon starting from past null infinity of
region~1 or 7 is nothing more than the initial angular frequency,
$\omega$ of the corresponding wave. Therefore, the frequency shift of
the wave is given by
\begin{equation}
\left.\frac{\omega_\FO}{\omega} \right|_{1, 7}
 = \frac{\omega_\FO}{E}
 = \frac{1}{1 + s \sqrt{1 - A(r)} \cos \delta} ,
\end{equation}
where $s = -1$ corresponds to the ingoing phase of the wormhole
crossing, and $s = 1$ to the outgoing one. When considering photons
coming from region~3, then, because $t$ is a past-oriented timelike
coordinate in region~3, the constant of motion $E$ corresponds to the
opposite of the angular frequency. Therefore, the frequency shift of
photons coming from region~3 is
\begin{equation}
\left.\frac{\omega_\FO}{\omega} \right|_3
 = \frac{\omega_\FO}{- E}
 = \frac{1}{- s \sqrt{1 - A(r)} \cos \delta - 1} ,
\end{equation}
with the same value for $s$ as above.

\begin{figure}[htbp]
\includegraphics*[angle=270,width=3.2in]{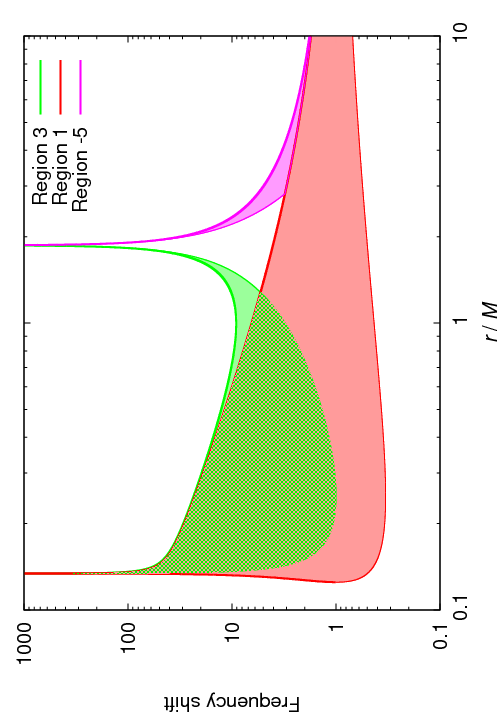}
\includegraphics*[angle=270,width=3.2in]{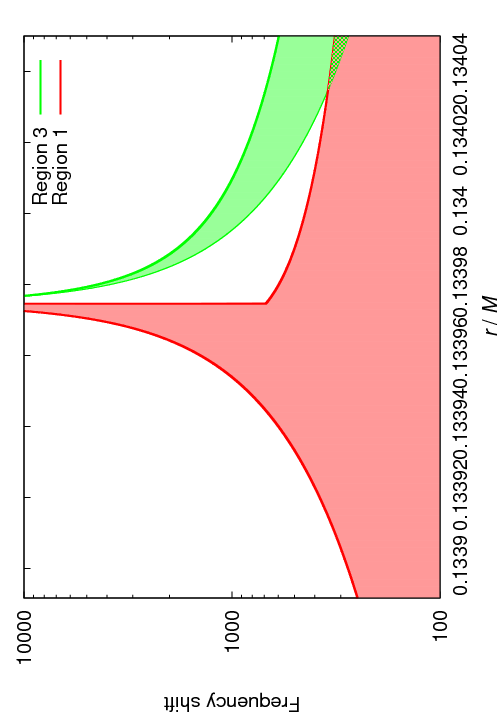}
\caption{Range of frequency shifts of regions~1 and 3 during the
  infalling phase into the wormhole (left panel). Frequency shift is
  infinite for region 3 at both horizon crossings, as well as for
  region~-5 when it disappears at outer horizon crossing, however it
  is large but finite for region~1 at inner horizon crossing and
  reaches its maximum value soon after, when the dark shell
  disappears. Right panel shows shows a zoomed-in view of the
  frequency shift in the vicinity of $r_\SSS, r_-$. Both panels are
  computed for $|Q| = M / 2$.}
\label{fig_zin}
\end{figure}

\begin{figure}[htbp]
\includegraphics*[angle=270,width=4.0in]{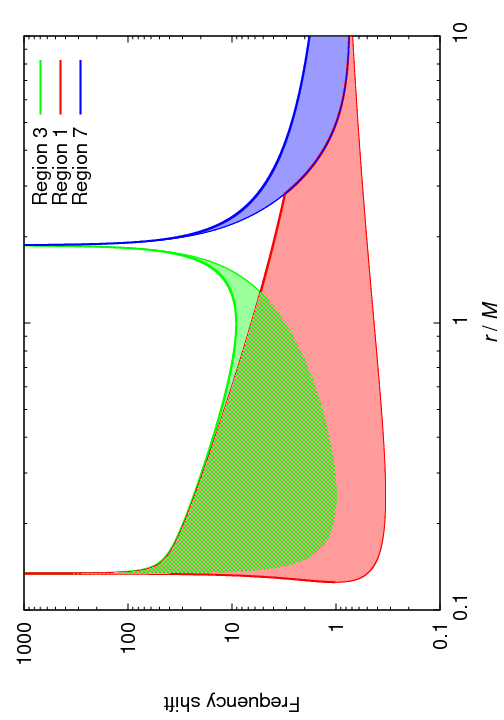}
\caption{Same as Fig.~\ref{fig_zin}, but for the outgoing phase. The
  fact that, at fixed $r$, the frequency shift range is the same for
  most regions during the infalling and outgoing phases is explained
  in the text.}
\label{fig_zout}
\end{figure}

Several features can be seen from these two formulae:

\begin{itemize}

\item One sees immediately that for each event of the wormhole
  crossing, the frequency shift is always equal to 1 in any direction
  perpendicular to the radial one, a result that matches Newtonian
  physics, but not special relativity.

\item From the first equation, we see that the radial null geodesics
  that catch the observer ``from behind'' ($\delta = \pi$) are, as can
  be guessed intuitively, redshifted, by a factor equal to
  $1 + \sqrt{1 - A (r)}$.  The redshift is negligible at large
  distance since the observer velocity with respect to distant sources
  is weak and so is the gravitation blueshift. Then, the redshift
  reaches $1$ when entering the outer horizon (since $A (r) = 0$
  there, so that $\sqrt{1 - A} = 1$), increases further till the
  minimum negative value of $A(r)$ which occurs at $r = Q^2 / M$, for
  which the redshift is $M / |Q|$. It then decreases to $1$ at inner
  horizon crossing and reaches 0 when the observer bounces at
  $r = r_\MIN^\FF$.  Once the observer is in his/her outgoing phase,
  radiation catching him/her from behind (which, this time, correspond
  to $\delta = 0$) is observed at a redshift which follows the same
  variation, so that once the observer exits and goes far from the
  wormhole, region~1 is seen without significant redshift.

\item The case of radiation coming from region~1 seen by the observer
  in front of him/her is more interesting. During the ingoing phase,
  there may by some radiation seen by the observer from direction
  $\delta = 0$, which sits in the middle of the wormhole silhouette;
  if one assumes that radiation comes from region~-5. In this
  case\footnote{Strictly speaking such geodesics do not exist in the
    sense that pure radial null geodesics coming from region~-5 or,
    later, region~1, hit the singularity rather than bouncing on it,
    however, null geodesics with a vanishingly small impact parameter
    can be seen. We shall therefore consider that geodesics are seen
    toward the $\delta = 0$ direction.}, the radiation is blueshifted
  and reaches an infinite blueshift as region~-5 disappears, when the
  observer crosses the outer horizon. At this instant, the observer
  sees radiation from region~3 which first appears infinitely
  blueshifted as well, but further, the blueshift decreases as the
  observer travels along region~2. The frequency shift reaches its
  minimum value at $r = Q^2 / M$, and is equal to $[Q| / (M - |Q|)$;
  which corresponds to either a redshift or a blueshift, depending on
  whether $|Q|$ is below of above $M / 2$. Further, the frequency
  shift diverges again when the observer reaches the inner
  horizon. Then, the flipped copy of region~1 appears with an infinite
  blueshift in its center, and the frequency shift decreases rather
  abruptly since, when the observer approaches the bouncing point at
  $r = r_\MIN^\FF$, it goes to $1$ and no blueshift or redshift is
  seen in any direction.

\item The outgoing phase give the same results are before except that
  they occur in opposite ($\delta' = \pi - \delta$) direction.

\item One may wonder whether there are some other sets of $r, \delta$
  which show infinite blueshift, at horizon crossing toward
  $\delta = 0$ in the ingoing phase and $\delta = \pi$ in the outgoing
  phase, but this is not case. Indeed, such infinite blueshift would
  occur when $\cos \delta = \pm (1 - A(r))^{-\frac{1}{2}}$, a
  situation that can only occur when $1 - A(r) > 1$, that is, when $A$
  is negative, i.e., between the two horizons, in regions~2 and
  10. Further, solving the equation
  $\cos \delta = \pm (1 - A(r))^{-\frac{1}{2}}$ in term of
  $\tan \delta / 2$ gives the same second order equation as in
  Eq.~(\ref{meq_hor}) except that the term in front of
  $\tan^1 \delta / 2$ is zero, which ensures that the curves
  $\cos \delta = \pm (1 - A(r))^{-\frac{1}{2}}$ never cross any of the
  $\delta_{1, 3}^{\IN, \OUT}$ except when $\delta = 0$ in the ingoing
  phase and $\delta = \pi$ in the outgoing phase, so that the curve
  $\cos \delta = \pm (1 - A(r))^{-\frac{1}{2}}$ is either entirely in
  region~1, or entirely in region~3 or in the dark shell region. It
  then suffices to compute all the $\delta$'s for a single value of
  $r$ (the simplest ones being $r = M$ and $r = Q^2 / M$) to show that
  the curve of infinite blueshift lies in the dark shell region, so
  that no infinite blueshift are seen, except, as we said, at the
  disappearance and appearance of regions~-5, 1, 3 and 7 at either
  $r = r_-$ or $r = r_+$.

\end{itemize}
The maximum and minimum frequency shift of each region observable
during wormhole crossing is shown in Figures~\ref{fig_zin}
and~\ref{fig_zout}.

A natural question that arises is which of the fluxes from these
regions (except when then actually diverge) is the strongest, at some
small but finite coordinate distance $|\Delta r|$ from the
horizons. In this respect, region~1 differs from all the others by the
fact that all the latter exhibit an infinite blueshift at horizon
crossing in the same time they have a vanishingly small angular size,
whereas region~1 has a macroscopic angular size at inner horizon
crossing, which suggests that the overall flux received from it will
be larger. Also, angular size of region~3 does not vary at the same
rate at inner and outer horizon crossings, so that its flux at
$r_+ - \Delta r$ should differ from that at $r_- + \Delta r$. A crude
estimate of these two results can be performed as follows. 

Close to horizon crossings, the frequency shift of radiation coming
from any of the region visible from a direction close to $\delta = 0$
is given by
\begin{eqnarray}
\label{eq_approx_flux_1}
\left.\frac{\omega_\FO}{\omega} \right|_1 & \simeq &
\frac{2}{A(r) + \delta^2} , \\
\label{eq_approx_flux_3}
\left.\frac{\omega_\FO}{\omega} \right|_1 & \simeq &
\frac{2}{|A(r)| - \delta^2} .
\end{eqnarray}
The total fluxes ${\mathcal F}_1, {\mathcal F}_3$ coming from
regions~1 or 3 is given by the fourth power of the above equation
integrated on the total angular area of the corresponding
region. Since Eq.~(\ref{eq_approx_flux_1}) is large only when
$\delta \lesssim \sqrt{A(r)}$, the total flux ${\mathcal F}_1$ is of
order
\begin{equation}
\label{eq_F1}
{\mathcal F}_1
 \simeq \int_0^{\sqrt{A}} \frac{32 \pi \delta \ddd \delta}{A^4}
 \simeq \frac{16 \pi}{A^3} .
\end{equation}
The same reasoning holds for region~3, except that the integral is
now limited by the actual angular size of this region, which is obtained
by expanding Eq.~(\ref{delta_3_in}), which gives
\begin{equation}
\delta_3^\IN \simeq \frac{|A| \beta}{2 x_\HOR} ,
\end{equation}
where $x_\HOR$ corresponds to either $r_\pm / M$, depending on whether
one consider inner or outer horizon. Consequently, the flux from
region~3 is proportional to
\begin{equation}
\label{eq_F3}
{\mathcal F}_3
 \simeq \int_0^{\delta_3^\IN} \frac{32 \pi \delta \ddd \delta}{A^4}
 \simeq \frac{4 \beta^2 \pi}{x_\HOR^2 A^2} .
\end{equation}
Further expanding $A (r)$ close to $r_\HOR$, gives
\begin{equation}
\left|A (r_\pm \mp M \Delta x) \right|
 = \frac{x_+ - x_-}{x_\pm^2} | \Delta x | ,
\end{equation}
so that the ratio between the two fluxes from region~3 at outer and
inner horizon crossing is
\begin{equation}
\frac{\left.{\mathcal F}_3\right|_{x= x_+ - \Delta x}}
     {\left.{\mathcal F}_3\right|_{x= x_- + \Delta x}}
 \simeq \frac{x_+^2}{x_-^2} ,
\end{equation}
a result that is in fact qualitatively opposite to the naive
expectation for the angular size of region~3 in both cases.

Regarding region~-5 and region~3 at outer horizon crossing, the
angular size vary in the same way, so that their fluxes diverge in a
very similar way.

\begin{figure}[htbp]
\includegraphics*[angle=270,width=3.2in]{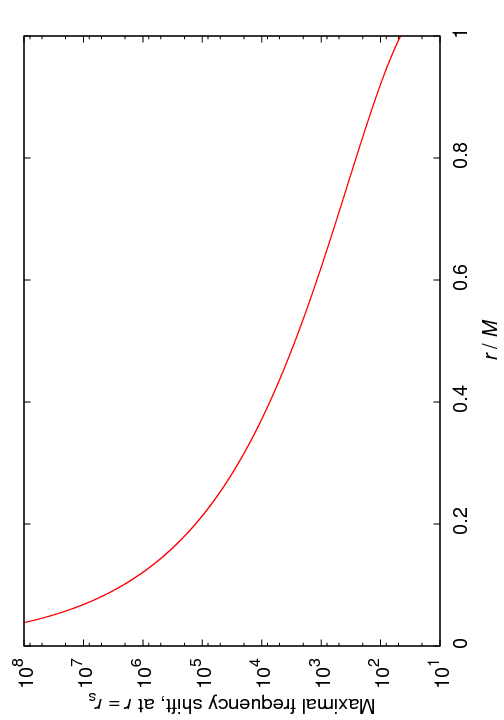}
\caption{Maximal frequency shift see of region~1 when crossing a
  Reissner-Nordstr\"om wormhole. This frequency shift is obtained when
  $r = r_\SSS$ and is always finite, although very large, even in the
  most favourable case of an extremal wormhole.}
\label{fig_shiftmax}
\end{figure}

The last case of interest is how the flux from region~1 behaves close
to inner horizon crossing with respect to that of region~3. Using all
the results above, we have
\begin{equation}
\frac{\left.{\mathcal F}_1\right|_{x= x_- - \Delta x}}
     {\left.{\mathcal F}_3\right|_{x= x_- + \Delta x}}
 \simeq \frac{4 x_-^4}{\beta^2 (x_+ - x_-) \Delta x} ,
\end{equation}
so that the flux from region~1 just after inner horizon crossing is
much larger than that of region~3 just before. It is then possible to
compute for which interval of $\Delta r$ (or, equivalently,
$\Delta x$), this holds, and it happens that it corresponds to
$\Delta x = x_- - x_\SSS$. Indeed, as we already said, $r_-$ and
$r_\SSS$ are very close to each other, as can be seen by expanding
both $x_\SSS = r_\SSS / M$ and $x_- = r_- / M$ as a function of
$|Q| / M = q$. This gives
\begin{eqnarray}
\label{dl_xm}
x_-
 & = &   \frac{q^2}{2} + \frac{q^4}{8} + \frac{q^6}{16} + \frac{5}{128} q^8 
       + \frac{7}{256} q^{10} + O(q^{12}) , \\
\label{dl_xs}
x_\SSS
 & = &   \frac{q^2}{2} + \frac{q^4}{8} + \frac{q^6}{16}
       + \frac{131}{27 \times 128} q^8 + \frac{523}{81 \times 256} q^{10}
       + O(q^{12}) .
\end{eqnarray}
The two quantities differ only at the $q^8$ level, and even then, by a
narrow amount since their respective coefficients differ by
$\frac{5}{128} - \frac{131}{27 \times 128} = \frac{1}{27 \times 32}$,
that is a $\sim 3\%$ difference, a situation that, incidentally, also
applies for the $q^{10}$ terms, whose coefficients are also of similar
amplitude ($0.2734$ and $0.02522$, respectively, a $5.5\%$
difference). Considering $\Delta x = x_- - x_\SSS$ leads to
\begin{equation}
\frac{\left.{\mathcal F}_1\right|_{x= x_- - \Delta x}}
     {\left.{\mathcal F}_3\right|_{x= x_- + \Delta x}}
 \simeq \frac{4 x_-^4}{\beta^2 (x_+ - x_-) \frac{q^8}{27 \times 32}} .
\end{equation}
One can simplify further this expression by (somehow crudely)
approximating $x_-$, as well as $x_+ - x_-$ and $\beta$ by their
values at lowest order in $q$, i.e., $q^2/2$, $2$ and $3 \sqrt{3}$,
respectively. One then obtains
\begin{equation}
  \frac{\left.{\mathcal F}_1\right|_{x= x_- - \Delta x}}
  {\left.{\mathcal F}_3\right|_{x= x_- + \Delta x}}
  \simeq 4 + O (q^2) .
\end{equation}
This means that after $x = x_\SSS$, the flux dissymmetry between
before and after inner horizon crossing is weak, although the overall
flux is fairly large, as can be seen by computing the maximal (i.e.,
toward $\delta = 0$) frequency shift at $x_\SSS$, which is given by
\begin{equation}
\label{omfo_om}
\left.\frac{\omega_\FO}{\omega}\right|^\MAX_{1, r = r_\SSS}
 = \frac{1}{1 - \sqrt{1 - A (r_\SSS)}}
 \simeq \frac{432}{q^4} - \frac{180}{q^2}+ 14 + O (q^2) .
\end{equation}
Even when $q$ is large (i.e., close to 1), the above expression is
large. In this limit, one has $r_\EE (q = 1) = 2 M$, $r_\pm = M$ and
$r_\SSS = (2 \sqrt{2} - 1) M$, which gives
$( 1 - \sqrt{1 - A (r_\SSS)})^{-1} \sim 46.12$. The frequency shift in
the front direction at $r = r_\SSS$ is shown in
Figure~\ref{fig_shiftmax}.

\subsection{Solving the geodesic equation}

An observer starting from region~1 will of course see (i.e., be
intersected by null geodesics originating from) region~1, but will
also eventually see the region from which null geodesics originate
when they exit the past $r = r_+$ horizon. With our conventions, such
region is labelled -5. Depending on whether there is anything in -5,
the metric will either appear as in the standard (i.e., black hole
type) Reissner-Nordstr\"om case when the silhouette delineating of the
central part of the metric is perfectly black and where region~1 is
distorted, or, alternatively, the same distortion in region~1 but this
time a highly distorted view of region~-5.

In order to compute this distortion, we have to adapt the set of
equations~(\ref{rrn_deb}--\ref{rrn_fin}) to the case where horizons
are crossed, i.e., when the function $A(r)$ is zero. As well-known,
one has to switch to another system of coordinates as the $(t, r)$
coordinates exhibit a coordinate singularity at horizon
crossing. Although the coordinate change is outlined in several
textbooks, it is almost never done with sufficient detail to allow a
straight implementation, therefore we shall give a more-than-usual
detailed derivation of it.

The first step consists into transforming the radial coordinate $r$
into the so-called tortoise coordinate $r^*$ so that that line element
can be written under the form
$A(r^* (r)) (\ddd t^2 - \ddd r^*{}^2) - ...$\,.
Writing $\ddd r^* = \ddd r / A(r)$ leads to
\begin{equation}
\ddd r^* = \ddd r \left(1 + \frac{X_+}{r - r_+} + \frac{X_-}{r - r_-} \right) ,
\end{equation}
where we have set
\begin{equation}
X_+ = \frac{r_+^2}{r_+ - r_-} , \qquad X_- = \frac{r_-^2}{r_- - r_+} .
\end{equation}
We integrate this into
\begin{equation}
r^* = r + X_+ \log \left|\frac{r}{r_+} - 1 \right| 
        + X_- \log \left|\frac{r}{r_-} - 1 \right| ,
\end{equation}
from which one sees that $r^*$ is a growing function of $r$ for
$r < r_-$ or $r > r_+$, and a decreasing function for $r_- < r < r_+$.
This coordinate system can be made regular around one of the values
$r_+$, $r_-$ (but not both simultaneously) by exponentiating it and
performing a careful choice of signs.  We therefore define
\begin{eqnarray}
\label{debdefUV}
U_- & = & \varepsilon_- \exp \left(\frac{t - t_0 + r^*}{2 X_-} \right) , \\
V_- & = & \eta_- \exp \left(\frac{-t + t_0 + r^*}{2 X_-} \right) , \\
U_+ & = & \varepsilon_+ \exp \left(\frac{t - t_0 + r^*}{2 X_+} \right) , \\
V_+ & = & \eta_+ \exp \left(\frac{-t + t_0 + r^*}{2 X_+} \right) ,
\label{findefUV}
\end{eqnarray}
and we need to choose the values of
$\epsilon_+, \varepsilon_-, \eta_+, \eta_- = \pm 1$ so as to ensure
that coordinate system $(U_-, V_-)$ is regular around $r_-$ and so is
$(U_+, V_+)$ around $r_+$. These signs of course depend on the region
one dwells in. After some tinkering, one obtain the values given in
Table~\ref{signes}, which insure that in any region where they need to
be defined, $U_\pm$ and $V_\pm$ are future-oriented null
coordinates. These coordinates are only of interest when one is in one
region which has one edge in common with the $r = r_+$ (for
$U_+, V_+$) or $r = r_-$ (for $U_-, V_-$) lines, so that it is
unnecessary to define, say, $U_-$ and $V_-$ in region~1.
\begin{table}
\begin{center}
\begin{tabular}{|c|c|c|}
\hline 
Region neighbouring $r_+$ & $\varepsilon_+$ & $\eta_+$  \\ \hline
1 & 1 & -1 \\ \hline
2 & 1 & 1 \\ \hline
3 & -1 & 1 \\ \hline
4 & -1 & -1 \\ \hline
\end{tabular}
\qquad
\begin{tabular}{|c|c|c|}
\hline 
Region neighbouring $r_-$ & $\varepsilon_-$ & $\eta_-$  \\ \hline
5 & 1 & -1 \\ \hline
10 & 1 & 1 \\ \hline
6 & -1 & 1 \\ \hline
2 & -1 & -1 \\ \hline
\end{tabular}
\end{center}
\caption{Sign conventions that make coordinate changes of
  Eqns.~(\ref{debdefUV}--\ref{findefUV}) consistent. The region numbers 
  that are given are to be understood modulo~6 since any region in the 
  next set of six patches of the Carter-Penrose diagram behaves in the 
  same way as its upward or downward neighbour with respect to the 
  coordinates we use throughout this paper. }
\label{signes}
\end{table}
Most if not all paper which give this coordinate transform do not
introduce the time constant $t_0$ in
Eqns.~(\ref{debdefUV}--\ref{findefUV}), however it is crucial to do so
when one actually solves this set of equations with any standard
numerical method.  This is because without the adjunction of this
constant in the coordinate transform, one usually has, just after
performing the coordinate change, a large ratio between $U$ and $V$, a
situation that very significantly hinders the precision of any
numerical methods we have tested for solving this set of equation. In
practice, we choose the $t_0$ constant as the value of $t$ at the
moment we perform the coordinate change.

When one performs the coordinate change, one also must do it with the
coordinate derivatives of the geodesics we are interested in. They are
written as, by using a dot to denote a derivative with respect to the
geodesic affine parameter,
\begin{eqnarray}
\dot U & = & \frac{U}{2 X} \left(\dot t + \frac{\dot r}{A (r)} \right), \\
\dot V & = & \frac{U}{2 X} \left(- \dot t + \frac{\dot r}{A (r)} \right) ,
\end{eqnarray}
where we have dropped the $+$, $-$ subscripts in front of $U, V,
X$, keeping in mind that they are the same everywhere.

Regarding the inverse transform, the case of $t$ is rather easy since
one has immediately
\begin{eqnarray}
t & = & t_0 + X \log \left|\frac{u}{v}\right| , \\
\dot t & = & X \left(\frac{\dot U}{U} - \frac{\dot V}{V} \right) .
\end{eqnarray}
One then has
\begin{equation}
\dot r^* = X \left(\frac{\dot U}{U} + \frac{\dot V}{V} \right) .
\end{equation}
(In this equation as well as the previous ones, we dropped the $+$ or
$-$ sign which must be the same everywhere.) The next step is to notice
that the product $U V$ can be written
\begin{eqnarray}
\label{eq_UVp}
U_+ V_+ & = & - \exp \left(\frac{r}{X_+}\right) 
               \times \left( \frac{r}{r_+} - 1\right)
               \left| \frac{r}{r_-} - 1\right|^\frac{X_-}{X_+} ,  \\
\label{eq_UVm}
U_- V_- & = & - \exp \left(\frac{r}{X_-}\right) 
               \times \left( \frac{r}{r_-} - 1\right)
               \left| \frac{r}{r_+} - 1\right|^\frac{X_+}{X_-} . 
\end{eqnarray}
(In the first equation, one gets rid of the absolute value of
$(r / r_+ - 1)$ because the sign of the product $\varepsilon_+ \eta_+$
is the opposite of that of $(r / r_+ - 1)$, hence the minus sign in
front of the result; the same holds with $(r / r_- - 1)$ for the second
equation.) None of these equations allow an analytical solution,
however, they can be solved by any standard method (Newton-Raphson,
etc.). The last step is then
\begin{equation}
\dot r = A(r) \dot r^* .
\end{equation}
This being set, we need to use the coordinates $U$ and $V$ to solve
the geodesic equation. It is not possible to write these equations in
a fully closed form that does not depend on $r$, therefore $r$ must be
understood as a function of the product $U V$, see
Eqns.~(\ref{eq_UVp},\ref{eq_UVm}). Dropping the $+, -$ subscript
which, again, are the same everywhere, one has
\begin{eqnarray}
\ddot U
 & = &   \frac{V F'}{2 X} \dot U^2
       + \frac{r U}{2 X} (\dot \theta^2 + \sin^2 \theta \dot \varphi^2 ) , \\
\ddot V
 & = &   \frac{U F'}{2 X} \dot V^2
       + \frac{r V}{2 X} (\dot \theta^2 + \sin^2 \theta \dot \varphi^2 ) , \\
\ddot \theta
 & = &   \frac{F}{X r} (\dot U V + U \dot V) \dot \theta
       + \sin \theta \cos \theta \dot \varphi^2 , \\
\ddot \varphi
 & = &   \frac{F}{X r} (\dot U V + U \dot V) \dot \varphi
       - 2 \frac{\cos \theta}{\sin \theta}  \dot \theta \dot \varphi ,
\end{eqnarray}
where we have used the function $F$, which has to be understood as a
function of $r$ only (and hence $U V$), defined as
\begin{equation}
\label{def_F}
F = - \frac{2 X^2 A(r)}{U V} ,
\end{equation}
its derivative $F'$ with respect to $r$ being expressed under the most
compact form
\begin{equation}
\label{def_Fp}
F' = \frac{F}{A} \left(A'(r) - \frac{1}{X} \right) .
\end{equation}
Despite the way they are defined, $F$ or $F'$ are not singular, either
when $U V = 0$ in~Eq.(\ref{def_F}) or $A = 0$ in Eq.~(\ref{def_Fp}),
because in both cases, their numerator of $F$ or $F'$ is also zero and
all the quantities we deal with are continuous here (except of course
for the usual coordinate singularity of the spherical
coordinates). Indeed, this is exactly this set of equations we solve
using some standard integrators~\cite{press92} without encountering
significant issues (except when we made an improper choice of $t_0$).

\subsection{Computing images}

Showing all the features that arise during the crossing of a wormhole
is quite demanding as many worthwhile details are seen in very
different directions. We therefore chose to perform a fish-eye view of
such journey, and made our computations in Domemaster format images,
that is the current standard for digital planetariums. Images computed
that way fill a half sphere, that is $2 \pi$ steradians, a value which
is obviously barely sufficient for our purpose, since the most
interesting features of the infalling part of the crossing are toward
$r = 0$ whereas the equally interesting features during the outgoing
part of the journey are in the opposite direction. Therefore, we
rotate the view during the journey. During all the infalling phase,
the center of the coordinate system, where the wormhole lies, is shown
45~degrees above the front of the audience. When the observer bounces
in region~6 the interesting part of the view corresponds to the
opposite direction, and therefore would be behind the audience,
45~degrees offscreen under the edge of the screen. We therefore slowly
rotate the view by 90~degrees during the outgoing phase of region~6,
so that when exiting region~6, the outgoing direction is onscren,
45~degrees below the upper edge of the screen, which means behind the
audience. Then, as the observer exits the wormhole and enters
region~7, we rotate back the view by 90~degrees so that the
center of the coordinate systems which still shows the wormhole and
the region from which the observer comes from is again in front of the
audience, 45~degrees above the edge of the screen. We have computed a
whole movie of 2500 individual frames, some of which are shown here.
In order to more easily distinguish the different regions, we used the
following data for each of these:
\begin{itemize}
\item Region~1 corresponds to the Milky Way seen from the Solar System
  (without the Sun nor the planets). We use 2MASS survey starless
  celestial sphere, to whom we add a 200k star catalog (more details
  are given in Ref.~\cite{riazuelo15}).

\item Region~-5 is simulated by a random star catalog whose properties
  in term of number of stars vs. magnitude are similar to that of
  region~1. However, no pixellized celestial sphere is included, which
  is sufficient to distinguish between the two as region~-5 is
  somewhat darker than region~1.

\item Region~3 is simulated using another mock star catalog with the
  same properties as above. The background pixellized celestial sphere
  is (rather arbitrarily) a full-sky false color Cosmic Microwave
  Background map which is colourful enough to be distinguished from
  region~1 and~-5.

\item Region~7 is simulated with, again, a third mock star catalog,
  and with a coordinate grid playing the role of the celestial sphere.
\end{itemize}
Careful implementation of all the special and general relativistic
effects proved to be almost impossible because of the large blueshift
experienced in some parts of the trajectory. We therefore significantly
attenuated the flux increase associated with frequency shift by
implementing a flux increase proportional to $\omega_\FO / \omega$
instead of $(\omega_\FO / \omega)^4$, a choice which gave a satisfying
rendering, although at the cost of some qualitative differences with
respect to the analytical calculations we performed earlier: (i) all
the integrated fluxes are finite at horizon crossing; (ii) the flux
from region~3 soon after outer horizon crossing is now smaller than
that from region~3 just before inner horizon crossing. Those two
differences come from the fact that the powers of $A(r)$ in
Eqns.~(\ref{eq_F1}, \ref{eq_F3}) are different due to our attenuation
of the flux.

\begin{figure}[htbp]
\includegraphics*[width=3.2in]{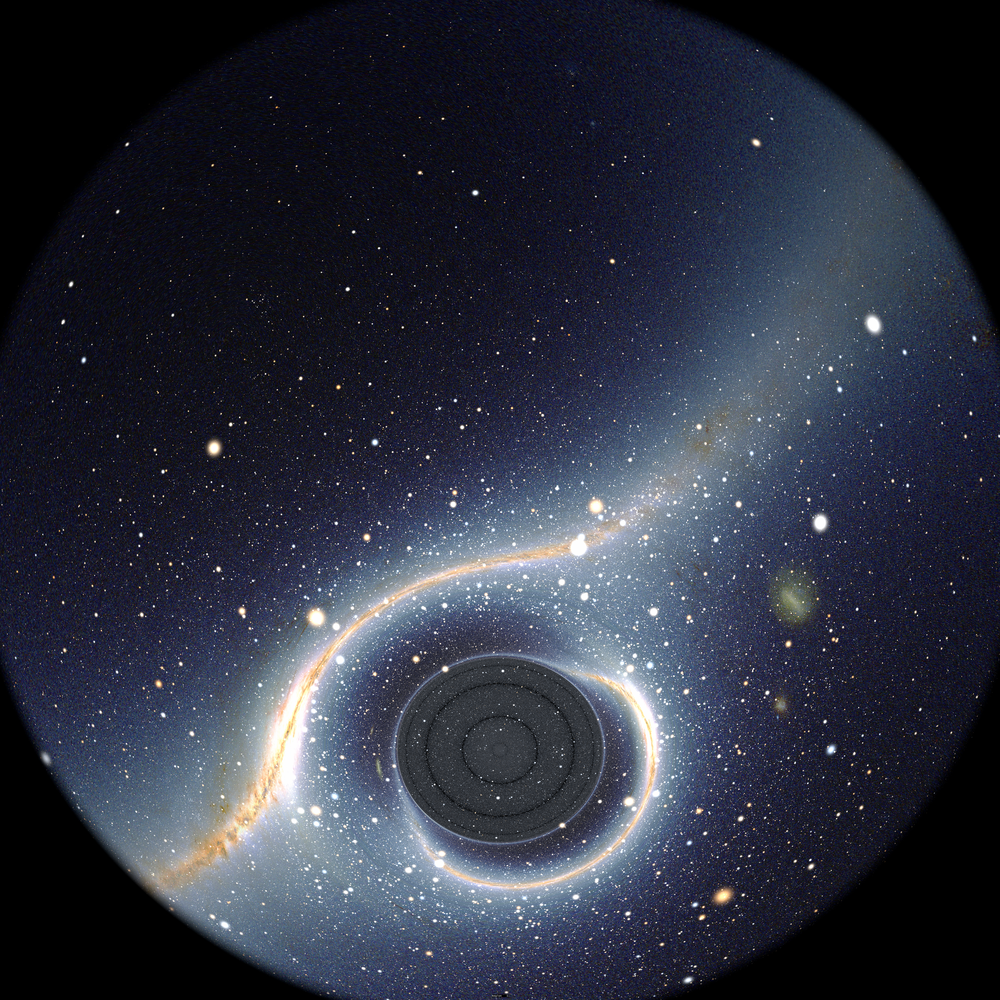}
\includegraphics*[width=3.2in]{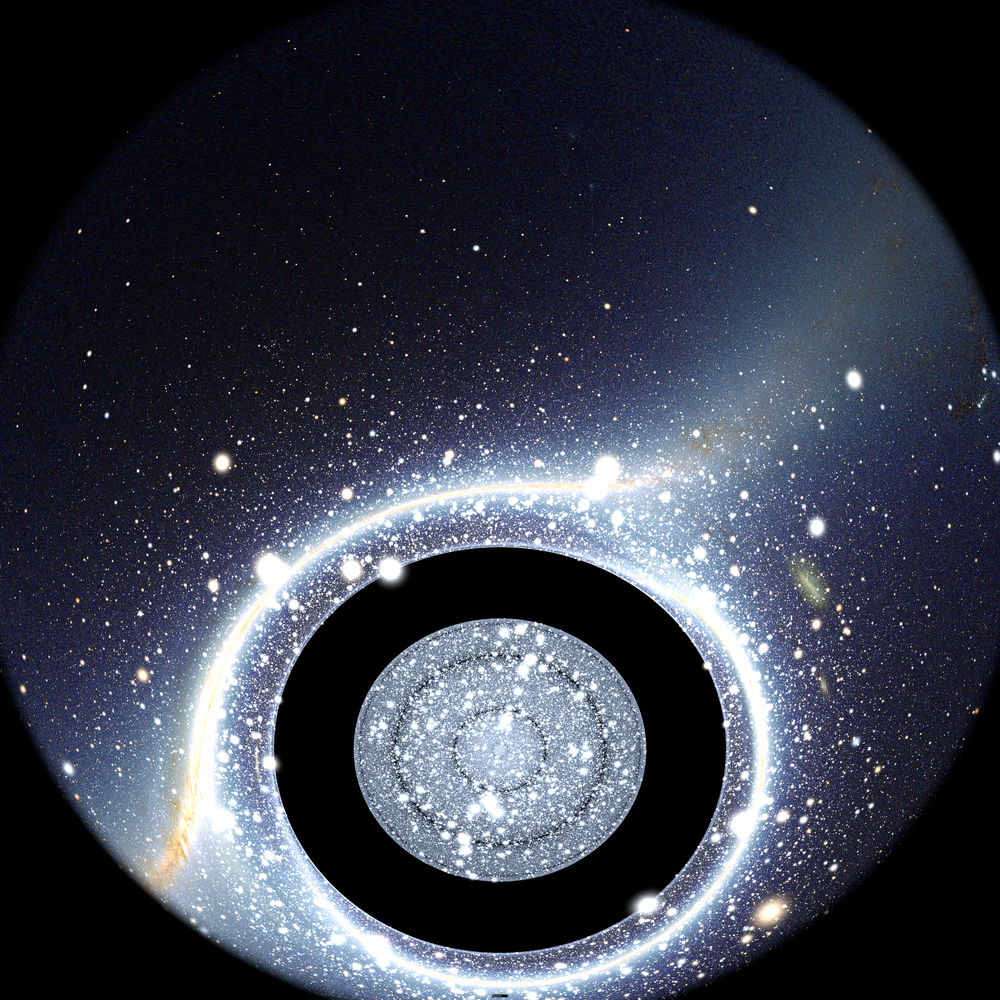}
\vskip 0.12cm
\includegraphics*[angle=270,width=3.2in]{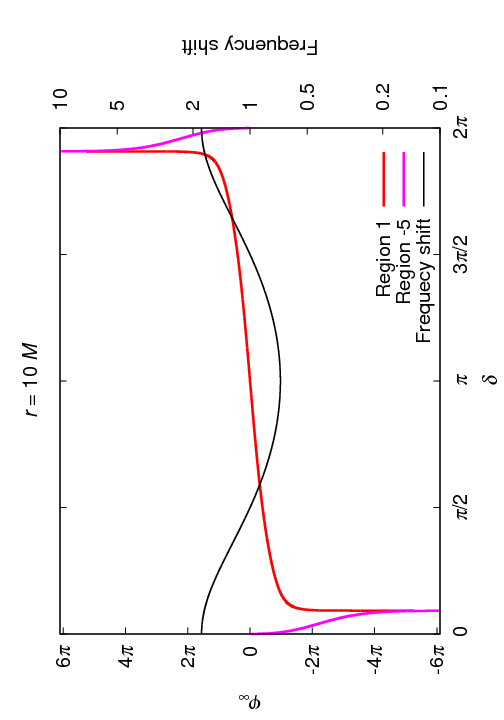}
\includegraphics*[angle=270,width=3.2in]{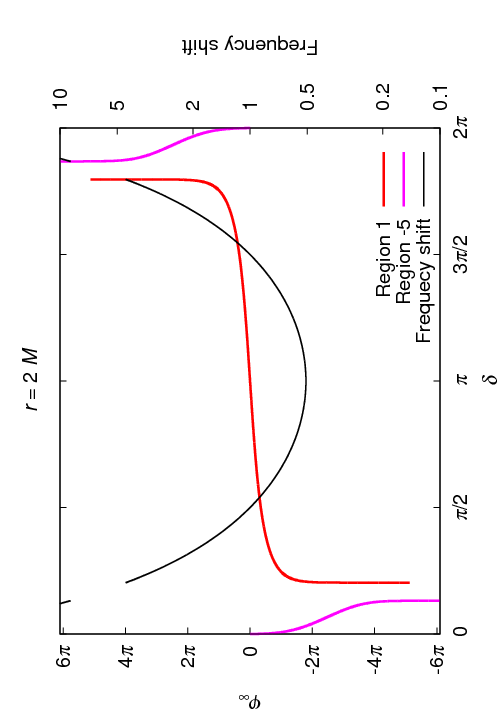}
\caption{First part of the infalling phase. The observer lies at
  $r = 10 M$ (left) and $r = 2 M$ (right). Region~1 occupies most of
  the celestial sphere except for the wormhole silhouette. When the
  observer is sufficiently for from the wormhole ($r > r_\EE$, which
  is the case of left image), image of region~-5, which is in the
  observer's causal past, touches that of region~1. When the observer
  is closer to the wormhole, bound null geodesics translate into the
  dark shell that separate both regions. Angular size of the wormhole
  (including the dark shell) increases as the observers travels closer
  to it, but that of region~-5 decreases once the dark shell
  appears. The fact that the observer's velocity increases translates
  into brighter stars with a bluer hue toward the wormhole.}
\label{fig_cross_1}
\end{figure}

\begin{figure}[htbp]
\includegraphics*[width=3.2in]{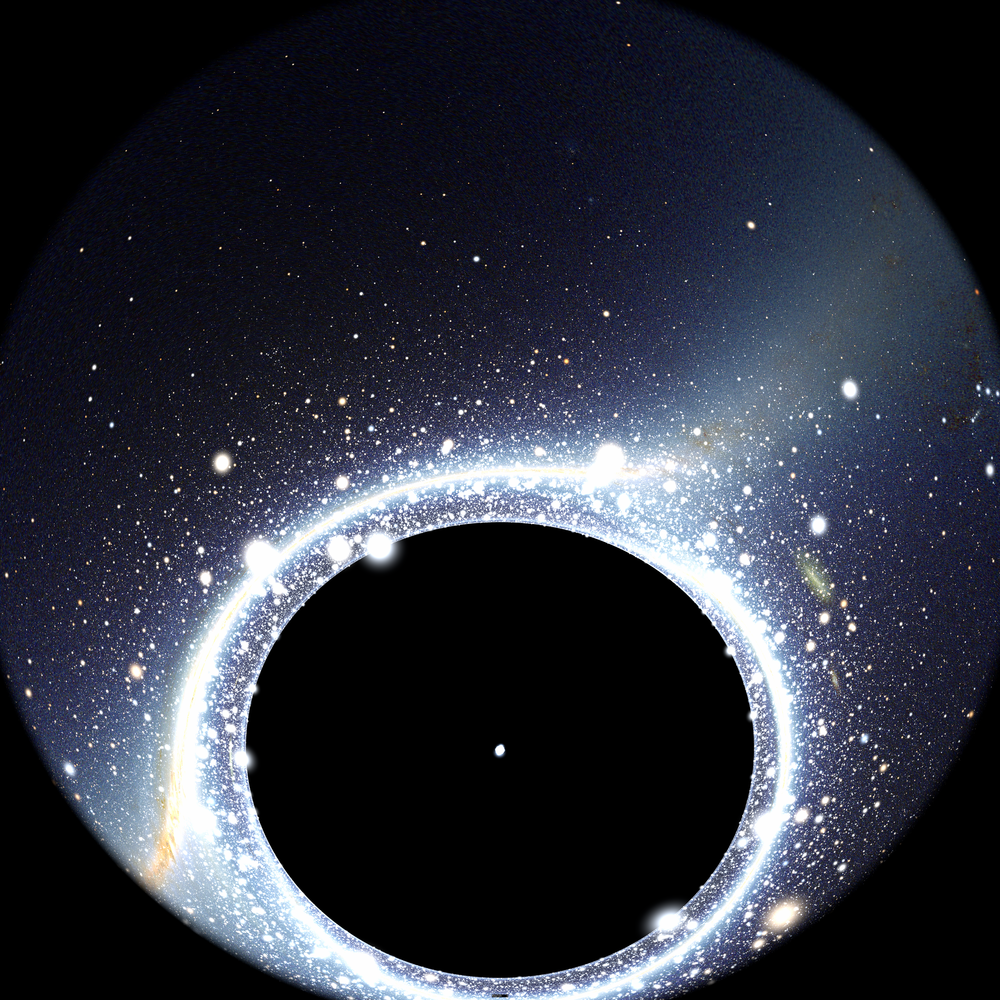}
\includegraphics*[width=3.2in]{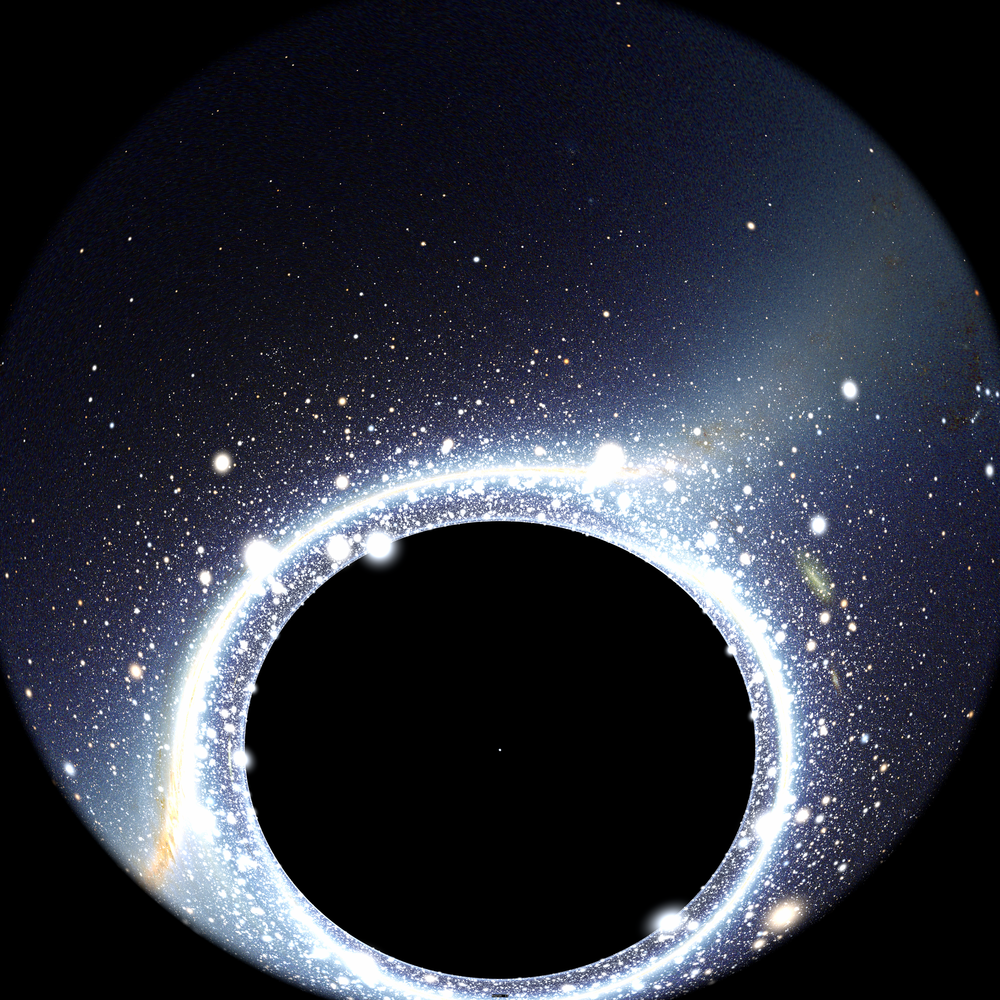}
\vskip 0.12cm
\includegraphics*[angle=270,width=3.2in]{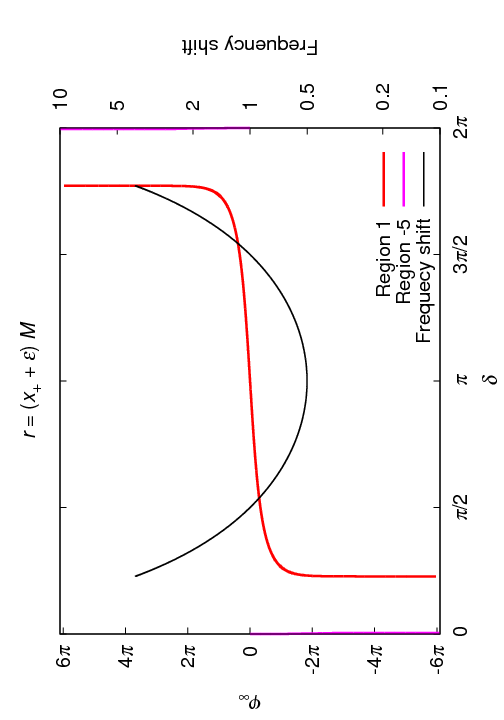}
\includegraphics*[angle=270,width=3.2in]{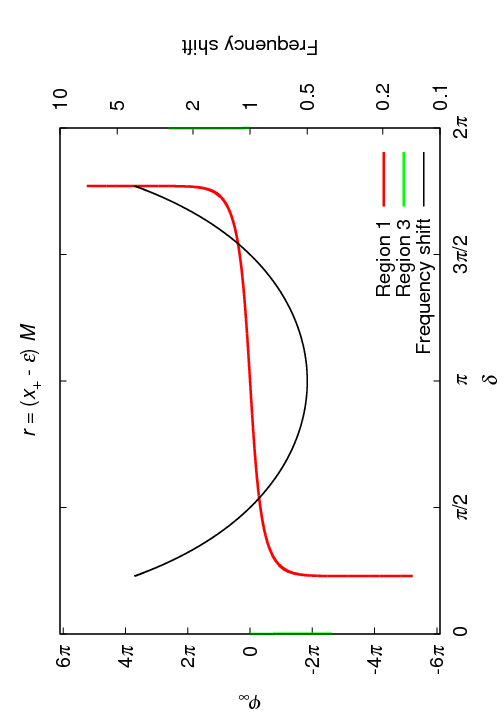}
\caption{As the observer reaches (left view) and enters (right view)
  the outer horizon, the size of region~-5 shrinks to 0 and disappears
  forever, and region~3 appears for the first time. Both regions,
  although of vanishingly small angular size, should be seen at an
  infinite blueshift which we did not depict here. Regarding region~1,
  the fact that it becomes part of the observer''s causal past when
  entering region~2 does not have any incidence on its visual
  aspect. Maximum blueshift of the edge of region~1 is large, although
  finite. }
\label{fig_cross_2}
\end{figure}

\begin{figure}[htbp]
\includegraphics*[width=3.2in]{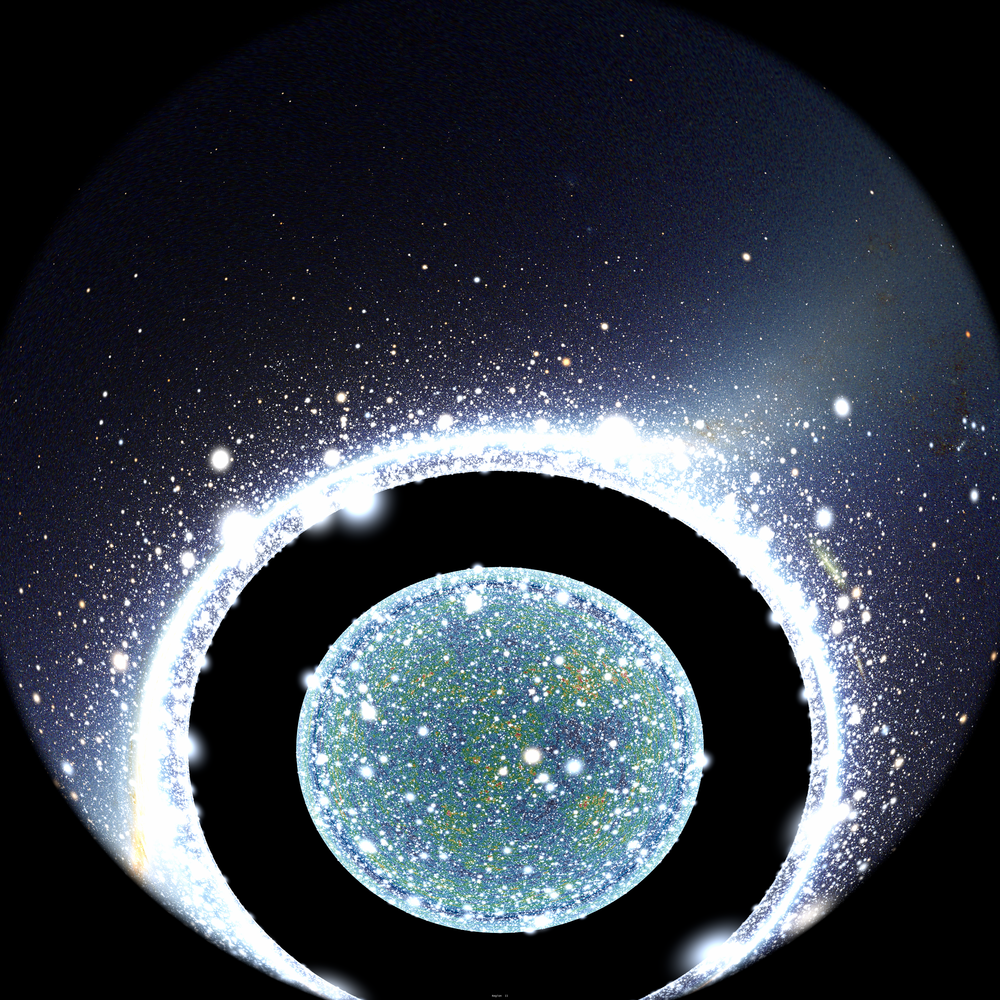}
\includegraphics*[width=3.2in]{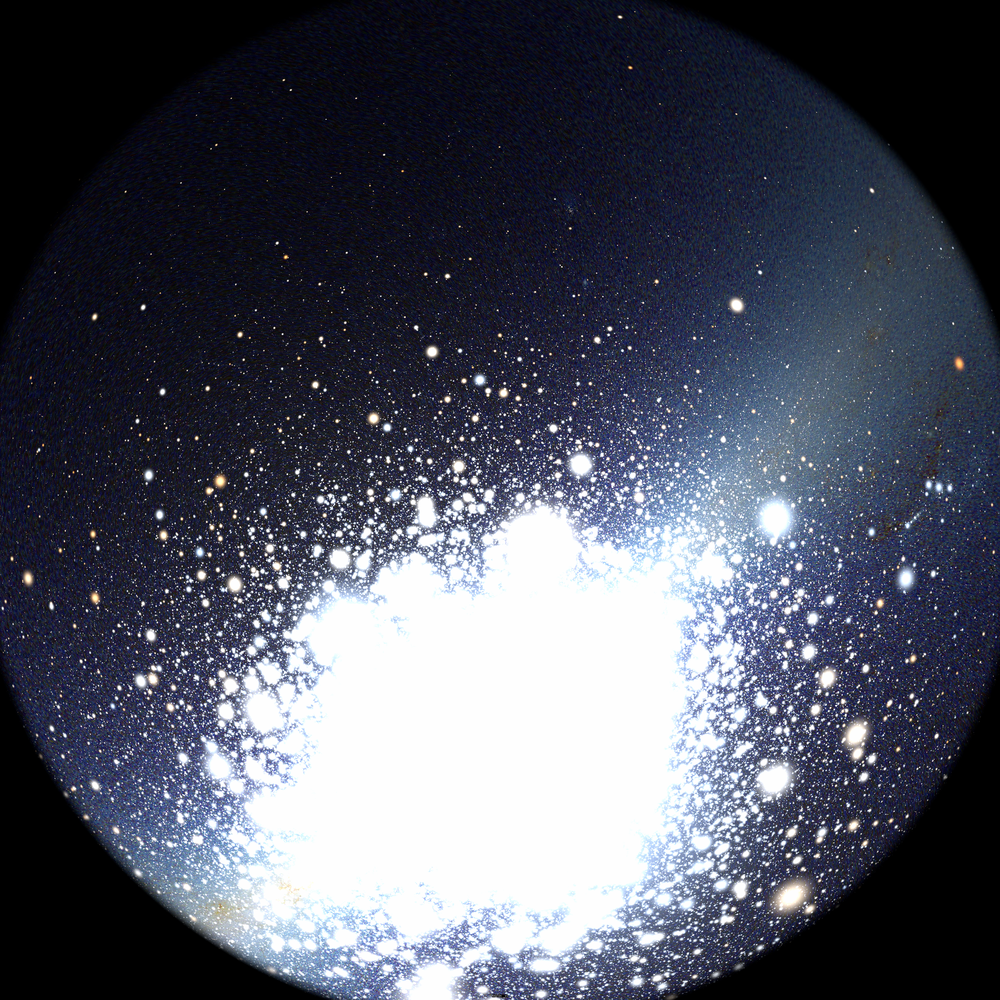}
\vskip 0.12cm
\includegraphics*[angle=270,width=3.2in]{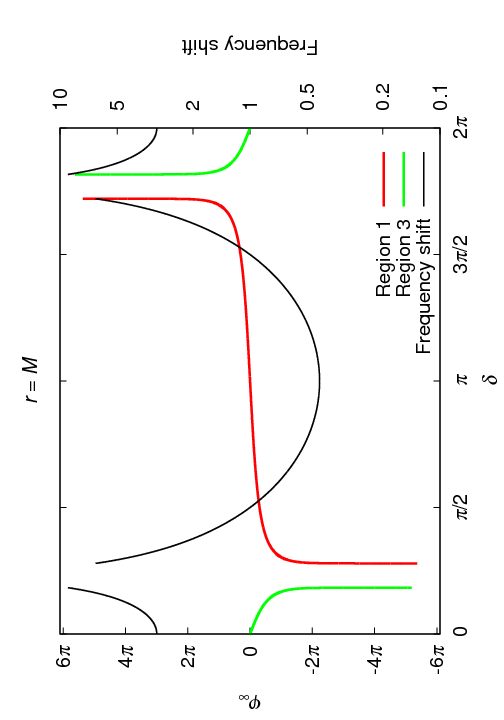}
\includegraphics*[angle=270,width=3.2in]{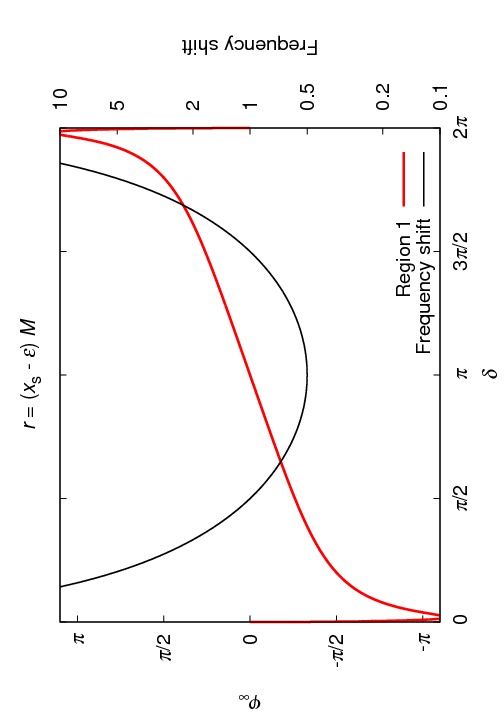}
\caption{As the observer travels along region~2, the size of region~3
  starts increasing (left image). Its edge is significantly
  blueshifted, although now by a finite amount. Is center is
  comparatively dimmer. For some values of $|Q|/M$, it might even be
  redshifted even though it seems to be in the direction the observer
  is heading to. Close to the inner horizon (right image), angular
  size of region~3 shrinks rapidly to 0 and then disappears, although
  temporarily. At this moment, a second image of region~1 appears in
  the same direction with an infinite blueshift and almost immediately
  fuses with the rest of region~1, thus encompassing almost all the
  celestial sphere with, toward the center of the coordinate system an
  enormous although finite blueshift. Notice Orion constellation at
  the right of right image. It was actually visible earlier, although
  harder to notice because it was redshifted and usually bright red
  $\alpha$~Orionis (Betelgeuse) was barely visible.}
\label{fig_cross_3}
\end{figure}

\begin{figure}[htbp]
\includegraphics*[width=3.2in]{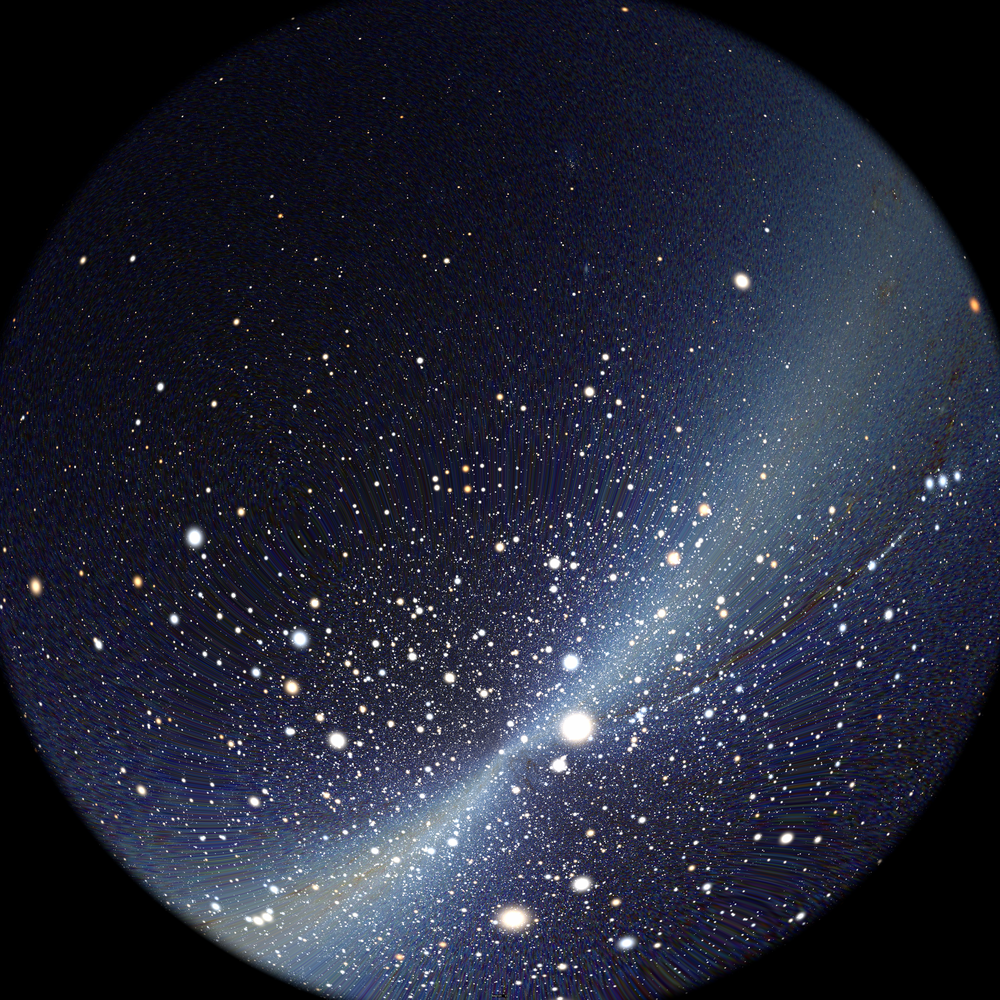}
\includegraphics*[width=3.2in]{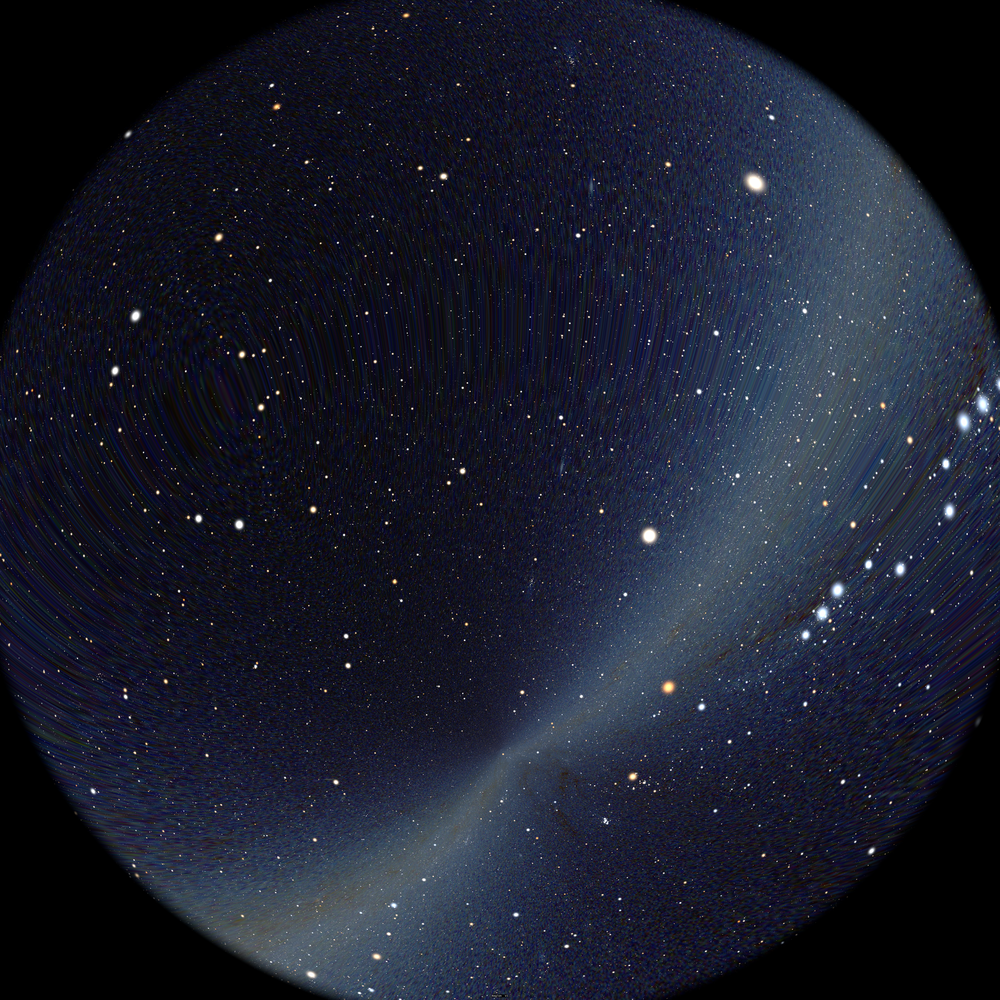}
\vskip 0.12cm
\includegraphics*[angle=270,width=3.2in]{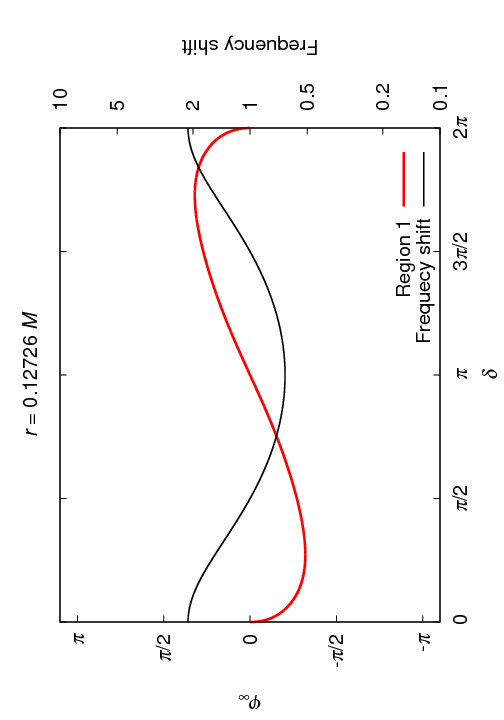}
\includegraphics*[angle=270,width=3.2in]{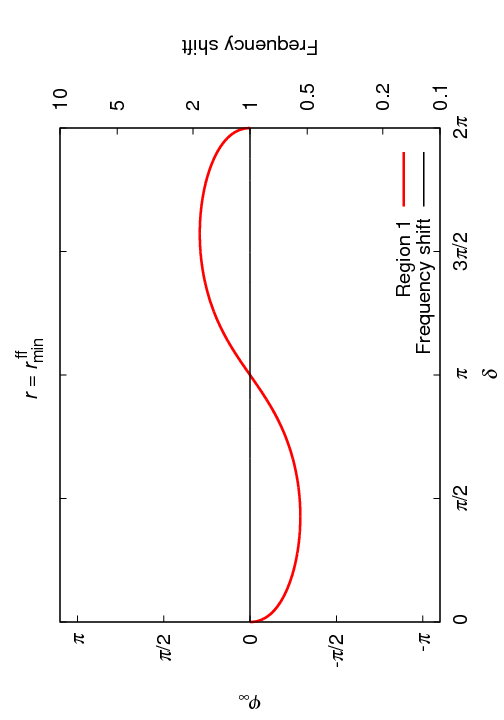}
\caption{Within region~6, region~1 occupies the whole celestial
  sphere, up to a unique direction corresponding to radial outgoing
  null geodesics originating from the singularity (which we assume
  does not emit anything). Moreover, in this region, the metric is
  static and one can define static observers. With respect to those,
  the infalling observer's velocity drops rapidly, significantly
  attenuating the frequency shift (left image). Meanwhile, a second,
  flipped view of region~1 spreads outward where region~3 has
  disappeared (it corresponds to the decreasing part of the deviation
  function below). The flipped view is initially of negligible angular
  size, but shows the whole celestial sphere (and actually several
  copies of it since the deviation function spans more than
  $2\pi$). Further, it increases in angular size but the part of
  region~1 it shows decreases. From a visual point of view, this
  translates into to the fact (not easy to notice with a few frames
  and more readily visible in a movie) that stars seem to disappear by
  pairs there. More explicitly, this is because the extrema of the
  deviation function are closer to $0$ and occur at angles further and
  further from $0$ (and $2 \pi$).  As the observer stops at
  $r_\MIN^\FF$ (right image) there is no frequency shift anywhere and
  the flipped view of region~1 occupies the same size
  ($2\pi$~steradians) as the normal view and both show a limited part
  of the initial celestial sphere. One can barely recognize a flipped,
  strongly elongated and partial view of Orion ($\alpha$ and
  $\beta$~Ori, plus Orion's belt) in the right of the picture. }
\label{fig_cross_4}
\end{figure}

\begin{figure}[htbp]
\includegraphics*[width=3.2in]{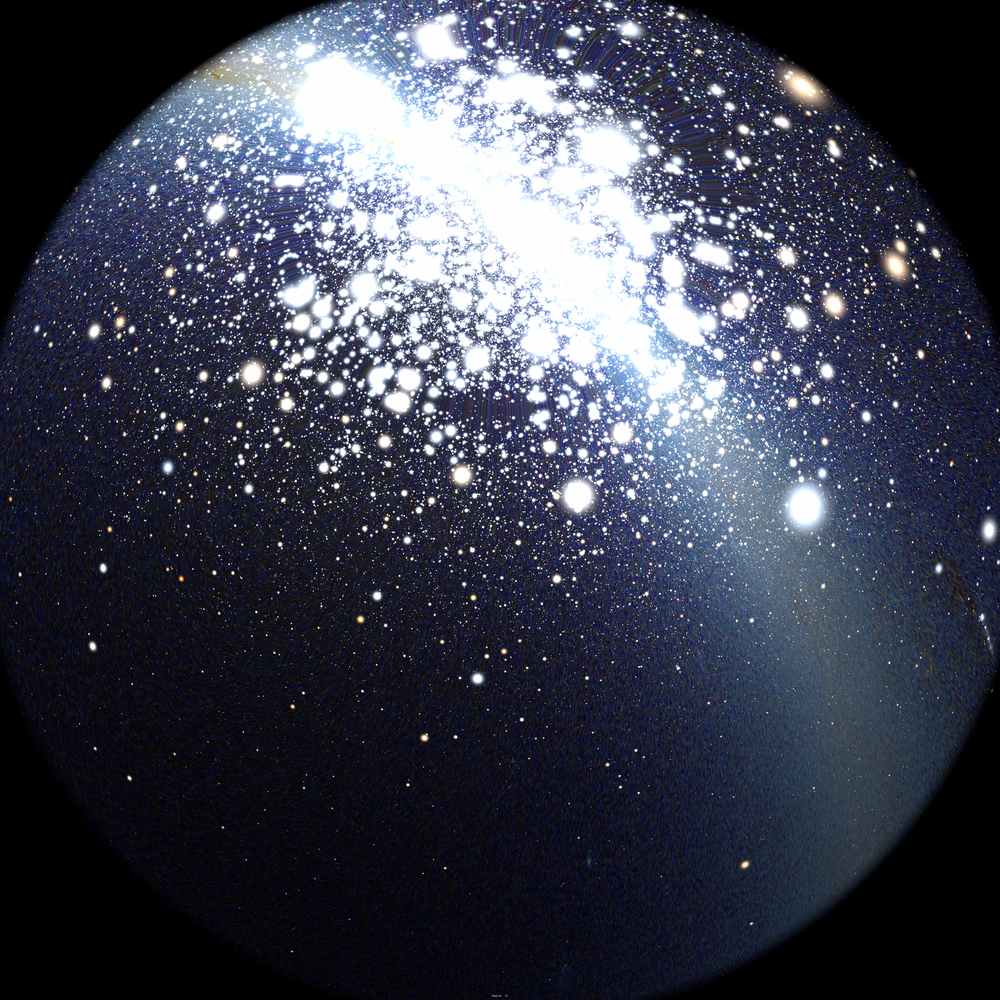}
\includegraphics*[width=3.2in]{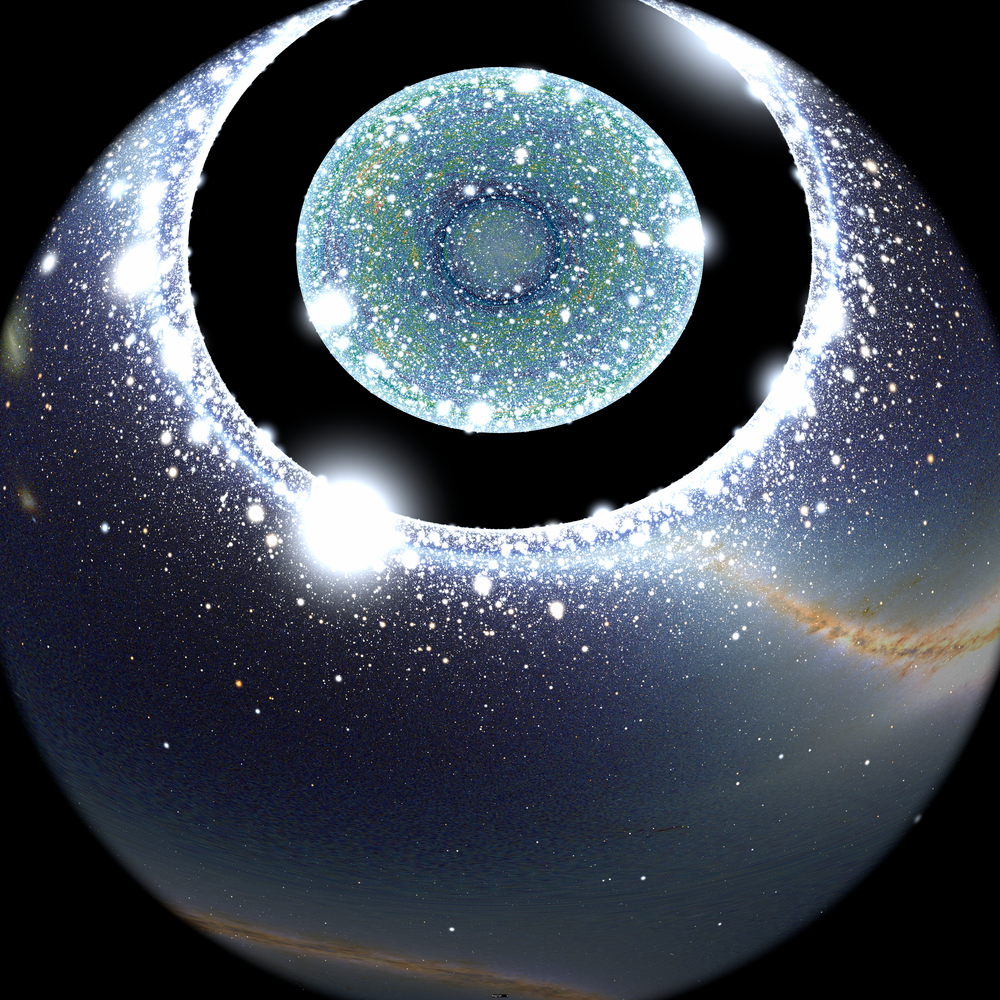}
\vskip 0.12cm
\includegraphics*[angle=270,width=3.2in]{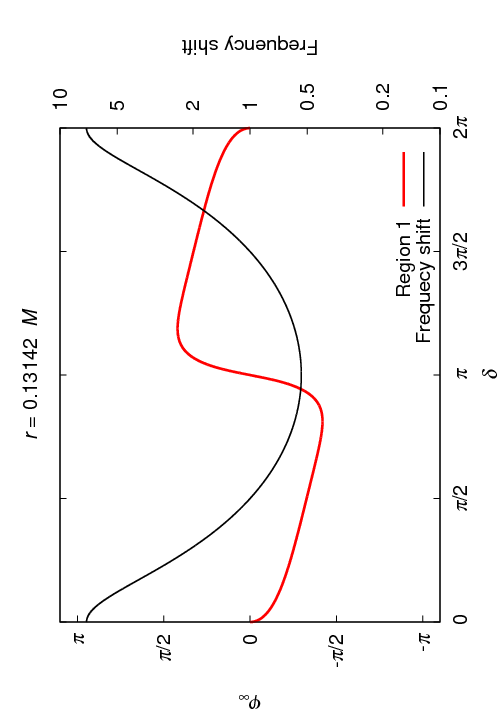}
\includegraphics*[angle=270,width=3.2in]{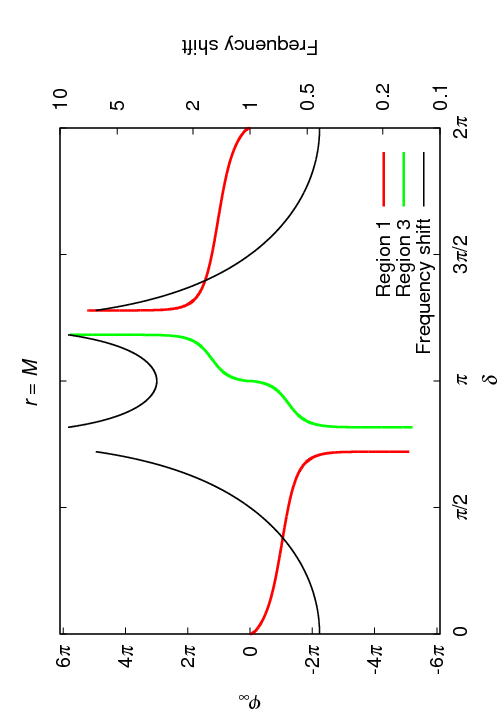}
\caption{As the observer starts the outgoing phase of the travel, we
  rotate the view by 90~degrees upward, so that the outgoing direction
  now appears in the upper part of the screen. The flipped view of
  region~1 now occupies a larger and larger part of the view (left
  image) and the blueshift of the decreasing unflipped view of
  region~1 strongly increases and even diverges along the ingoing
  direction at horizon crossing. Immediately after having crossed back
  the inner horizon, region~3 becomes visible again with a rapidly
  increasing angular size. Unflipped view of region~1 has completely
  disappeared and (flipped) view of region~1 is separated from that of
  region~3 by the dark shell. Incidentally, view of region~3 is flipped
  with respect to that of the ingoing phase.}
\label{fig_cross_5}
\end{figure}

\begin{figure}[htbp]
\includegraphics*[width=3.2in]{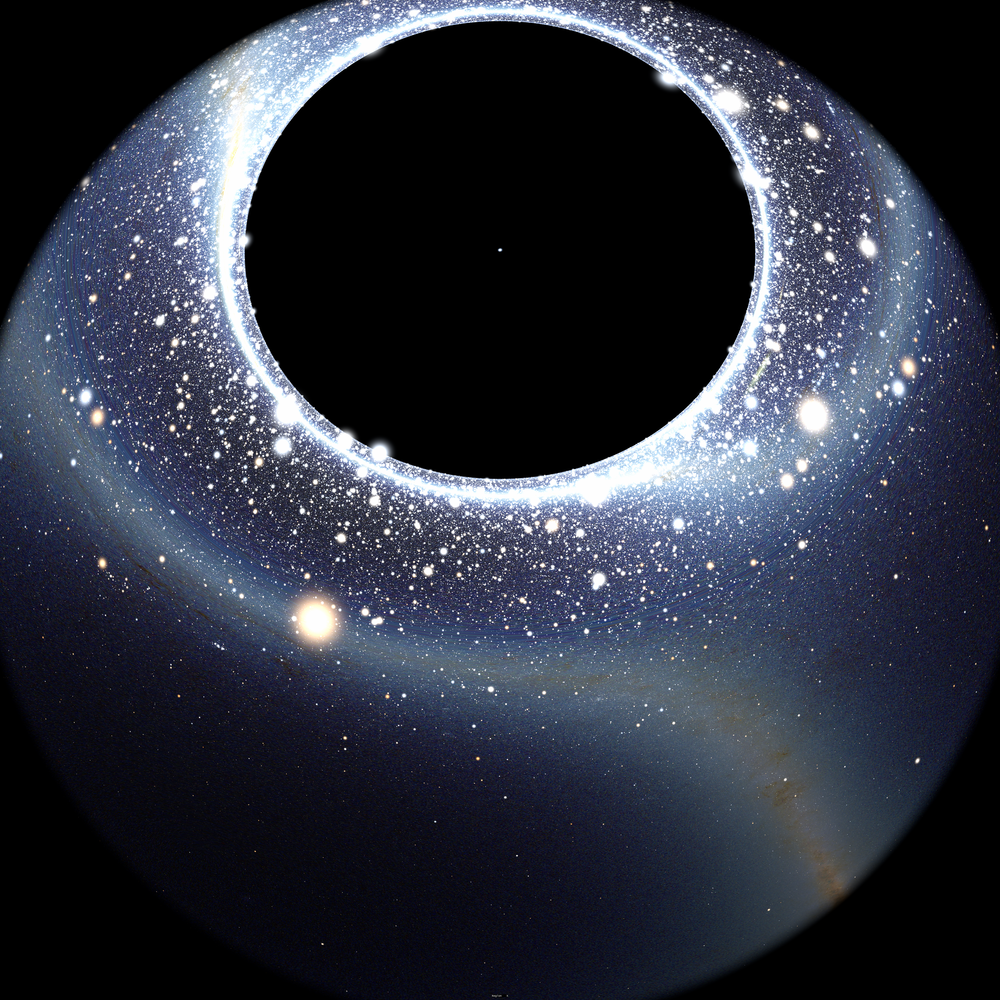}
\includegraphics*[width=3.2in]{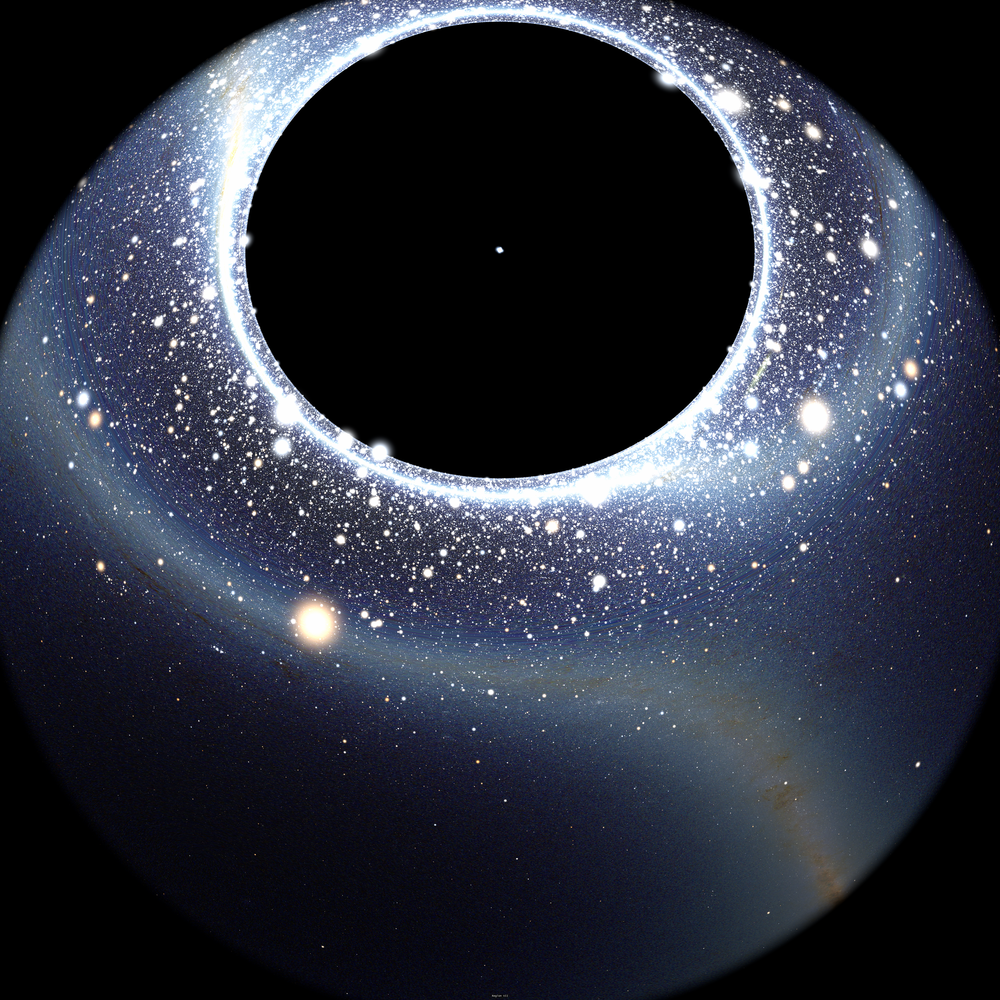}
\vskip 0.12cm
\includegraphics*[angle=270,width=3.2in]{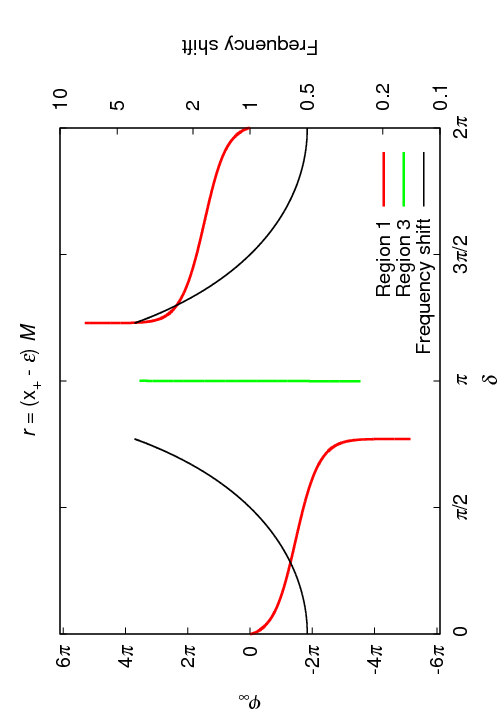}
\includegraphics*[angle=270,width=3.2in]{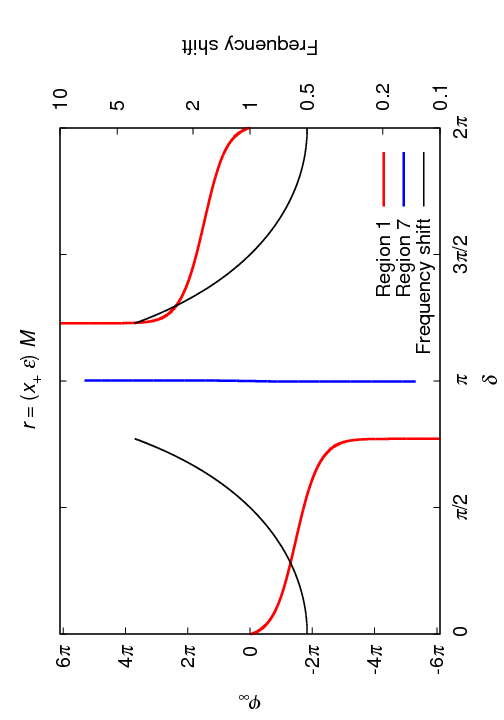}
\caption{When crossing the outer horizon, region~3 size shrinks to 0
  (left image) and is immediately replaced by a vanishingly small
  region~7 (right image), both of which seen with infinite blueshift
  (poorly depicted here). This horizon crossing does not induce
  significant changes in the aspect of region~1, which is blueshifted
  toward the dark shell and redshifted (both by a finite amount) in
  the opposite direction.}
\label{fig_cross_6}
\end{figure}

\begin{figure}[htbp]
\includegraphics*[width=3.2in]{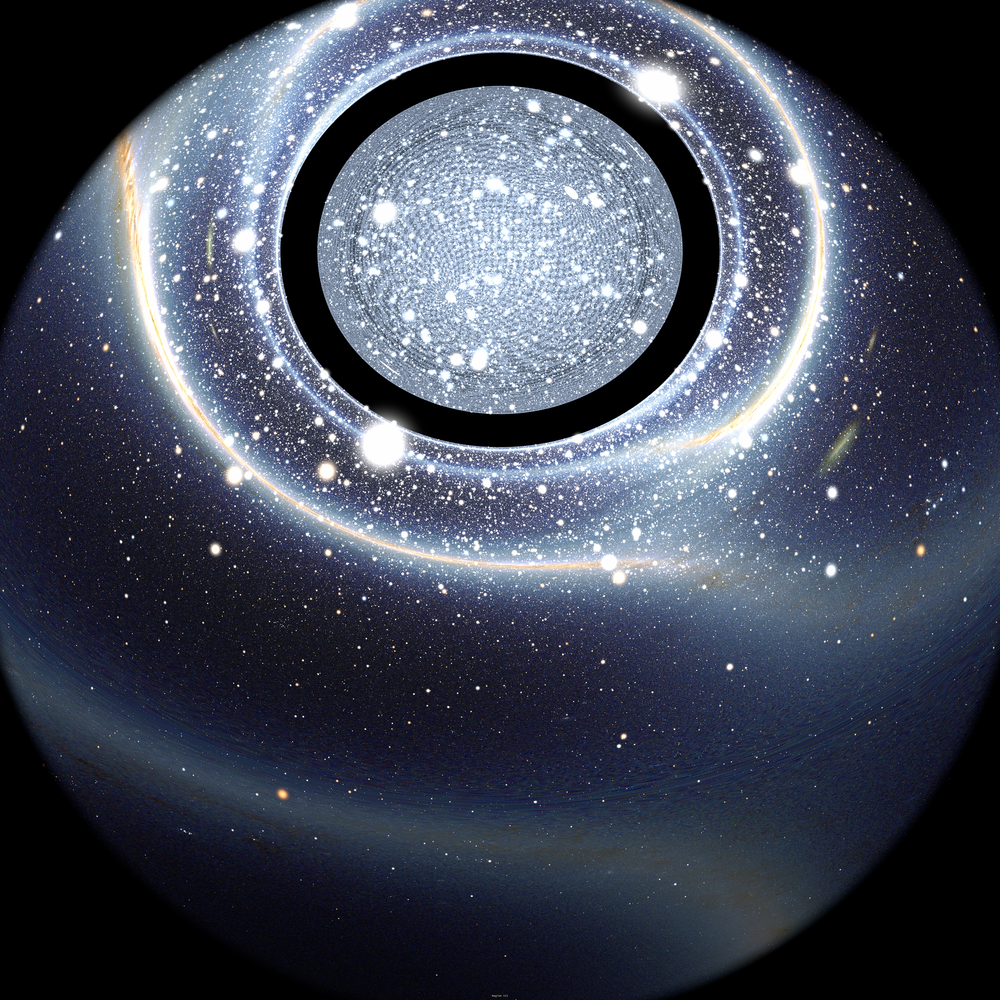}
\includegraphics*[width=3.2in]{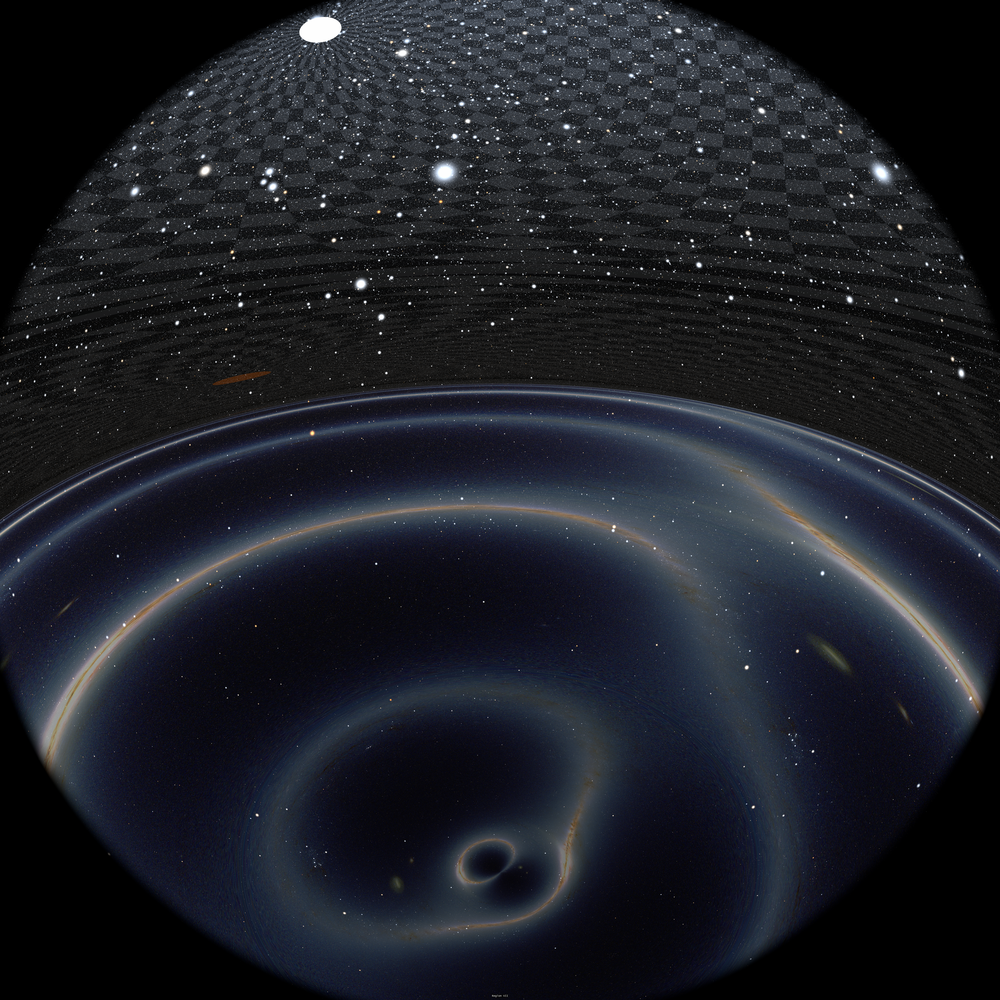}
\vskip 0.12cm
\includegraphics*[angle=270,width=3.2in]{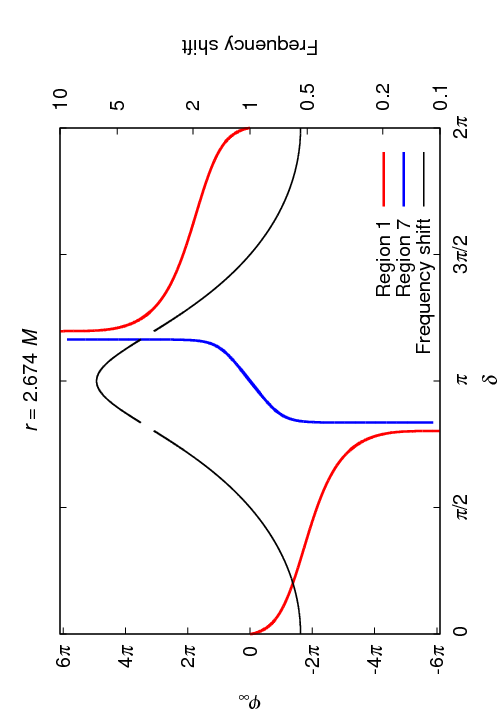}
\includegraphics*[angle=270,width=3.2in]{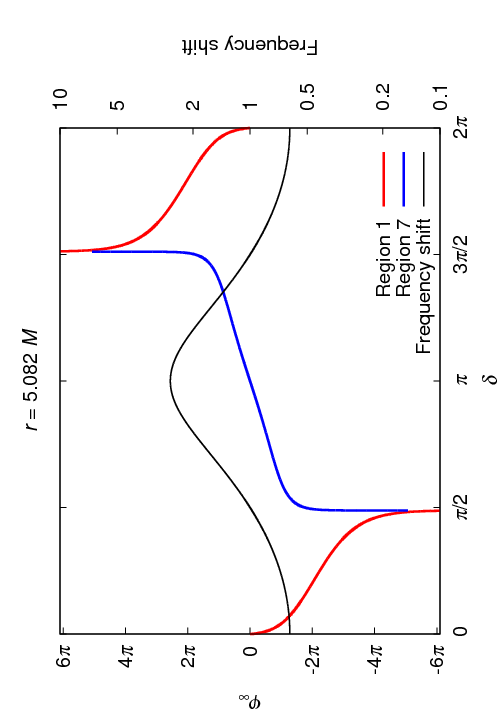}
\caption{As the observer cruises away from the wormhole, angular size
  of region~7 increases and its blueshift decreases, slightly faster
  at its edge as in its center. The dark shell gets thinner (left
  image) and disappears when $r > r_\EE$. The, region~1 angular size
  begins to decreases and region~7 slowly encompasses the whole view.}
\label{fig_cross_7}
\end{figure}

\begin{figure}[htbp]
\includegraphics*[width=4.8in]{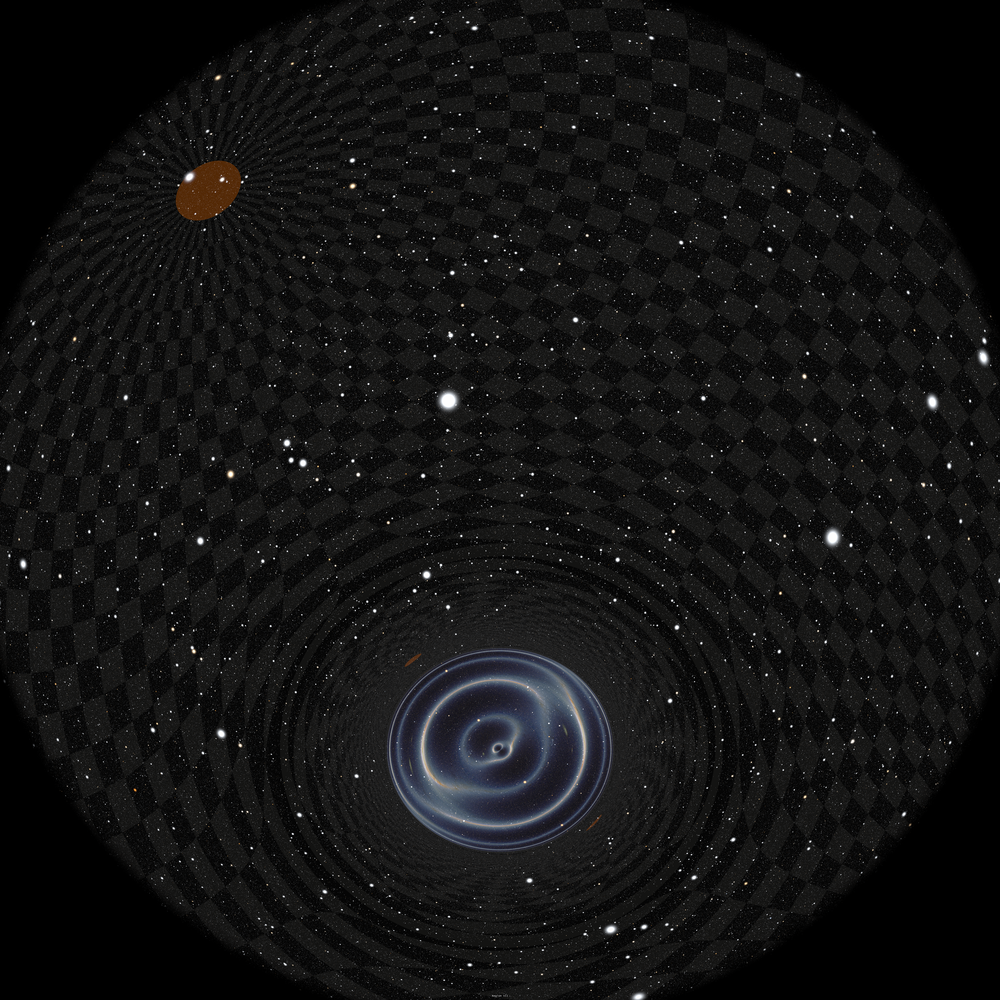}
\vskip 0.12cm
\includegraphics*[angle=270,width=3.2in]{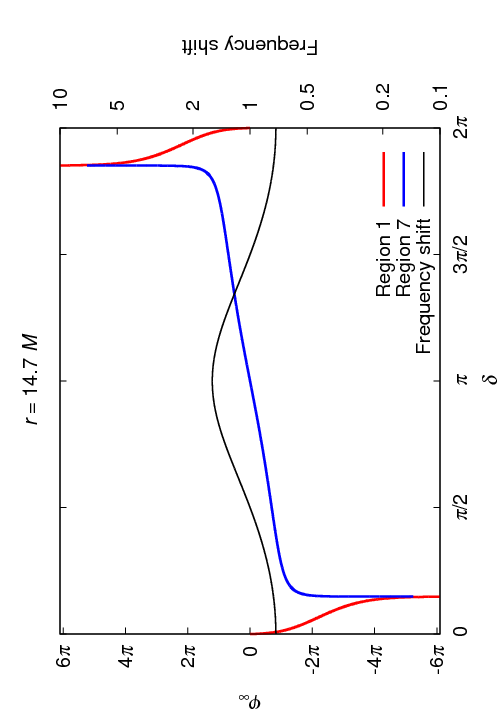}
\caption{After having rotated the view by 90~degrees downward so as to
  recover the same orientation as in the beginning of the sequence,
  one now sees region~1 within the silhouette of the wormhole. Its
  center still shows of flipped view of region~1, as well as several
  copies outward, resulting here in a series of ringlike structures
  which correspond to several copies of the (barely recognizable)
  Milky Way. }
\label{fig_cross_8}
\end{figure}

The sequence of the main features of the wormhole crossing is outlined
in fifteen frames shown in
Figs.\ref{fig_cross_1}--\ref{fig_cross_8}. For each frame, we included
the corresponding deviation functions as well as frequency
shift. Those are the exact values, independently of our (possibly
disputable) rendering choice.

\section{Conclusion}
\label{sec_conc}

In this paper, we have performed as thoroughly as possible a visual
description of the Reissner-Nordstr\"om metric, whose richness with
respect to the Schwarzschild metric induces several novel,
counter-intuitive effects that could hardly be guessed from a purely
formal study of the metric. In particular, we have described the
observable consequences of bounded null geodesics, which led to what
we called the dark shell phenomenon. We have also studied some
features that arise in the case the singularity is naked. More
importantly, we have made a careful, step-by-step analysis of a
wormhole crossing which we simulated with very high resolution
(8k$\times$8k) full dome frames.

This does not cover all the possibilities offered by the metric. For
example, the wormhole study focus on a single value of $|Q| / M$ and a
more quantitative study of what happens for different values of this
ratio was not performed. Also, we focused on an observer crossing the
wormhole following a radial geodesic, but non radial and even non
geodesic motion were not shown, the latter allowing for a difference
sequence of region seen during and after wormhole crossing. All these
aspects could possibly deserve being studied in a future work.

\appendix

\section{Proof of Eq.~(\ref{rshellmoins})}

Surprisingly, Eq.~(\ref{rshellmoins}) is not given in Chandrasekhar's
book~\cite{chandrasekhar83}. We therefore shall derive it here. We are
interested in finding the value of $r_\SHELL^- \neq r_\EXTR^+$ such
that $V_\NULL(r_\SHELL^-) = V_\NULL(r_\EXTR^+)$. This amounts to find
the solution of this equation~:
\begin{equation}
\frac{1}{r^4} (r^2 - 2 M r + Q^2)
 = \frac{1}{r_\EXTR^+} \left((r_\EXTR^+)^2 - 2 M r_\EXTR^+ + Q^2 \right) .
\end{equation}
Defining $u \eqdef 1 / r$ and $u_\EXTR^+ = 1 / r_\EXTR^+$, we obtain
\begin{equation}
\label{Vu}
0 =   (u - u_\EXTR^+) (u^3 Q^2 + u^2 (u_\EXTR^+ Q^2 - 2 M)
    + u (1 - 2 M u_\EXTR^+ + ^2 (u_\EXTR^+)^2)
    + u_\EXTR^+ - 2 M (u_\EXTR^+)^2 + Q^2 (u_\EXTR^+)^3 ) .
\end{equation}
We therefore need to find the root of the third order polynomial in
$u$ corresponding to the right part of right hand side of above
equation. However, we already know that equation
$V_\NULL(r) = V_\NULL(r_\EXTR^+)$ admits $u_\EXTR^+$ as a double root
because it is both a root of $V_\NULL(r) = V_\NULL(r_\EXTR^+)$ and a
local extremum (so that the derivative of either $V_\NULL(r)$ or
$V_\NULL(r) - V_\NULL(r_\EXTR^+)$ is zero there). Consequently,
because we have for this extremum
\begin{equation}
(r_\EXTR^+)^2 - 3 M r_\EXTR^+ + 2 Q^2 = 0 ,
\end{equation}
or, conversely in term of $u$
\begin{equation}
\label{rel_uextr}
1 - 3 M u_\EXTR^+ + 2 Q^2 (u_\EXTR^+)^2 = 0 ,
\end{equation}
we obtain from Eq.~(\ref{Vu})
\begin{equation}
0 =   (u - u_\EXTR^+)^2 (u^2 Q^2 + 2 u (u_\EXTR^+ Q^2 - M)
    + 1 - 4 M u_\EXTR^+ + 3 Q^2 (u_\EXTR^+)^2) .
\end{equation}
We therefore are left with the resolution of a second order equation,
which, going back to the $r$ variable, writes
\begin{equation}
r^2 (1 - 4 M u_\EXTR^+ + 3 Q^2 (u_\EXTR^+)^2) + 2 r (u_\EXTR^+ Q^2 - M) + Q^2 = 0 .
\end{equation}
Using again Eq.~(\ref{rel_uextr}), we have
\begin{equation}
r^2 (- M u_\EXTR^+ + Q^2 (u_\EXTR^+)^2) + 2 r (u_\EXTR^+ Q^2 - M) + Q^2 = 0 , 
\end{equation}
the solution of which is
\begin{equation}
r_\SHELL^- = \frac{M - u_\EXTR^+ Q^2 \pm 
                  \sqrt{  (u_\EXTR^+ Q^2 - M)^2
                        - Q^2 (- M u_\EXTR^+ + Q^2 (u_\EXTR^+)^2)}}
                 {- M u_\EXTR^+ + Q^2 (u_\EXTR^+)^2} .
\end{equation}
Multiplying everything by $u_\EXTR^+$ then gives
\begin{equation}
\frac{r_\SHELL^-}{r_\EXTR^+}
  = - 1 \pm \frac{\sqrt{  ((u_\EXTR^+)^2 Q^2 - M u_\EXTR^+)^2
                        - Q^2 (u_\EXTR^+)^2 (- M u_\EXTR^+ + Q^2 (u_\EXTR^+)^2)}}
                 {- M u_\EXTR^+ + Q^2 (u_\EXTR^+)^2} .
\end{equation}
The root we are interested in is the positive one, so that this reduces to 
\begin{equation}
\frac{r_\SHELL^-}{r_\EXTR^+}
  = - 1 + \sqrt{1 - \frac{Q^2 (u_\EXTR^+)^2 (- M u_\EXTR^+ + Q^2 (u_\EXTR^+)^2)}
                         {(- M u_\EXTR^+ + Q^2 (u_\EXTR^+)^2)^2}} .
\end{equation}
Expanding the term within the square root then leads to
\begin{equation}
\frac{r_\SHELL^-}{r_\EXTR^+}
  = - 1 + \sqrt{\frac{M u_\EXTR^+}
                     {M u_\EXTR^+ - Q^2 (u_\EXTR^+)^2}} ,
\end{equation}
which can further be simplified into the compact form
\begin{equation}
\frac{r_\SHELL^-}{r_\EXTR^+}
  = - 1 + \frac{1}{\sqrt{1 - \frac{Q^2}{M r_\EXTR^+}}} .
\end{equation}
This indeed corresponds to Eq.~(\ref{rshellmoins}). As a last note, we
add that for low values of $|Q|$, the potential is extremely steep
close to $r_\SHELL^- \simeq r_-$ since one has
$V_\NULL'(r_\SHELL^-) \simeq (V_\NULL(r_\SHELL^-) - V_\NULL(r_-)) /
(r_\SHELL^- - r_-) \simeq O(L^2 M^5/Q^8)$,
the last result coming from Eqns.~(\ref{dl_xm},
\ref{dl_xs}). Consequently, numerical errors sometimes give to
$V_\NULL(r_\SHELL^-)$ a value that significantly differs from the
correct one, i.e., $\sim L^2 / 54 M^2$ for low values of $|Q|$ (see
Eq.(\ref{V_null_re})).

\end{document}